\documentclass[letterpaper,11pt,oneside]{book}

\font\Bxfnt=cmsy10 scaled\magstep1
\def\Box{\mathord{\hbox{\Bxfnt\char116}\llap{\Bxfnt\char117}}}

\font\titlefont=cmbx10 scaled \magstep3

\begin{document}
\input{epsf}

\begin{titlepage}
\vspace*{-0.5in}
\begin{flushright}
gr-qc/9805037\hspace{0.5in}\\
April 23, 1998\hspace{0.5in}\\
\end{flushright}
\begin{center}
\vskip 0.25in
{\titlefont Quantum Inequality Restrictions on Negative\\
\vspace{0.1in}
Energy Densities in Curved Spacetimes\\}
\vskip 0.25in 
 A dissertation submitted by\\
{\bf Michael John Pfenning}\footnote{email: mitchel@cosmos2.phy.tufts.edu}\\
In partial fulfillment of the requirements\\
for the degree of Doctor of Philosophy in the \\
Department of Physics and Astronomy,\\
 TUFTS UNIVERSITY\\
Medford,  Massachusetts  02155\\
\vskip 0.3in
Advisor: {\bf Lawrence H. Ford}\footnote{email: ford@cosmos2.phy.tufts.edu}\\
\end{center}
\vskip 0.3in
\begin{center}
{\bf Abstract}
\end{center}
In quantum field theory, 
there exist states in which the expectation value of the energy
density for a quantized field is negative.  These negative energy
densities lead to several problems such as the failure of
the classical energy conditions, the production of closed timelike
curves and faster than light travel, violations of the second law
of thermodynamics, and the possible production of naked singularities.

Although quantum field theory introduces negative energies, it
also provides constraints in the form of quantum inequalities
(QI's).  These uncertainty principle-type relations limit the
magnitude and duration of any negative energy.
We derive a general form of the QI on the energy density 
for both the quantized scalar and electromagnetic fields in
static curved spacetimes. In the case of the scalar field,
the QI can be written as the Euclidean wave operator acting
on the Euclidean Green's function.  Additionally, a small distance
expansion on the Green's function is used to derive the QI in the
short sampling time limit.  It is found that the QI in this limit
reduces to the flat space form with subdominant correction terms
which depend on the spacetime geometry.   

Several example spacetimes are  studied in which exact forms of
the QI's can be found.  These include the three- and four-dimensional
static Robertson-Walker spacetimes, flat space with perfectly reflecting
mirrors,  Rindler and static de~Sitter space, and the spacetime outside
a black hole.  In all of the above cases, we find that the quantum
inequalities give  a lower limit on how much negative energy may
be observed relative to the vacuum energy density of the spacetime.
For the particular case of the black hole, it is found that the
quantum inequality on the energy density is measured relative to the
Boulware vacuum.  

Finally, the application of the quantum inequalities to the
Alcubierre warp drive spacetime leads to strict constraints
on the thickness of the negative energy region needed to maintain
the warp drive.  Under these constraints, we  discover that
the total negative energy required exceeds the total mass of the
visible universe by a hundred billion times, making warp drive
an impractical form of transportation.

\end{titlepage}
\pagenumbering{roman}
\pagestyle{empty}

\vspace*{.75in}
\begin{center}
{\Large \bfseries Acknowledgments}
\end{center}

There are many people to whom I would like to express my gratitude.
First and foremost is my advisor and mentor, Dr.~Lawrence H.~Ford.
Without his encouragement, patience and support, I would never have
been able to complete this dissertation.  A special thanks also goes
to Dr.~Thomas A.~Roman for his unfailing advice and encouragement. 
I also wish to thank Dr.~Allen Everett for all of the input over the
course of my research.  The Institute of Cosmology and the Department
of Physics at Tufts University also deserve credit for making all of
my research possible.  I would also like to thank all of the 
people in the Department Office who's humor and assistance  were
invaluable.   

On a more personal level, an extra special
thank you goes to Dorothy Drennen, who gave up her personal time to
edit this manuscript from cover to cover. The same heart felt gratitude
goes to Vicki Ford, who was also there to keep me on course.
Finally, I would like to thank my family for all of their love throughout
the years. A special wish goes to my two sons, Henry and Ren\'{e}, who
have brought so much joy into my life.  To Fran, Herbert, Heidi,
John, and my Grandmother Mary, I dedicate this manuscript. 

The financial support of NSF Grant No.~Phy-9507351 and the
John F.~Burlingame Physics Fellowship Fund is gratefully acknowledged.
  
\newpage
\pagestyle{headings}
\tableofcontents
\listoffigures
\newpage
\pagenumbering{arabic}

\chapter{Energy Conditions in General Relativity}

\section{The Classical Energy Conditions}

The most powerful and intriguing tool in cosmology is Einstein's equation,
\begin{equation}
G_{\mu\nu} \equiv R_{\mu\nu}-{1\over 2}R g_{\mu\nu} = 8\pi G T_{\mu\nu}.
\end{equation}
Here we have chosen the spacetime metric to have a $(-,+,+,+)$ signature
and the definitions of the Riemann and Ricci tensors in the convention
of Misner, Thorne and Wheeler \cite{MTW}. Einstein's
equation relates the spacetime geometry on the left with the matter
source terms on the right.  The equation is interesting in
that it admits a large number of solutions. If there are no physical
constraints on the metric or the stress-tensor, then Einstein's equation
is vacuous since any metric would
be a ``solution'' corresponding to some distribution of the stress-energy.
We could then have ``designer  spacetimes'' which could exhibit any 
behavior that one likes.  Therefore, if we can place any limitations upon
the equation, or on terms in the equation, we could limit the search
for physically realizable solutions.

The most trivial condition comes from the Bianchi identities, that
the covariant divergence of the Einstein tensor must vanish:
\begin{equation}
\nabla_\mu G^{\mu\nu} = 0.
\end{equation}
From Einstein's equation, this leads to the condition on the 
stress-energy tensor of
\begin{equation}
\nabla_\mu T^{\mu\nu} = 0.
\end{equation}
This is the covariant form of the conservation of energy in general
relativity.

The next condition which is generally considered reasonable for all
classical matter is that the energy density seen by an observer
should be non-negative.  Consider an observer who moves along the
geodesic $\gamma^\mu (\tau)$. Then his four-velocity,
\begin{equation}
u^\mu(\tau) \equiv {d\over d\tau}\gamma^\mu (\tau),
\end{equation}
is everywhere timelike and tangent to the curve $\gamma^\mu(\tau)$.
In the observer's frame, the statement that the energy density
should be non-negative is
\begin{equation}
T_{\mu\nu}u^\mu(\tau)u^\nu(\tau) \geq 0, \qquad\mbox{ for all timelike } 
u^\nu(\tau).\label{eq:WEC}
\end{equation}
This is typically called the {\it Weak Energy Condition} (WEC).
This condition was an important assumption in the proof of the
singularity theorem developed by Penrose \cite{Penrose}.
In the early part of this century, it was known that a spherically
symmetric object that underwent gravitational collapse would form
a black hole, with a singularity at the center.  However, relativists
were unsure if the same would be true for a more generic distribution
of matter.  Penrose's original singularity theorem showed
that the end product of gravitational collapse would always be a
singularity without assuming any special symmetries of the matter
distribution.
The WEC is used in the proof of the singularity theorem in the
Raychaudhuri equation to ensure focusing of a congruence of null
geodesics.  An additional assumption  sometimes used in the proof
of the singularity theorems is the {\it Strong Energy Condition} (SEC),
\begin{equation}
\left(T_{\mu\nu} - {1\over 2} T g_{\mu\nu} \right)
u^\mu(\tau)u^\nu(\tau) \geq 0, \qquad\mbox{ for all timelike } u^\nu(\tau).
\label{eq:SEC}
\end{equation} 
Here $T = {T_\mu}^\mu$ is the contraction of the stress-tensor.
If the SEC is satisfied, then it ensures that gravity is always an
attractive force.  The SEC is ``stronger'' than the WEC only in the
sense that it is a more restrictive condition, and therefore it is
much easier to violate the SEC than  the WEC. 

Immediately following the successes of Penrose's singularity
theorem for black holes, Hawking \cite{Hawking,Ha&El} developed a similar
theorem for the open Friedman-Robertson-Walker universe which showed that
there had to exist an essential, initial singularity from which the
universe was born.  Nearly thirty years later, singularity theorems
are still being proven for various spacetimes.  For
example, Borde and Vilenkin have proven the existence of initial
singularities for inflationary cosmologies and for certain classes
of closed universes \cite{Borde94,Bo&Vi94}.

For completeness, we also mention two other classical energy
conditions.  The first is the {\it Null Energy Condition} (NEC), 
\begin{equation}
T_{\mu\nu}K^\mu K^\nu \geq 0, \qquad\mbox{ for all } K^\nu,
\label{eq:NEC}
\end{equation}
where $K^\mu$ is the tangent to a null curve.  The NEC follows
by continuity from both the WEC~(\ref{eq:WEC}) and the SEC~(\ref{eq:SEC})
in the limit when $u^\mu$ becomes a null vector.
The second condition is a little different from the previous ones.
Let us assume, as we did above, that $u^\nu$ is a future-directed
timelike vector. It follows that the product, $-{T^\mu}_\nu u^\nu$,
should also be a future-directed timelike or null vector.  This is
known as the {\it Dominant Energy Condition} (DEC) and can be
interpreted to mean that the speed of energy flow of matter
should always be less than the speed of light. 

We see that the energy conditions at the classical level are an
extremely useful tool that have yielded some remarkable results in 
the form of the singularity theorems and positive mass theorems.
However, many of the energy conditions are doomed to failure when 
quantum field theoretic matter is introduced as the source of gravity.
A first approximation to a ``quantum theory of gravity'' is achieved
via the semiclassical equation
\begin{equation}
G_{\mu\nu} = 8\pi G \langle T_{\mu\nu} \rangle .
\label{eq:semiclassical}
\end{equation}
We are still treating the spacetime as a ``smooth'' classical background,
but we have replaced the classical stress-tensor with its renormalized
quantum expectation value.  It is presumed that the field equation should
be valid in the test field limit where the mass and/or the energy of the
quantized field is not so large as to cause significant back-reaction
on the spacetime. However, the replacement of the classical stress-tensor
with its quantum expectation value is not without difficulty.  
At the same time that Penrose and Hawking were introducing the 
singularity theorems, Epstein, Glaser and Jaffe \cite{Epstein} were
demonstrating that the local energy density in a quantized field theory
was not always positive definite.  We will see in Chapter~\ref{Chap:QFT}
that it is possible to construct specific quantum states in which the
energy density is negative along some part of an observer's worldline.
The simplest example, discussed in
Section~\ref{sec:vac+2particle}, is a state in which the particle
content of the theory is a superposition of the vacuum plus two
particle state.  Another example is  the squeezed states of light
which have been produced and observed experimentally \cite{Wu_etal},
although it is unclear at this time whether it is possible to 
measure the negative energy densities that exist for these two cases.
However, there is the interesting case of the Casimir vacuum energy
density for the quantized electromagnetic field between two perfectly 
conducting parallel plates \cite{Casimir}. Here the vacuum energy
between the plates is found to be a constant and everywhere negative.
Indirect effects of the negative vacuum energy, in
the form of the Casimir force which causes the plates to be
attracted to one another, has been measured experimentally 
\cite{Lamo97,Onofrio}.  
It has even been suggested that the Casimir force may play a role
in the measurement limitations in electron force microscopy \cite{Sidles}.

Related to the Casimir vacuum energy is the vacuum energy of a
quantized  field in curved spacetime, which we will discuss further
in Section~\ref{sec:vac_energy}.  An example is the static closed
universe, where the vacuum energy for the conformally coupled field
is positive \cite{Ford75b,Ford76}, while that of the minimally coupled
field is negative \cite{Elizalde}.  Instances of non-zero vacuum energies
in the semiclassical theory of gravity seem to be extremely widespread.
The most well known, and probably most carefully studied, is the 
vacuum energy around a black hole. Hawking has shown that this causes
the black hole to evaporate, with a flux of positive energy particles
being radiated outward.  In this case  both the
WEC and the NEC are violated \cite{F&Ro92,F&Ro96b,Viss96a,Viss96b,
Viss96c,Viss97a,Viss97c}. 

In terms of a consistent mathematical theory of semiclassical gravity,
the existence of negative energy densities is not necessarily a problem.
On the contrary, it could be viewed as beneficial because it would allow
spacetimes in which time travel would be possible, or where mankind
could construct spacecraft which could warp spacetime around them in
order to travel around the universe faster than the speed of light.
Even wormholes would be allowed as self-consistent solutions to 
Einstein's equation.

Objections to negative energies are raised for several reasons such
as violations of the second law of thermodynamics,  the possibility
of creating naked singularities by violating cosmic censorship, and
the loss of the positive mass theorems.   On the more philosophical
side, if time travel, space or time warps, wormholes and any other
exotic solution that physicists can dream up were allowed, then we
would have to deal with a new set of complexities, such as problems
with causality and the notion of free will.  Visser's \cite{Visser}
description of four possible schools of thought on how to build an
entirely self-consistent theory of physics  with regard to time travel
is quoted here:
\vspace{-0.25in}
\begin{quotation}
\noindent
\begin{enumerate}
\item	{\bf The radical rewrite conjecture:} One might make one's
		peace with the notion of time travel and proceed to rewrite
		all of physics (and logic) from the ground up. This is a 
		very painful procedure, not to be undertaken lightly.
		
\item	{\bf Novikov's consistency conjecture:} This is a slightly
		more modest way of making one's peace with the notion of time
		travel.  One simply {\it asserts} that the universe is consistent,
		so that whatever temporal transpositions and trips one undertakes,
		events must conspire in such a way that the overall result is
		consistent.  (A more aggressive version of the consistency
		conjecture attempts to {\it derive} this ``principle of
		self-consistency'' from some appropriate micro-physical assumptions.)
		
\item	{\bf Hawking's chronology protection conjecture:}  Hawking has
		conjectured that the cosmos works in such a way that time travel
		is completely and utterly forbidden.  Loosely speaking, ``thou
		shalt not travel in time,'' or ``suffer not a time machine exist.''
		The chronology protection conjecture permits space\-warps/worm\-holes
		but forbids time\-warps/time machines.
		
\item	{\bf The boring physics conjecture:}  This conjecture states,
		roughly: ``A pox upon all nonstandard speculative physics.
		There are no worm\-holes and/or space\-warps.  There are no time
		ma\-chines/time\-warps.  Stop speculating.  Get back to something
		we know something about.''
\end{enumerate}
\end{quotation}

Besides the philosophical reasons, there is one particularly important
mathematical result of the classical theory of gravity that would fail
if negative energies did exist and were widespread.  These are the
singularity theorems which establish the existence of singularities
in general relativity, particularly for gravitational collapse and at
the initial ``big bang.''  

\section{The Averaged Energy Conditions}

In the 1970's and 1980's, several approaches were proposed to study
the extent to which quantum fields may violate the weak energy
condition, primarily with the hope of restoring some, if not all,
of the classical general relativity results. In 1978, Tipler suggested
averaging the pointwise conditions over an observer's geodesic 
\cite{Tipler}. Subsequent work by various authors \cite{ Chicone, 
Galloway, Borde87, Roman86, Roman88, Klinkhammer, Wald91} eventually
led to what is presently called the {\it Averaged Weak Energy Condition}
(AWEC), \index{averaged weak energy condition}
\begin{equation}
\int_{-\infty}^\infty \langle T_{\mu\nu}u^\mu u^\nu\rangle
  d\tau \geq 0, \label{eq:awec}
\end{equation}
where $u^\mu$ is the tangent to a timelike geodesic and $\tau$ is
the observer's proper time.  Here the energy density of the observer
is not measured at a single point, but is averaged over the entire
observer's worldline. Unlike the WEC, the AWEC allows for the
existence of negative energy densities so long as there is
compensating positive energy elsewhere along the observer's worldline.  

There is also an {\it Averaged Null Energy Condition} (ANEC),
\begin{equation}
\int_{-\infty}^\infty \langle T_{\mu\nu}K^\mu K^\nu \rangle d\lambda \geq 0,
\end{equation} 
where $K^\mu$ is the tangent to a null geodesic, and $\lambda$ is the
affine parameter along that null curve. Roman \cite{Roman86, Roman88}
has demonstrated in a series of papers that the
singularity theorem developed by Penrose will still hold if the
weak energy condition is replaced by the averaged null energy
condition. Therefore, within the limits of the semiclassical theory
there exists the possibility of recovering some, if not all, of the
classical singularity theorems. 

The ANEC is somewhat unique in that it follows directly from
quantum field theory.  Wald and Yurtsever \cite{Wald91} showed
that the ANEC could be proven to hold in two dimensions for any 
Hadamard state along a complete, achronal null geodesic.  They
also showed that the same result held in four-dimensional
Minkowski spacetime as well.  The results in two-dimensional
flat spacetime were confirmed by way of an entirely different
method by Ford and Roman \cite{F&Ro95}.  The question of 
whether back-reaction
would enforce the averaged null
energy condition in semiclassical gravity has been addressed
recently by Flanagan and Wald \cite{Fl&W96}.  Their work seems
to indicate that if violations of the ANEC do occur for
self-consistent solutions of the semiclassical equations, then
they cannot be macroscopic, but are limited to the Planck scale.
On such a tiny scale, we do not know if the semiclassical theory 
is reliable. It is expected that at the Planck scale the effects
of a full quantum theory of gravity would be important.

At first the averaged energy conditions seem like a vast
improvement.  If the averaged energy conditions hold in
general spacetimes, there is hope that we can restore the singularity
theorems and gain a new class of energy conditions better
suited for dealing with quantum fields.  However, it is
straightforward to show that both the AWEC and the ANEC are violated
for any observer when the quantum field is in the vacuum state in
certain spacetimes. The simplest example is an observer who sits at
rest between two uncharged, perfectly conducting plates.  Casimir
\cite{Casimir} showed for the quantized electromagnetic field that
the vacuum energy  between the plates is 
negative. Therefore, over any time interval which we average,
the AWEC  will fail.  Similar effects can occur in curved spacetime
for the vacuum state.  Examples are the various vacuum states outside
a black hole \cite{F&Ro96b,Viss96a,Viss96b,Viss96c,Viss97a,Viss97c}.
It appears that a difference between an arbitrary state and the
vacuum state often obeys a type of AWEC, which can be written as
\begin{equation}
\int_{-\infty}^\infty \left[\langle\psi| T_{\mu\nu}u^\mu u^\nu |\psi\rangle
-\langle 0_c | T_{\mu\nu}u^\mu u^\nu | 0_c \rangle \right] d\tau \geq 0.
\label{eq:QAWEC} 
\end{equation}
This equation, which has been called the {\it Quantum Averaged Weak
Energy Condition} (QAWEC) is derived in Section~\ref{sec:QAWEC}.  It
holds for a certain class of observers in any static spacetime
\cite{F&Ro95,Yurtsever,Pfen97a,Pfen98a}
and is the result of the infinite  sampling time limit of the quantum
inequalities, which will be defined below.

Another aspect of the averaged energy conditions is that there are
a number of ``designer'' spacetimes where they are necessarily
violated. Three examples of particular interest are:  the traversable 
wormhole \cite{Morris88a,Morris88b}, the Alcubierre ``warp drive'' 
\cite{Alcu94,Pfen97b}, and the Krasnikov tube \cite{Kras95,E&Ro97}.
All three require negative energy densities 
to maintain the spacetime.  Likewise, all three have similar problems
in that slight modifications of these spacetimes would lead to closed
timelike curves, and would allow backward time travel \cite{Ever94}.
Thus, these spacetimes may be entirely unphysical in the first place.
It has also been shown by various researchers that if warp drives and
wormholes exist, then they must be limited to the Planck size 
\cite{Pfen97b,E&Ro97,F&Ro96}; otherwise extreme differences in the 
scales which characterize them would exist.  In Chapter~\ref{Chapt:Warp}
we will discuss the example of the macroscopic warp drive, where
we find that the warp bubble's radius can be on the order of tens or hundreds
of meters, but the bubble's walls are constrained to be near the Planck
scale.  Similar results have been found for wormholes and the Krasnikov
tube.

There is also a limitation in the averaged energy conditions. 
The AWEC and the ANEC say nothing about the distribution or
magnitude of the negative energy that may exist in a spacetime.
It is possible to have extremely large quantities of negative
energy over a very large region, but the AWEC would still hold if the
observer's worldline passes through a region of compensating positive
energy. If we can have large negative energies for an arbitrary time,
it would be possible to violate the second law of thermodynamics or
causality over a macroscopic region \cite{Ford78}.  Also, within the
negative energy regions the stress-tensor has been shown to have
quantum fluctuations on the order of the magnitude of the negative
energy density \cite{Kuo93}. This could mean that the metric of the
spacetime in the negative energy region would also be fluctuating,
and we may lose the notion of a ``smooth'' classical background.

\section{The Quantum Inequalities}

A new set of energy constraints were pioneered by Ford in the late
1970's \cite{Ford78}, eventually leading to constraints on negative
energy fluxes in 1991 \cite{Ford91}. He derived, directly from quantum
field theory, a strict constraint on the magnitude of the negative
energy flux that an observer might measure.  It was shown that if 
one does not average over the entire worldline of the observer as is
the case with the AWEC, but weights the integral with a sampling 
function of characteristic width $t_0$, then the observer could at
most see negative energy fluxes, $\hat F_x$, in four-dimensional
Minkowski space bounded below by
\begin{equation}
\hat F_x \equiv {t_0 \over \pi} \int_{-\infty}^\infty 
{\langle T^{xt} \rangle \over {t^2+t_0^2}} dt  \geq -{3\over 32 \pi t_0^4}.
\end{equation}    
Roughly speaking, if a negative energy flux exists for a time $t_0$,
then its magnitude must be less than about $t_0^{-4}$. We see that
the specific sampling function chosen in this case, $t_0/\pi(t^2+t_0^2)$,
is a Lorentzian with a characteristic width $t_0$ and a time
integral of unity.  Ford and Roman also
probed the possibility of creating naked singularities by injecting
a flux of negative energy into a maximally charged Reissner-Nordstr\"{o}m
black hole \cite{F&Ro92}.  The idea was to use the negative energy
flux to reduce the black hole's mass below the magnitude of the charge.
They found that if the black hole's mass was reduced by an
amount $\Delta M$, then the naked singularity produced
could only exist for a time $\Delta T < |\Delta M|^{-1}$.

Ford and Roman introduced the {\it Quantum Inequalities} (QI's)
on the energy density in 1995 \cite{F&Ro95}. These uncertainty
principle-type relations  constrain the magnitude and duration
of negative energy densities in the same manner as negative energy
fluxes. Originally proven in four-dimensional Minkowski space, 
\index{Minkowski spacetime} the quantum inequality for free, quantized,
massless scalar fields can be written in its covariant form as
\begin{equation}
\hat\rho = {\tau_0 \over \pi} \int_{-\infty}^\infty 
{\langle T_{\mu\nu}u^\mu u^\nu \rangle \over {\tau^2 + \tau_0^2}}\, d\tau
\geq -{3\over 32\pi^2 \tau_0^4},   \qquad\mbox{ for all } \tau_0.
\label{eq:qi1}
\end{equation}
Here $u^\mu$ is the tangent to a geodesic observer's worldline and
$\tau$ is the observer's proper time. The expectation value $\langle
\rangle$ is taken with respect to an arbitrary state $|\psi\rangle$,
and $\tau_0$ is the characteristic width of the Lorentzian sampling
function.  Such inequalities limit the magnitude of the negative
energy violations and the time for which they are allowed to exist.
In addition, such QI relations reduce to the usual AWEC type conditions
in the infinite sampling time limit.

Flat space quantum inequality-type relations of this form were 
first applied to curved spacetime by Ford and Roman \cite{F&Ro96}
in a paper which discusses Morris-Thorne wormhole geometries.
\index{wormhole spacetime}  
It was argued that if the sampling time was kept small, then the 
spacetime region over which an observer samples the energy density
would  be approximately flat and the flat space quantum inequality
should hold.  Under this restriction they showed that wormholes are
either on the order of a few thousand Planck lengths
in size or there is a great disparity between the length scales
that characterize the wormhole.  Inherent in this argument is what 
should characterize a  ``small'' sampling time?  In their case,
they restricted the sampling time to be less than the minimal 
``characteristic'' curvature radius of the geometry and/or the proper
distance to any boundaries. When a small
sampling time expansion of the quantum inequality is performed in
a general curved spacetime, such an assumption is indeed born out
mathematically.  This short sampling time technique is particularly
useful in nonstatic spacetimes where exact quantum inequalities
have yet to be proven exactly.   

\section{Curved Spacetime Quantum Inequalities}
Although the method of small sampling times is useful, it does not
address the question of how the curvature will enter into the quantum
inequalities for arbitrarily long sampling times.  This is the main
thrust behind the present work.  In Chapter~\ref{chapt:QI_proof} we
will derive, for an arbitrarily curved static spacetime, a formulation
for finding exact quantum inequalities which will be valid for all
sampling times. For the scalar field, we find that the quantum
inequality can be written in one of two forms, either as a sum of
mode functions defined with positive frequency on the curved background,
\begin{equation}
\Delta\hat\rho \equiv {t_0 \over \pi} \int_{-\infty}^\infty 
{\langle :T_{tt}:/g_{tt} \rangle  \over {t^2 + t_0^2}}\,dt
\geq - \sum_\lambda \left({\omega_\lambda^2\over |g_{tt}|}
 + {1\over 4}\nabla^j\nabla_j\right) |U_\lambda({\bf x})|^2 {\rm e}^{-2
\omega_\lambda t_0}\, ,
\end{equation}
or as the Euclidean wave operator acting on
the Euclidean Green's function\index{Euclidean Green's function}
(two-point functions)
\begin{equation}
\Delta\hat \rho \geq - {1\over 4}\Box_E \, G_E({\bf x},-t_0;{\bf x},+t_0).
\end{equation}
Both forms are functionally  equivalent, but the Euclidean Green's 
function method is often much easier to use if the Feynman or Wightman Green's
function is already known in a given spacetime.  If the Green's function
is inserted into the expression above, we can find the general
form of the quantum inequality.  In three dimensions it
is given by
\begin{equation}
\Delta\hat\rho \geq - {1\over 16\pi \tau_0^3}\; 
^{(3)}{\cal S}(m, \tau_0, g_{\alpha\beta})\, ,
\label{eq:QI_3D}
\end{equation}
and in four dimensions,
\begin{equation}
\Delta\hat\rho \geq - {3\over 32\pi^2 \tau_0^4}\, 
^{(4)}{\cal S}(m, \tau_0,  g_{\alpha\beta})\, .
\end{equation}
Here, $\tau_0 = |g_{tt}|^{1/2}\,t_0$ is the proper sampling time of a static
observer,  $g_{\alpha\beta}$ is the metric for the static spacetime,
$m$ is the mass of the scalar particle, and $^{(3)}{\cal S}$ and 
$^{(4)}{\cal S}$ are called the ``scale'' functions.  They determine 
how the flat space quantum inequalities are modified in curved 
spacetimes and/or for massive particles.
  
In Section~\ref{sec:QAWEC} we will show that in the infinite sampling
time limit, the quantum inequality will in general reduce to the condition 
\begin{equation}
\lim_{\tau_0\rightarrow \infty} {\tau_0\over \pi} \int_{-\infty}^\infty 
{\langle T_{\mu\nu}u^\mu u^\nu\rangle_{Ren.} \over \tau^2 + \tau_0^2}
 d\tau \geq \rho_{vacuum}. \label{eq:qawec}
\end{equation}
Here we are taking the expectation value of the renormalized
energy density on the left-hand side with respect to some arbitrary
particle state $|\psi\rangle$, and $\rho_{vacuum}$ is the energy density
of the vacuum defined by the timelike Killing vector. This is a
modification of the classical AWEC inequality by the addition of
the vacuum energy term.

In Section~\ref{sec:expans} we will proceed with a  small sampling
time expansion of the Green's function to obtain an asymptotic form
for the quantum inequalities.  In this limit we find that the
expansion of the quantum inequality reduces to the flat space form,
with subdominant terms dependent upon the curvature.

In Section~\ref{sec:EM_QI}, we will develop the quantum inequality
for the quantized electromagnetic field in a curved spacetime.  The
flat space quantum inequality was originally derived by Ford and Roman 
\cite{F&Ro97}, and takes an especially simple form
\begin{equation}
\hat\rho \geq -{3\over 16 \pi^2 \tau_0^4}.
\end{equation}
The difference of a factor of two between the scalar field and 
electromagnetic field quantum inequalities occurs because the 
photons have two spin degrees of freedom, while the scalar particles
have only one.  As we will see, the electromagnetic field quantum 
inequality in static curved spacetimes can also be written as a 
mode sum, very similar to the mode sum form of the quantum inequality
for the scalar field.  At present, it is not known if there also
exists a Green's function approach for the formulation of the 
quantum inequalities for the quantized electromagnetic field.

In Chapter~\ref{Chapt:examp} we will apply the quantum inequality
to several examples of curved spacetimes in two, three, and four
dimensions.  In two dimensions, the quantum inequality on the scalar
field for static observers will be shown to have the general form
\begin{equation}
\Delta\hat\rho  \geq - {1\over 8\pi \tau_0^2}.
\end{equation}
Again, $\tau_0$ is the proper sampling time of the observer.
This is the inequality in all two-dimensional spacetimes.  In
Section~\ref{sec:2D_conform} we will see that this occurs
because of the conformal equivalence between all two-dimensional
spacetimes. 

In Section~\ref{sec:3D_spaces} we will find the exact quantum 
inequalities for the three-dimensional equivalent of the closed
and flat static Robertson-Walker spacetimes. In this way, the explicit
form for the scale function in Eq.~(\ref{eq:QI_3D}) will be
determined, and we will examine its behavior.  This will be
followed by the application of the quantum inequalities in
four-dimensional static Robertson-Walker spacetimes. In all of the
above cases, the process of renormalization will be important
for the determination of the renormalized quantum inequality.
In four-dimensional static spacetimes, such as those mentioned
above, the renormalized quantum inequality is given by 
\begin{equation}
{t_0 \over \pi} \int_{-\infty}^\infty 
{\langle T_{tt}/g_{tt} \rangle_{Ren.}  \over {t^2 + t_0^2}}\,dt
\geq - {3\over 32\pi^2 \tau_0^4}\, ^{(4)}{\cal S}
(\mu, \tau_0,  g_{\alpha\beta})\, + \rho_{vacuum}.
\end{equation}
Here the vacuum energy, $\rho_{vacuum}$, is important in determining
what the absolute lower bound may be on any measurement of the 
renormalized energy density.

In Section~\ref{sec:planar} we will consider the particular case of
the quantum inequalities in flat spacetimes in which there
exist perfectly reflecting mirrors.  Both for a single mirror and
for parallel plates, we will find that the quantum inequality is modified 
from the flat space form due to boundary effects.  Not only is
the spacetime curvature important in determining the form of the
quantum inequality, but so is the presence of a boundary.  We will
continue along these lines by looking at spacetimes in which
there are horizons. In Section~\ref{sec:horizons} we will look at
quantum inequalities in Rindler and de~Sitter spacetimes.

In Section~\ref{sec:black_holes} quantum inequalities
will be found for two- and four-dimensional black holes. For the
black hole in two dimensions, we find the exact renormalized quantum
inequality to be 
\begin{equation}
\hat\rho_{Ren.}  \geq - {1\over 8\pi (1-2M/r) t_0^2} + \rho_{Boulware}.
\end{equation}
The first term is the standard result for two-dimensional spacetime
where the proper time is given by the relation $\tau = (1-2M/r)^{1/2}t$.
The remarkable thing about the above expression is the second term.
It is known that there are three possible vacuums around a black hole
that could be chosen for renormalization: the Boulware, Unruh and
Hartle-Hawking vacuum states.  However, the way in which the quantum
inequality is developed picks out the Boulware vacuum.  The mode
functions that are used in deriving the quantum inequality are defined
to have positive frequency with respect to the timelike Killing vector.
This is also the condition for defining the Boulware vacuum state. The
same is true for the four-dimensional black hole where the renormalized
quantum inequality again selects the four-dimensional Boulware vacuum.

Finally, in Chapter~\ref{Chapt:Warp} we will look at the dynamic
spacetime of the Alcubierre ``warp drive'' \cite{Alcu94} in the
short sampling time limit.  We will
show for superluminal bubbles of macroscopically useful size that
the bubble wall thickness must be on the order of only a few
hundred Planck lengths \cite{Pfen97b}. Under this restriction, the
total required negative energy (the energy density integrated over
all space) is 100 billion times larger than the entire mass of the
visible universe. This would seem to dash any hope of using such a
spacetime for superluminal travel.

It comes as no surprise that similar results are found for the 
Krasnikov spacetime\index{Krasnikov spacetime} \cite{Kras95, E&Ro97}.
In this spacetime, a starship traveling to a distant star at subluminal
velocity, creates a tube of negative energy behind it, altering the
spacetime as it goes. A good analogy here is the digging of a subway
tunnel. When the ship reaches its destination it can then turn around
and return home through the tube.  To observers who remain stationary
at the point of departure, the amount of time that elapses  between the
departure and return of the spaceship can be made  arbitrarily small, or
even zero.  Everett and Roman \cite{E&Ro97} showed that when the quantum
inequality is applied to this spacetime in the short sampling time limit,
the negative energy is constrained in a wall whose thickness is again
only on the order of a few Planck lengths.

\chapter{How Negative Energies Arise in QFT} \label{Chap:QFT}

In quantum field theory, there are certain states in which the local
energy density is negative \cite{Epstein}, thus violating the weak
energy condition \cite{Kuo97}. One example is the vacuum plus two
particle state for the quantized scalar field.  Another is the squeezed
states of light for the quantized electromagnetic field. We will
look closely at both cases in Section~\ref{sec:vac+2particle}.  

Specific states in which the quantized fermion field can yield
negative energy densities are just beginning to emerge.   Vollick
has recently demonstrated that the energy density can become
negative when the particle content is a superposition of two
single particle states traveling in different directions \cite{Vollick}.
In Section~\ref{sec:fermions}, we will demonstrate that the 
energy density for the fermion field is negative when the
state is a superposition of the vacuum and a particle-antiparticle
pair. 

In Section~\ref{sec:vac_energy}, the negative vacuum energy
density associated with the Casimir effect for a quantized field
near perfectly reflecting boundaries will be discussed.  For the
quantized electromagnetic field, this gives rise to an attractive
force between two parallel perfectly conducting planar plates. We 
will also look at how non-zero energy densities arise for the
vacuum state in curved spacetimes.   

\section{Quantized Scalar Field}\label{sec:vac+2particle}

We begin our discussion of negative energy densities with the
quantized scalar field in Minkowski spacetime.  This model
demonstrates many of the significant points about negative
energy in quantum field theory.

Let us assume that we have a massive, quantized scalar field $\phi(x)$
which resides in Minkowski spacetime and satisfies the Klein-Gordon
equation
\begin{equation}
-\partial_t^2 \phi(x) + \nabla^2\phi(x) - m^2 \phi(x) = 0.
\label{eq:flat_WE}
\end{equation}
The field $\phi(x)$ can be Fourier expanded in terms of a complete
set of plane wave mode functions 
\begin{equation}
f_{\bf k}(x) = \left( 2 \omega L^3 \right)^{-1/2} e^{i{\bf k}
\cdot{\bf x} - i\omega t},
\end{equation}
each of which individually satisfies the wave equation (\ref{eq:flat_WE})
for any choice of the mode label ${\bf k} = (k_x, k_y, k_z)$.  Note,
we are using box normalization of the mode functions.  The general
solution to the wave equation above is given by
\begin{equation}
\phi(x) = \sum_{\bf k} \left[ a_{\bf k} f_{\bf k}(x) + 
a_{\bf k}^\dagger f_{\bf k}^*(x) \right].
\end{equation}

In the rest frame of an observer, the energy density is the
time component of the stress-energy tensor,
\begin{equation}
\rho = T_{tt} = {1\over 2} \left[ \left( \partial_t \phi(x)\right)^2
+ \nabla\phi(x)\cdot\nabla\phi(x) + m^2 \phi^2(x) \right],
\end{equation}
where we have chosen to use minimal coupling of the scalar field.
Similar effects could be proven in general for any form of the 
coupling in Minkowski space. If we now insert the mode function
expansion of the scalar field, we arrive at
\begin{eqnarray}
T_{tt} & = & {1\over 2} \sum_{{\bf k}{\bf k}'} \left\{ 
\left(\omega\omega' +{\bf k}\cdot{\bf k}'\right)
\left[ f_{\bf k}^* f_{{\bf k}'} a_{\bf k}^\dagger a_{{\bf k}'}
- f_{\bf k}^* f_{{\bf k}'}^* a_{\bf k}^\dagger a_{{\bf k}'}^\dagger
- f_{\bf k} f_{{\bf k}'} a_{\bf k} a_{{\bf k}'}
+ f_{\bf k} f_{{\bf k}'}^* a_{\bf k} a_{{\bf k}'}^\dagger\right]\right.
\nonumber\\
&& \qquad + m^2 
\left.\left[ f_{\bf k}^* f_{{\bf k}'} a_{\bf k}^\dagger a_{{\bf k}'}
+ f_{\bf k}^* f_{{\bf k}'}^* a_{\bf k}^\dagger a_{{\bf k}'}^\dagger
+ f_{\bf k} f_{{\bf k}'} a_{\bf k} a_{{\bf k}'}
+ f_{\bf k} f_{{\bf k}'}^* a_{\bf k} a_{{\bf k}'}^\dagger\right]\right\}.
\label{eq:Mink_energydensity}
\end{eqnarray}
This particular equation has a divergence for any quantum state
for which we may choose to evaluate 
its expectation value. The problem stems from the $a_{\bf k}
a_{{\bf k}'}^\dagger$ terms of the above expression.  For example,
in the vacuum $| 0\rangle$, the stress-tensor
reduces to
\begin{equation}
\langle 0 | T_{tt} | 0 \rangle  =  {1\over 2} \sum_{{\bf k}{\bf k}'}
\left( \omega\omega' + {\bf k}\cdot{\bf k}' + m^2 \right)
f_{\bf k}^{} f_{{\bf k}'}^* \,
\langle 0 |a_{\bf k}^{} a_{{\bf k}'}^\dagger | 0 \rangle
 = {1\over 2L^3} \sum_{\bf k} \left( |{\bf k}|^2 + m^2 \right)^{1/2}\, .
\label{eq:diverg_vac}
\end{equation}
Here we have made use of the standard definitions of the creation
and annihilation operators such that when they act on the vacuum state,
they have the values
\begin{equation}
a_{\bf k}^\dagger | 0 \rangle = | 1_{\bf k} \rangle
\qquad {\rm and} \qquad a_{\bf k} | 0 \rangle = 0,
\end{equation}
respectively.  In addition, when they act on a number eigenstate they
become
\begin{equation}
a_{\bf k}^\dagger \,| n_{\bf k} \rangle = \sqrt{n+1}\, | n+1_{\bf k} \rangle
\qquad {\rm and} \qquad
a_{\bf k} \,| n_{\bf k} \rangle = \sqrt{n}\, | n-1_{\bf k} \rangle.
\end{equation}
We see  once the summation is carried out in Eq.~(\ref{eq:diverg_vac})
that the answer is infinite.  This divergence seems to indicate that
the vacuum itself contains an infinite density of energy.  In 
non-gravitational physics, absolute values of the energy are not
measurable, only energy differences.  It has become standard
practice to {\it renormalize} the energy density (as well as the energy)
by measuring all energies relative to the infinite background energy.
Typically this is achieved by normal ordering such that all of the
annihilation operators stand to the right of the creation operators
within the expectation value. Then, we subtract away all of the
divergent vacuum energy terms.  For example, if we make use of the
commutation relation for the $a_{\bf k} a_{{\bf k}'}^\dagger$ terms in 
Eq.~(\ref{eq:Mink_energydensity}), we find
\begin{eqnarray}
T_{tt} & = & {1\over 2} \sum_{{\bf k}{\bf k}'} \left\{ 
\left(\omega\omega'+{\bf k}\cdot{\bf k}'\right)
\left[ f_{\bf k}^* f_{{\bf k}'} a_{\bf k}^\dagger a_{{\bf k}'}
- f_{\bf k}^* f_{{\bf k}'}^* a_{\bf k}^\dagger a_{{\bf k}'}^\dagger
- f_{\bf k} f_{{\bf k}'} a_{\bf k} a_{{\bf k}'}
+ f_{\bf k} f_{{\bf k}'}^*  a_{{\bf k}'}^\dagger a_{\bf k}\right]\right.
\nonumber\\
&& \qquad\quad + m^2 
\left.\left[ f_{\bf k}^* f_{{\bf k}'} a_{\bf k}^\dagger a_{{\bf k}'}
+ f_{\bf k}^* f_{{\bf k}'}^* a_{\bf k}^\dagger a_{{\bf k}'}^\dagger
+ f_{\bf k} f_{{\bf k}'} a_{\bf k} a_{{\bf k}'}
+ f_{\bf k} f_{{\bf k}'}^* a_{{\bf k}'}^\dagger a_{\bf k} \right]\right\}
+ {1\over 2L^3} \sum_{\bf k} \omega.
\end{eqnarray}
The last term is the infinite vacuum energy, and is subtracted away
to complete normal ordering.  We define the normal ordered energy
density as
\begin{equation}
:T_{tt}: = T_{tt} - \langle 0 | T_{tt} |0 \rangle,
\end{equation}
which produces a finite expectation value for physically allowable states.

\subsection{Vacuum Plus Two Particle State}
 
\begin{figure}[hbtp]
\begin{center}
\leavevmode\epsfxsize=\textwidth\epsffile{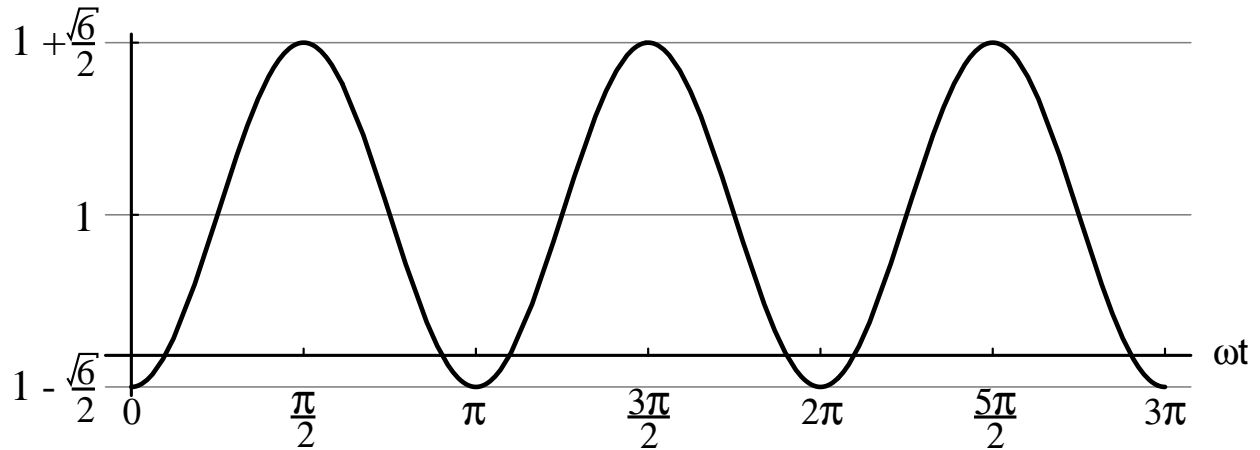}
\end{center}
\caption[Time evolution of negative energy density]
{ Time evolution of the energy density for the vacuum plus
two particle state.  Note that during every
oscillation, the energy density dips below zero for a brief portion of
the cycle. By manipulating the quantum state, it is possible to drive
the energy density to increasingly negative values.  Units on the vertical
axis are in multiples of $\omega/2L^3$. }
\label{fig:NegEner}
\end{figure}

We can now address how negative energy densities can be found.  Let
us consider the particle content to be defined by the superposition
of two number eigenstates, {\it e.g.},
\begin{equation}
|\psi \rangle = {\sqrt{3} \over 2} | 0 \rangle + {1\over 2} 
| 2_{\bf k} \rangle .
\end{equation}
If we look at the expectation value of the energy density we find,
\begin{equation}
\langle \psi | : T_{tt} : | \psi \rangle = {\omega \over 2 L^3} \left[
1 - {\sqrt{6} \over 2} \cos 2 ({\bf k}\cdot{\bf x}-\omega t) \right]\, ,
\end{equation}
where we have also assumed that the field is massless.  We see that
the energy density oscillates at twice the frequency of the mode 
that composes the state.  We also see that once during every cycle
the energy density becomes negative for a brief period of time.
A plot of this is shown in Figure~\ref{fig:NegEner}.
\begin{figure}
\begin{center}
\leavevmode\epsffile{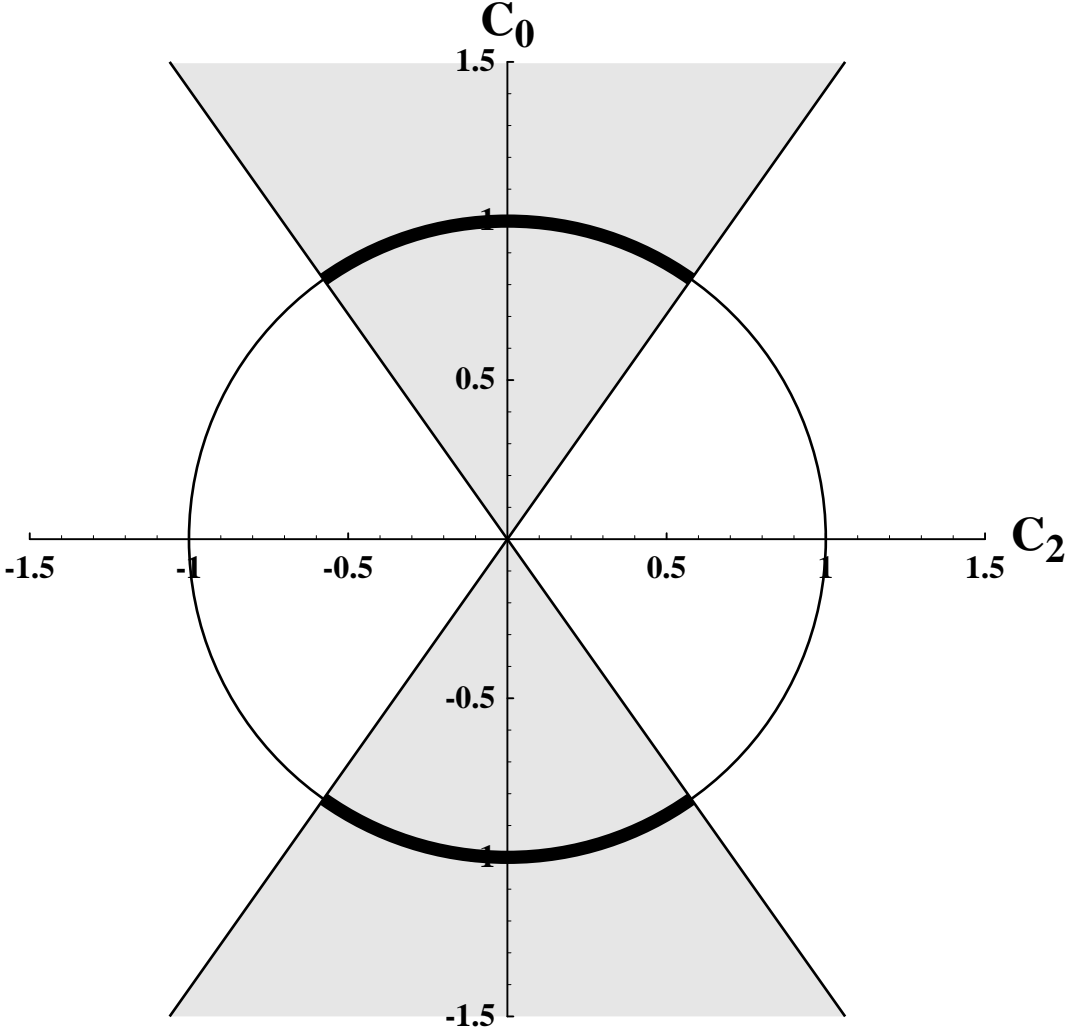}
\end{center}
\caption[Plot of allowed negative energy states]
{ A plot showing the two conditions necessary to obtain negative
energy densities for the vacuum plus two particle state.  The normalization
condition for the coefficients $c_0$ and $c_2$ is the unit circle.  The
necessary condition for negative energy $|c_0| > \sqrt{2}\,|c_2|$ is 
represented by the light gray region.  The set of all possible states
that meet both of these conditions is denoted by the bold curve at the
top and bottom of the circle. The states with $|c_0| = 1$ must be excluded
because they correspond to the vacuum state, where the energy density
is zero.}
\label{fig:circle_plot}
\end{figure}

Actually, there are an infinite set of vacuum plus two particle 
states that give rise to negative energy densities.  To show this,
let us begin with a general superposition of an $n$ and an $n+2$
state, 
\begin{equation}
|\psi \rangle = c_n | n_{\bf k} \rangle + c_{n+2} | n+2_{\bf k} \rangle ,
\end{equation}
where the normalization of the state is expressed by the condition
\begin{equation}
|c_n|^2 + | c_{n+2} |^2 = 1. \label{eq:normalized}
\end{equation}
There is an additional restriction that $ |c_n| \neq 1$.
This corresponds to a pure number eigenstate because the
unit normalization would cause $c_{n+2} = 0$. 

For a superposition of the $n$ and  $n+2$ particle states, the energy
density is given by
\begin{eqnarray}
\langle \psi | : T_{tt} : | \psi \rangle &=& {\omega \over  L^3}
\left( n + 2|c_{n+2}|^2 \right) - {\omega^2 - m^2 \over \omega L^3}
\sqrt{(n+1)(n+2)}\times\nonumber\\
&& \left[ {\rm Re}(c_n^* c_{n+2}^{}) \cos 
2 ({\bf k}\cdot{\bf x}-\omega t) - | {\rm Im}(c_n^* c_{n+2}^{}) |
\sin 2 ({\bf k}\cdot{\bf x}-\omega t) \right].\nonumber\\
\end{eqnarray}
In order to simplify our calculations, let us take $c_n$ and $c_{n+2}$ to
be real.  Then, in order to have negative energy densities,
we must satisfy the condition
\begin{equation}
| c_n | > {\omega^2\over \omega^2 - m^2}{1\over \sqrt{(n+1)(n+2)}}
\left( 2 |c_{n+2}| + {n\over|c_{n+2}|}\right)\, .\label{eq:neg_condit}
\end{equation}
In addition, we must satisfy the normalization condition
(\ref{eq:normalized}), which for real coefficients is the equation
for a unit circle in the ($c_n,\,c_{n+2}$) plane.

For the massless ($m=0$) scalar field  in the vacuum plus two particle
state ($n=0$), the condition to obtain negative energy densities,
Eq.~(\ref{eq:neg_condit}), reduces to
\begin{equation}
|c_0| > \sqrt{2}\,|c_2|\, .
\end{equation}
We can see from Figure~\ref{fig:circle_plot} that any state, except
$|c_0| = 1$,  that lies on the  bold sections of the circle will satisfy
both the above condition to obtain negative energy densities and the
unit normalization condition. For real coefficients, $c_n$ and $c_{n+2}$,
only the vacuum plus two particle state can generate negative energy
densities. If we plot the condition to find negative energy densities, 
Eq.~(\ref{eq:neg_condit}), for $c_n$ and $c_{n+2}$ real and $n\geq 1$, we
find that the range of values that would allow negative energy densities
is incompatible with the unit normalization condition.

\subsection{Squeezed States}

The squeezed states in quantum field theory can also produce
negative energies.  The squeezed states of light have been
extensively investigated in the field of quantum optics and
are realized experimentally. We will confine ourselves to the
squeezed states of the quantized scalar field, although the
treatment would be the same for the electromagnetic field.

We begin with the definition of two additional operators.  The first,
introduced by Glauber \cite{Glau63}, is the displacement operator
\begin{equation}
D_{\bf k} (z) \equiv \exp \left( z a_{\bf k}^\dagger - z^* a_{\bf k}\right)
= e^{-|z|^2 /2} e^{z a_{\bf k}^\dagger}  e^{-z^*a_{\bf k}},
\end{equation}
which satisfies the commutation relations
\begin{eqnarray}
\left[ a_{\bf k'}^{}, \, D_{\bf k}(z) \right] &=&  a_{\bf k'} + z 
\delta_{\bf k k'}, \\
\left[ a_{\bf k'}^\dagger , \, D_{\bf k}(z) \right] &=& 
a_{\bf k'}^\dagger  + z^*\delta_{\bf k k'}.  
\end{eqnarray}
The second is the squeeze operator
\begin{equation}
S_{\bf k}(\zeta)\equiv \exp \left[ {1\over 2} \zeta^* a_{\bf k}^2 - 
{1\over 2} \zeta (a_{\bf k}^\dagger)^2 \right].
\end{equation}
which satisfies the relations
\begin{eqnarray}
S_{\bf k}^\dagger(\zeta) a_{\bf k} S_{\bf k}(\zeta) &=&  a_{\bf k}
\cosh r -  a_{\bf k}^\dagger e^{i\delta} \sinh r,\\
S_{\bf k}^\dagger(\zeta) a_{\bf k}^\dagger S_{\bf k}(\zeta) &=& 
a_{\bf k}^\dagger \cosh r -  a_{\bf k} e^{-i\delta} \sinh r. 
\end{eqnarray}
The complex parameters $z$ and $\zeta$ may be written in terms of
their magnitude and phases as
\begin{equation}
z = s e^{i\gamma} \qquad\mbox{ and }\qquad \zeta = r e^{i\delta}.
\end{equation}

We can write a general squeezed state for a single mode as \cite{Caves81}
\begin{equation}
|\{z,\zeta\}_{\bf k}\rangle = D_{\bf k} (z) S_{\bf k}(\zeta) |0\rangle.
\end{equation}
It has been shown that for a massless scalar field the expectation
value of the energy density for the above squeezed state is given
by \cite{Kuo93}
\begin{eqnarray}
\rho &\equiv& \langle \{z,\zeta\}_{\bf k}| :T_{tt}: |\{z,\zeta\}
_{\bf k}\rangle, \nonumber\\
&=&{ \omega \over L^3}\left\{\sinh r \cosh r \cos(2\theta + \delta) +
\sinh^2 r + s^2\left[ 1-\cos 2(\theta + \gamma)\right]\right\},
\end{eqnarray}
where $\theta = {\bf k}\cdot{\bf x} - \omega t$.  When $\zeta = 0$,
the squeezed state reduces to a coherent state and the expectation
value of the energy density is always positive definite.  The coherent
states are interpreted as the quantum states that describe classical
field excitations.  Thus, the production of only positive energy in
coherent states is consistent with classical sources for gravity
obeying the WEC.

A different case, known as the squeezed vacuum state, is defined
when $z=0$.  Such states result from quantum mechanical particle
creation.  An example is second harmonic generation in non-linear
optical media.  For the squeezed vacuum state, the energy density
is given by
\begin{equation}
\langle \{0,\zeta\}_{\bf k}| :T_{tt}: |\{0,\zeta\}_{\bf k}\rangle =
{\omega\over L^3} \sinh r \left[ \sinh r + \cosh r \cos(2\theta + 
\delta)\right].
\end{equation}
The squeezed vacuum energy density will have the same negative energy
density behavior as the vacuum plus two particle state above, with
the energy density falling below zero once every cycle if the condition
\begin{equation}
\cosh r > \sinh r
\end{equation}
is met. This happens to be true for every nonzero value of $r$, so
the energy density becomes negative at some point in the
cycle for a general squeezed vacuum state.  In a more general state
which is both squeezed and displaced, the
energy density will become negative at some point in the cycle if
$s \ll r$.  

There are other interesting aspects of these states as well.  When
the quantum state is very close to a coherent state, we have seen
that the energy density is usually not negative.  In addition,
fluctuations in the stress-tensor for these states are small compared
to the expectation value of the stress-tensor in that state.  On the
other hand when the state is close to a squeezed vacuum state,
there will almost always be some negative energy densities present,
and the fluctuations in the expectation value of the stress-tensor
start to become nearly as large as the expectation value itself.
This seems to indicate that the semiclassical theory is beginning
to break down \cite{Kuo93}. 

\section{The Quantized Fermion Field}\label{sec:fermions}


In this section we will show that it is possible to generate
negative energy densities using the fermion field.  Therefore it 
does appear possible to use fermions to violate the weak energy
condition.  One example of a fermion state that produces negative
energies has been demonstrated by Vollick \cite{Vollick} for a
superposition of two single particle states traveling in different
directions.  The particular example that we will demonstrate has a
particle content which is a superposition of the vacuum plus a
particle-antiparticle pair.  However the occurrence of negative
energy densities for the above particle states can be a frame-dependent
phenomenon. If the particle-antiparticle pair is traveling in the
same direction, but with differing momenta, then it is possible
for the energy density to become negative.  However, in the center
of mass frame of the particle-antiparticle pair, the local energy
density always has a positive definite value.

The fermion field, denoted by $\psi(x)$, is the solution to the
Dirac equation,
\begin{equation}
i \gamma^\mu \nabla_\mu \psi(x) - m \psi(x) = 0,
\end{equation}
where $m$ is the mass of the spin-$1\over 2$ particle, and the
$\gamma^\mu$ are the the Dirac matrices in flat spacetime, which
satisfy the anticommutation relations
\begin{equation}
\left\{ \gamma^\mu , \, \gamma^\nu \right\} = 2 g^{\mu\nu}.
\end{equation}
The general solution of $\psi(x)$ can then be expanded in plane
wave modes as
\begin{equation}
\psi(x) = \sum_s \int {d^3 p \over (2\pi)^{3/2}} \sqrt{m\over E_p}
\left[ b(p,s) \, u(p,s)\, e^{-ip_\mu x^\mu} + 
d^\dagger(p,s) \, v(p,s)\, e^{ip_\mu x^\mu} \right]\, ,
\end{equation}
Here $b(p,s)$ and its hermitian conjugate $b^\dagger (p,s)$ are the
annihilation and creation operators for the electron (fermion),
respectively, while  $d(p,s)$ and its hermitian conjugate $d^\dagger(p,s)$
are the respective annihilation and creation operators for the positron.
All four operators anticommute except in the case
\begin{equation}
\left\{ b(p,s), \, b^\dagger(p',s')\right\} = \delta_{ss'} \delta^3(p-p')
=\left\{ d(p,s), \, d^\dagger(p',s')\right\}.
\end{equation}
The annihilation and creation operators for 
spin-${1\over2}$ particles satisfy the anticommutation relations as
opposed to the commutation relations for the scalar field because
the fermions must obey the Pauli exclusion principle.  

For the fermion field, the stress-tensor is given by \cite{Brl&Dv}
\begin{equation}
T_{\mu\nu} = {i\over 2} \left[ \bar\psi \gamma_{(\mu} \nabla_{\nu )}
\psi - \left( \nabla_{(\mu} \bar\psi \right) \gamma_{\nu)} \psi\right]\, .
\end{equation}
Inserting the mode function
expansion into the energy density, and then normal ordering 
by making use of the anticommutation relations, we find
\goodbreak 
\begin{eqnarray}
T_{tt} &=& {1\over 2(2\pi)^3}\sum_{ss'}\int d^3p\, d^3p' {m\over
\sqrt{E_p\, E_{p'}}} (E_p + E_{p'}) \times\nonumber\\
&&\;\times\left[ b^\dagger(p',s') b(p,s) u^\dagger(p',s') u(p,s)
e^{-i(p_\mu - p_\mu') x^\mu}  +  d^\dagger(p,s) d(p',s') v^\dagger(p',s') v(p,s)
e^{+i(p_\mu - p_\mu') x^\mu}\right]\nonumber \\
&& +{1\over 2(2\pi)^3}\sum_{ss'}\int d^3p \,d^3p' {m\over \sqrt{E_p\, E_{p'}}}
(E_p - E_{p'}) \times\nonumber\\
&&\; \times\left[d(p',s') b(p,s) v^\dagger(p',s') u(p,s)
e^{-i(p_\mu + p_\mu') x^\mu}  -  b^\dagger(p,s) d^\dagger(p',s') 
u^\dagger(p',s') v(p,s) e^{+i(p_\mu + p_\mu') x^\mu}\right]\nonumber \\
&&-{1\over (2\pi)^3} \sum_s\int d^3p \, E_p \, .
\end{eqnarray}
Unlike the vacuum energy density for the scalar field, we see that
the fermion field has an infinite {\it negative} energy density that must
be removed by renormalization.  Thus we again look at the fully normal
ordered quantity,
\begin{equation}
:T_{tt}: = T_{tt} - \langle 0 | T_{tt} | 0 \rangle.
\end{equation}
While we are about to show that there are specific states for which
the energy density can become negative, one should note that the
Hamiltonian,
\begin{equation}
H = \int d^3x :T_{tt}: = \sum_s \int d^3p \, E_p \left[ b^\dagger(p,s)
b(p,s) + d^\dagger(p,s) d(p,s)\right]\, ,
\end{equation}
is always a positive definite quantity.

Now, we would like to show that the state
\begin{equation}
|\phi \rangle = c_{00} | 0 \rangle + c_{1\overline{1}} | k_1\, s_1 \, ;
\overline{ k_2\, s_2} \rangle\, ,
\end{equation}
admits negative energies for some values of $c_{00}$ and $c_{1\bar 1}$.
For purposes of clarity, we  have defined the above states by
\begin{eqnarray}
| k\, s\rangle &=& b^\dagger (k,s) |0\rangle\, , \\
| \overline{k\, s}\rangle &=& d^\dagger (k,s) |0\rangle\, , \\
| k_1 \, s_1 \, ; \overline{k_2 \, s_2} \rangle &=& d^\dagger(k_2,s_2)
b^\dagger(k_1,s_1) | 0 \rangle = - | \overline{k_2 \, s_2}\, ;
 k_1 \, s_1 \rangle\, .
\end{eqnarray}
If we take the expectation value of the energy density with
respect to this vacuum plus particle-antiparticle state, we find
\begin{eqnarray}
\langle :T_{tt}: \rangle &=& {1\over (2\pi)^3} m |c_{1\bar 1}|^2 \left[
u^\dagger(k_1, s_1) u(k_1, s_1) + v^\dagger(k_2, s_2) v(k_2, s_2)\right]
\nonumber\\
&+& {m\over 2(2\pi)^3} {(E_1 -E_2)\over \sqrt{E_1 E_2}} \left[
c_{00}^* c_{1\bar 1}\, v^\dagger(k_2, s_2) u(k_1,s_1) e^{-i(k_1^\mu + k_2^\mu)
x_\mu} + C.C. \right].\nonumber\\
\label{eq:Gen_EDensity}
\end{eqnarray}
This expression is similar to the scalar field energy density for
the vacuum plus two particle state.  In order to have negative energy
densities, we must find some combination 
of the mode functions and the coefficients $c_{00}$ and $c_{1\bar 1}$
such that the amplitude of the interference term in the expression 
above is larger than the magnitude of the first term, which is the
sum of the energy of the particle and antiparticle.  In order to 
proceed any further, we must choose a specific form for the mode
functions. This involves choosing a representation of the Dirac
matrices.  Let,
\begin{equation}
u(p,s) = \left[ 2m (E+m) \right]^{-1/2} \left( 
\matrix{
        (E+m)\, \phi(s) \cr
        (\mbox{\boldmath $\sigma$} \cdot {\bf p}) \, \phi(s) \cr
       } \right)\, ,
\end{equation}
and
\begin{equation}
v(p,s) = \left[ 2m (E+m) \right]^{-1/2} \left( 
\matrix{
        (\mbox{\boldmath $\sigma$} \cdot {\bf p}) \, \phi(s) \cr
        (E+m) \, \phi(s) \cr
       } \right)\, ,
\end{equation}
where ${\bf p}$ is the three momentum of the particle, {\boldmath $\sigma$} 
is the vector composed of the Pauli spin matrices, and $\phi(s)$ is
a two-spinor.  Using this basis, it is possible to show
\begin{equation}
u^\dagger(k,s) u(k, s') = {E_k\over m} \delta_{ss'} = 
v^\dagger(k,s) v(k, s').
\end{equation}
In addition, the interference term between the particle-antiparticle
states is
\begin{equation}
v^\dagger(k_2,s_2) u(k_1, s_1) = {{ \phi^\dagger(s_2) \mbox{\boldmath $\sigma$}
\cdot \left[ {\bf k}_1(E_2 + m) + {\bf k}_2 (E_1 + m) \right] \phi(s_1)}
\over\sqrt{2m (E_1+m)(E_2+m)}}.
\end{equation}
To evaluate the energy density explicitly at this point, let us
take $s_1 = s_2 =+1$, therefore
\begin{equation}
\phi(s_1) =\phi(s_2) = \left( \matrix{ 1\cr 0 \cr} \right).
\end{equation}
For simplicity let the propagation vectors ${\bf k}_1$ and ${\bf k}_2$ both
point in the $z$-direction, {\it i.e.},
\begin{equation}
k_1^\mu = (E_1, 0, 0, k_1) \qquad\mbox{ and }\qquad k_2^\mu =
(E_2, 0, 0, k_2)\, ,
\end{equation}
and let $c_{00}$ and $c_{1\bar{1}}$ be real.  Then the energy density
can be written as
\begin{equation}
\langle:T_{tt}:\rangle = {c_{1\bar{1}}^2 \over (2\pi)^3}(E_1+E_2) +
{c_{00}\, c_{1\bar{1}} \over 2(2\pi)^2 }{(E_1 - E_2)\,
\left[ k_1 (E_2+m) + k_2(E_1+m) \right] \over 
\sqrt{E_1 E_2 (E_1 +m) (E_2 + m)}} \, \cos (E_1 +E_2) t \, .
\label{eq:Fermion_Edensity}
\end{equation}
By an appropriate choice of the momenta and the coefficients $c_{00}$
and $c_{1\bar{1}}$ this can be made negative.  This is most easily
seen in the ultrarelativistic limit when $E\gg m$. The energy density is
\begin{equation}
\langle:T_{tt}:\rangle = { E_1+E_2 \over (2\pi)^3} c_{1\bar{1}}^2 +
{ E_1-E_2 \over (2\pi)^3} c_{00} c_{1\bar{1}} \cos(E_1+E_2)t. 
\end{equation}
We then find that the energy density is negative, apart from the
special case $c_{00} = 1$, $c_{1\bar{1}} = 0$ corresponding to the
vacuum,  when the condition
\begin{equation}
|c_{00}| \geq \left| {E_1+E_2 \over E_1 -E_2} c_{1\bar{1}} \right|
\end{equation}
is met. Thus, an infinite number of states for negative
energy densities could possibly exist.

It is interesting to note that in the center of mass frame of the
particle-antiparticle pair, the local energy density in this state
is a positive constant.  This can easily be seen from 
Eq.~(\ref{eq:Fermion_Edensity}), or more generally
from Eq.~(\ref{eq:Gen_EDensity}), when $E_1 = E_2$ and $k_1 = -k_2$.
Thus, for a given particle content of the state, whether it is
possible to detect negative energies is dependent upon the frame in
which any measurement is to be carried out.


\section{The Vacuum Energy}\label{sec:vac_energy}
So far we have discussed only the cases where the particle content of 
the quantum state makes the energy density negative. These could be 
called coherence effects for the fields.  As we have seen in the 
plane wave examples above, the negative energy densities are periodic
in both space and time. For more general mode functions this need
not be the case, and we could have various configurations of negative
energy densities.
However, it is also possible to generate negative energies without the
presence of particles.  The vacuum energy density of a spacetime can be
positive or negative with respect to the Minkowski space vacuum after
renormalization. Casimir originally proved this in 1948 for the 
electromagnetic field between two perfectly conducting planar plates 
\cite{Casimir}.  Here the energy density between the two plates is negative.
Similar effects can be shown in gravitational physics.  For example,
the vacuum energy is non-zero for the scalar field in Einstein's universe
and can be positive or negative, depending on our choice of either minimal or
conformal coupling for the interaction between gravity and the scalar
field in the wave equation,
\begin{equation}
g^{\mu\nu}\nabla_\mu \nabla_\nu\, \phi(x) -\left(m^2 +\xi R(x)\right)\, 
\phi(x) = 0.
\end{equation}
Minimal coupling is $\xi = 0$, while conformal coupling is given by
\begin{equation}
\xi(n) = {1\over 4} {(n-2)\over(n-1)}\, ,
\end{equation}
where $n$ is the dimensionality of the spacetime.  In four dimensions
$\xi = 1/6$, while in two dimensions, minimal and conformal coupling 
are the same.  The vacuum energy density for a massless scalar field
in the four-dimensional static Einstein universe \cite{Elizalde,Pfen97a} is
\begin{equation}
\rho_{vacuum} =  - {0.411505 \over 4 \pi^2 a^4}
\end{equation}
for minimal coupling.  Meanwhile, for conformal coupling the 
vacuum energy density is \cite{Ford75b,Ford76}
\begin{equation}
\rho_{vacuum} =  + {1 \over 480 \pi^2 a^4}.
\end{equation}
Here, $a$ is the scale factor of the closed universe.  We will
discuss more fully the derivation of the vacuum energy for minimal
coupling in Einstein's universe in Section~\ref{sec:massless}, when
we renormalize the quantum inequality in Einstein's universe.  At
this point we would like to discuss a slightly simpler example, 
the vacuum energy near a planar conductor in both two dimensions
and four dimensions. 

\subsection{Renormalization}
We begin with a massless scalar field in Minkowski spacetime with no
boundaries. The positive frequency mode function solutions for the
scalar wave equation are
\begin{equation}
f_{\bf k}({\bf x},t) = (4\pi\omega)^{-1/2}\, e^{i({\bf k\cdot x}-\omega t)}\, ,
\end{equation}
where $-\infty< {\bf k} <+\infty$ is the mode label, and $\omega =
|{\bf k}|$ is the energy. In the case of arbitrary coupling, the
stress-tensor is
\begin{equation}
T_{\mu\nu}  = (1-2\xi) \phi_{,\mu} \phi_{,\nu} + (2\xi - 1/2) \eta_{\mu\nu}
\eta^{\rho\sigma}\phi_{,\rho} \phi_{,\sigma} - 2\xi \phi_{,\mu\nu}\phi\,.
\end{equation} 
We could proceed by placing the mode function expansion for the field
$\phi({\bf x},t)$ into the expression for the stress-tensor above and then
taking the expectation value as we did in the preceding sections.  We
would find the infinite positive energy density as we have previously
and then subtract it away to find the renormalized energy density.
There is an alternative method using the Wightman function, defined as
\begin{equation}
D^+(x,x') = \langle 0 | \phi(x) \phi(x') | 0 \rangle .
\end{equation}
The expectation value of the stress-tensor above in the vacuum state can
be written equivalently as
\begin{equation}
\langle 0 | T_{\mu\nu} | 0 \rangle = \lim_{x'\rightarrow x} \left[
(1-2\xi) \partial_\mu \partial'_\nu + (2\xi - 1/2) g_{\mu\nu} g^{\rho\sigma}
\partial_\rho \partial'_\sigma - 2\xi \partial_\mu \partial_\nu\right] D^+(x,x'),
\label{eq:vacuum_stresstensor}
\end{equation}
where $\partial_\mu$ is the ordinary derivative in the unprimed
coordinates and $\partial'_\nu$ in the primed coordinates. The
process of renormalization involves removing any infinities by
subtracting away the equivalent infinity in the Minkowski vacuum.
This can be accomplished by renormalizing the Wightman function
before we act on it with the derivative operator to find the
expectation value of the stress-tensor. The renormalized Wightman
function is defined as
\begin{equation}
D^+_{Ren.}(x,x') = D^+(x,x') - D^+_{Minkowski}(x,x').
\end{equation}
We see that the renormalized vacuum energy in Minkowski space is
then set to zero. A similar formalism can be carried out for curved
spacetimes as well. To find the renormalized Wightman function, we
would first find the regular Wightman function. Then we subtract
away the Wightman function that would be found by taking its limit
when the spacetime becomes flat.  However, in curved spacetimes, 
this does not remove all of the infinities in the stress-tensor.
There may be logarithmic and/or curvature-dependent divergences
which must also be removed.

\subsection{Vacuum Energies for Mirrors}

Let us start by considering the case of a two-dimensional flat spacetime
in which we place a perfectly reflecting mirror at the origin. In
two-dimensional Minkowski spacetime, the Wightman function is found to be
\begin{equation}
D^+_{Minkowksi}(x,x') = - {1\over 4\pi}\ln\left[ (t-t')^2- (z-z')^2 \right] .
\end{equation}
However, we do not expect this to be the Wightman function when the
mirror is present.  That is because the mode functions will be altered
in such a way that they vanish on the surface of the mirror.
Because the spacetime is still flat, just with a perfectly reflecting
boundary, the new Wightman function can be found by the method of images,
with a source at spacetime point to be $x'$, and the observation point
at $x$.  The Wightman function with the mirror present is now
composed of two terms,
\begin{equation}
D^+(x,x') = - {1\over 4\pi}\ln\left[ (t-t')^2- (z-z')^2 \right] 
            + {1\over 4\pi}\ln\left[ (t-t')^2- (z+z')^2 \right].
\end{equation}
To renormalize the Wightman function, we now subtract away the 
Minkowski Wightman function yielding
\begin{equation}
D^+_{Ren.}(x,x') =  {1\over 4\pi}\ln\left[ (t-t')^2- (z+z')^2 \right].
\end{equation}
The vacuum energy density in this spacetime is then
\begin{equation}
\langle 0 | T_{tt} | 0 \rangle_{Ren.} = \lim_{x'\rightarrow x}
\left[ {1\over 2} \partial_t \partial_{t'} + ({1\over 2} -2\xi) \partial_z
\partial_{z'} - 2\xi \partial_t \partial_t\right] {1\over 4\pi}
\ln\left[ (t-t')^2- (z+z')^2 \right] = {\xi \over 2 \pi z^2}.
\end{equation}
Similar calculations can be carried out for the other components of
the stress-tensor to find
\begin{equation}
\langle 0 | T_{\mu\nu} | 0 \rangle_{Ren.} = {\xi \over 2 \pi z^2} 
\left[ \begin{array}{cc} 1&0\\0&0\\ \end{array} \right].
\end{equation}
We see that the vacuum energy density is non-zero for any value of
$\xi \neq 0$. In addition, the energy density  diverges as one
approaches the mirror.  The actual sign of the energy density depends
on the sign of the coupling constant, so we can have infinite positive
as well as negative energy densities.  However, at large distances away
from the mirror the vacuum energy density rapidly goes to zero.

Similar results can be found for a planar perfectly reflecting mirror
in four dimensions.  With the mirror located in the $xy$-plane at $z=0$
the renormalized Green's function for a massless scalar field is
\begin{equation}
D^+_{Ren.}(x,x') = - {1\over 4 \pi^2}\left[ (x-x')^2 + (y-y')^2 +
(z+z')^2 -(t-t')^2\right]^{-1}.
\end{equation}
Using Eq.~(\ref{eq:vacuum_stresstensor}) we find
\begin{equation}
\langle 0 | T_{\mu\nu} | 0 \rangle_{Ren.} = {1-6\xi \over 16 \pi^2 z^4}
\left[\begin{array}{rrrr}
 -1 & 0 & 0 & 0\\  0 & 1 & 0 & 0\\ 0 & 0 & 1 & 0\\ 0 & 0 & 0 & 0\\
 \end{array}\right].\label{eq:exterior_ST}
\end{equation}
As was the case with the two-dimensional plate, the four-dimensional
vacuum energy still diverges as one approaches the mirror surface.  
In addition, there  are different values of the coupling constant
$\xi$ that cause different effects on the stress-tensor.  For the
coupling constant less than $1/6$, the energy density is everywhere
negative and the transverse pressures are positive.  In addition, 
both are divergent on the mirror's surface. 
This includes the often used case of minimal coupling, $\xi = 0$.
When the coupling constant is equal to $1/6$, the conformally coupled
case, the energy density and transverse pressures vanish.  Finally,
when the coupling constant becomes greater than $1/6$, the energy
density becomes positive and is still divergent at the mirror. However,
the transverse pressures now become negative.  These are summarized in
Table~\ref{tab:vac_ener}.  One should note that for all values of
the coupling constant, the component of the pressure that is
perpendicular to the surface to the mirror, the $z$-direction,
is always zero.
\begin{table}[htb]
\renewcommand{\arraystretch}{1.3}
\begin{center}
\leavevmode
\begin{tabular}{|c|c|c|}\hline
\bf Value of Coupling Constant & \bf Energy Density & \bf 
Transverse Pressure\\
$\xi$ & $\langle T_{tt} \rangle$ &  $\langle T_{xx} \rangle =
\langle T_{yy} \rangle$ \\ \hline\hline
$\xi < {1/ 6}$ & $ < 0 $ & $ >0$ \\
$\xi = {1/ 6}$ & $   0 $ & $  0$ \\
$\xi > {1/ 6}$ & $ > 0 $ & $ <0$ \\ \hline
\end{tabular}\end{center}
\caption[Vacuum energy for a perfect mirror]
{Vacuum energy density and pressures for a perfectly
reflecting mirror in four dimensions as a function of the coupling
parameter $\xi$.}
\renewcommand{\arraystretch}{1}\label{tab:vac_ener}
\end{table}

\subsection{The Casimir Force}
Let us consider two perfectly conducting planar plates, parallel to
each other and separated by a distance L along the $z$-axis.
Between the two conducting plates, the stress-tensor of the scalar 
field for conformal coupling is given by \cite{DeWitt,Fulling}
\begin{equation}
\langle 0 | T_{\mu\nu} | 0 \rangle_{Conf.} = {\pi^2 \over 1440 L^4}
\left[\begin{array}{rrrr}
 -1 & 0 & 0 & 0\\  0 & 1 & 0 & 0\\ 0 & 0 & 1 & 0\\ 0 & 0 & 0 & -3\\
 \end{array}\right].
\end{equation}
In the case of minimal coupling it is given by
\begin{equation}
\langle 0 | T_{\mu\nu} | 0 \rangle_{Min.} =
\langle 0 | T_{\mu\nu} | 0 \rangle_{Conf.} + {\pi^2 \over 48 L^4}
{3 - 2 \sin^2(\pi z/L) \over \sin^4(\pi z /L) }
\left[\begin{array}{rrrr}
 -1 & 0 & 0 & 0\\  0 & 1 & 0 & 0\\ 0 & 0 & 1 & 0\\ 0 & 0 & 0 & 0\\
 \end{array}\right].
\end{equation}
In the region outside of the plates, the stress-tensor is given by 
Eq.~(\ref{eq:exterior_ST}).  The net force, $F$, per unit area, $A$, 
acting on the plates is then given by the difference of the
$T_{zz}$ component of the stress-tensor on opposite sides of the
plate.  For both minimal and conformal coupling, there is an 
attractive force between the two plates with a magnitude
\begin{equation}
|F(L)/A| = {\pi^2 \over 480\, L^4 }.
\end{equation}

Similarly, for the quantized electromagnetic field between two
conducting plates, the attractive force between the plates is
\begin{equation}
|F(L)/A| = {\pi^2 \over 240\, L^4 }.\label{eq:Parall_plates}
\end{equation}
The factor of two difference comes about because the electromagnetic
field has two polarization states.  Lamoreaux has confirmed the 
existence of this force experimentally \cite{Lamo97}.  
This was done by measuring  the force of attraction
between a planar disk and a spherical lens, both of which were
plated in gold.  The Casimir force is known to be independent
of the molecular structure of the conductors, provided the material
is a near perfect conductor, but sensitive to
the actual geometry of the plates.   Attempts to measure the
force between two parallel plates were unsuccessful because of
the difficulty in trying to maintain parallelism to the required
accuracy of $10^{-5}$ radians.  For the plate and sphere, there 
is no issue of parallelism, and the system is described by the
separation of the points of closest approach.  However the Casimir
force for this geometry is not given by (\ref{eq:Parall_plates}),
but has the form
\begin{equation}
F(a)  = {2\pi a R \over 3} \left({\pi^2\over 240 a^4}\right) ,
\end{equation}
where $R$ is the radius of curvature of the spherical surface and
$a$ is the distance of closest approach.  In addition, the force
is independent of the plate area. Lamoreaux finds that the experimentally
measured force agrees with the predicted theory at the level of 5\%.
With direct confirmation of the Casimir force, we must admit that
the vacuum effects in the stress-tensor are not merely a mathematical
curiosity, but have direct physical consequences.  In general relativity
this is extremely important because the vacuum stress-tensor can 
act as a source of gravity.




\subsection{Vacuum Stress-Tensor for Two-Dimensional Stars}

In all of the examples of the preceding section for vacuum energy
densities, we studied how the effects of boundary conditions on 
the scalar field can induce the vacuum to take on a non-zero value in
flat spacetime.  We saw that the energy density near a perfectly
reflecting mirror becomes divergent on the surface to the plate. 
It is interesting to note that non-zero vacuum energies can also
be found in gravitational physics.  In some sense, the non-zero value
of the stress-tensor is more important in gravitational physics 
because it can serve as a source in Einstein's equation.  Thus the 
vacuum energy can have non-trivial effects on the evolution of the
spacetime.  The most well known, and well studied example of this
is the evaporation of black holes, originally demonstrated by Hawking
\cite{Hawk75}. In this section we will study
a slightly different problem, the vacuum energy for a massless scalar
field on the background spacetime of a constant density star. 

We begin with the metric of a static spherically symmetric star
composed of a classical incompressible constant density fluid. 
The metric for this type of star is well known \cite{Weinberg},
\begin{equation}
ds^2 =- B(r) dt^2 + A(r) dr^2 + r^2 d\Omega^2,
\end{equation}
where $d\Omega^2$ is the length element on a unit sphere, and the
functions $A(r)$ and $B(r)$ are defined by
\begin{equation}
A(r) = \left\{ \begin{array}{ll}
               \left[ 1- {2M r^2 / R^3}\right]^{-1}&\mbox{ for }r\leq R,\\
               \left[ 1- {2M/ r}\right]^{-1}&\mbox{ for }r>R,
               \end{array}\right. 
\end{equation}
and
\begin{equation}
B(r) = \left\{ \begin{array}{ll}
               {1\over 4}\left[ 3\left( 1- {2M/ R}\right)^{1/2}-
               \left(1-{2M r^2 / R^3} \right)^{1/2}\right]^2
               &\mbox{ for }r\leq R,\\
               \left[ 1- {2M / r}\right]&\mbox{ for }r>R.
               \end{array}\right. 
\end{equation}
Here $M$ is the mass of the star and $R$ its radius.  This metric
has one constraint: The mass and radius must satisfy 
\begin{equation}
R > {9\over 8}\,(2M) \, .
\label{eq:radius_limit}
\end{equation}
Otherwise the pressure at the core of the star will become singular.
This radius is just slightly larger than the Schwarzschild
radius of a black hole of the same mass. When
a star becomes sufficiently compact and its radius approaches
the above limit, then the gravitational back-reaction should start
to have a significant effect on the spacetime.  For the rest of
our calculations we will presume that condition~(\ref{eq:radius_limit})
is satisfied.  It would be a rather arduous task to find the
vacuum stress-tensor for the four-dimensional spacetime.  However,
we can make significant progress for the  two-dimensional
equivalent with the metric
\begin{equation}
ds^2 =- B(r) dt^2 + A(r) dr^2 \, .
\end{equation}
This metric can be recast into its null form
\begin{equation}
ds^2 = - B(r)\, dU\, dV\, ,
\label{eq:flat_star}
\end{equation}
where the null coordinates are defined by
\begin{equation}
U = t - r^* \qquad \mbox{ and } \qquad V = t + r^*,
\end{equation}
with $r^*$ given by
\begin{equation}
r^* = \int \sqrt{A(r)\over B(r)}\, dr.
\end{equation}
It is clear from the form of the metric~(\ref{eq:flat_star}) that
this spacetime, like all two-dimensional spacetimes, is conformally flat.
Thus, we can apply the method of finding the vacuum stress-tensor in
two-dimensional spacetimes, \cite{Dv&Ful} and Section~8.2 of \cite{Brl&Dv}.
The stress-tensor components of the vacuum, obtained when positive
frequency modes are defined by the timelike Killing vector, 
in the $(U,V)$ coordinates are
\begin{equation}
\langle 0 | T_{UU} | 0 \rangle = \langle 0 | T_{VV} | 0 \rangle = 
{1\over 192\pi} \left[ 2 B B'' - (B')^2 \right]
\end{equation}
and
\begin{equation}
\langle 0 | T_{UV} | 0 \rangle = {1\over 96\pi} B B''\, ,
\end{equation}
where the prime denotes the derivative with respect to $r$.  By
a coordinate transformation, we can calculate the components of the
stress-tensor in the original $(r,t)$ coordinates,
\begin{equation}
\langle 0 | T_{\mu\nu}(t,r) | 0 \rangle = {1\over 24 \pi}
\left[ \begin{array}{cc}
B(r)B''(r) - {1\over 4} B'(r)^2 & 0\\
0& -{1\over 4} A(r) B'(r)^2 / B(r) 
\end{array}\right].
\end{equation}
We can now directly calculate the value of the vacuum stress-tensor
both outside and inside the star.  In the exterior region
of the star, the stress-tensor is given by 
\begin{equation}
\langle 0 | T_{\mu\nu} | 0 \rangle_{exterior} = {1\over 24 \pi}
\left[ \begin{array}{cc}
(7M^2/r^4 - 4M/r^3)&0\\[8pt]
0& - (M^2/r^4)(1-2M/r)^{-2} 
\end{array}\right].
\end{equation}
This is the well known result for the Boulware vacuum of a black hole
\cite{DF&U}.
We can see that for the stationary observer at a fixed radius outside
of the star that the energy density is 
\begin{equation}
\rho_{exterior}(r) = {(7M^2/r^4 - 4M/r^3)\over 24\pi (1-2M/r)}.
\end{equation}
For the case of the star, the exterior region does not reach all the
way to the Schwarzschild radius, so there is no divergence as there 
was for a black hole.  However, the energy density for an observer fixed
at some constant distance away from the star is still negative.

\begin{figure}[hbtp] 
\begin{center}
\leavevmode\epsfxsize=0.75\textwidth\epsffile{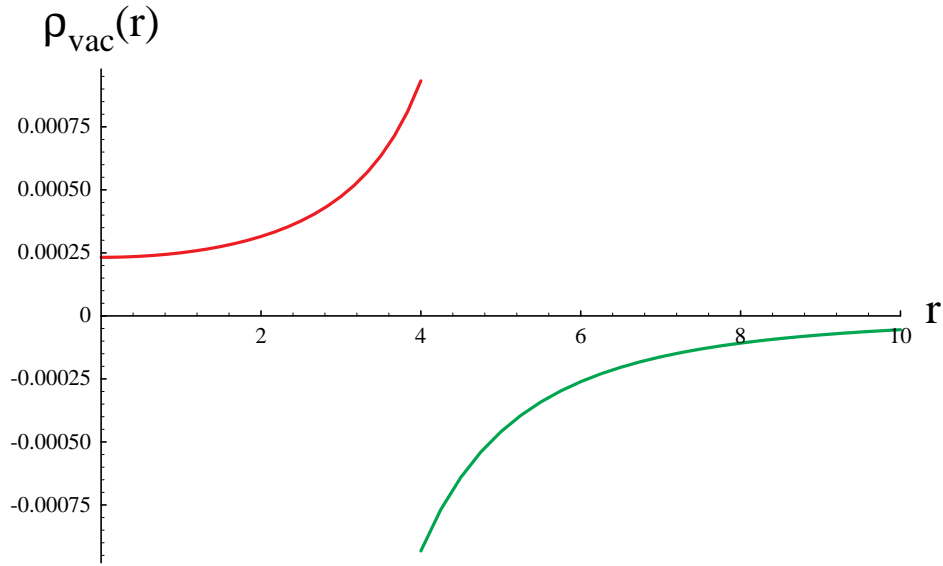}
\end{center}
\caption[Vacuum polarization for constant density stars]
{The vacuum energy density induced inside and outside
a star.  Here we have taken the star to be composed of some
form of classical matter that has a constant density throughout.
The vacuum energy is then calculated for a scalar field which
does not interact with the matter, except that it is subject to
the spacetime curvature.  Units have been normalized such that
the mass of the star is unity.  The radius of the star is 4 units.}
\label{Fig:star_vac}
\end{figure}

The situation is somewhat different in the interior of the star.
The stress-tensor components are given by
\begin{equation}
\langle 0 | T_{\mu\nu} | 0 \rangle_{interior} = {A(r)\over 24 \pi} 
\left[ \begin{array}{cc}
B(r){M \over R^3}\left[ 3 \sqrt{A(r) \left( 1 - {2M\over R}\right)}
-1+{M r^2\over R^3}\right]&0\\[8pt]
0& - A(r){M^2 r^2 \over R^6}
\end{array}\right].
\end{equation}
The vacuum energy density for the interior of the star as seen by a
stationary observer is
\begin{equation}
\rho_{interior}(r) = {1\over 24 \pi} \left( 1- {2Mr^2\over R^3} \right)^{-1}
\left({M\over R^3}\right)\left[ 3 \sqrt{A(r) \left( 1 - {2M\over R}\right)}
-1+{Mr^2\over R^3}\right].
\end{equation}
It is not immediately obvious, but the energy density in the interior
is everywhere positive, and grows in magnitude as we move outward from
the center of the star, as plotted in Figure~\ref{Fig:star_vac}.
In contrast, the vacuum energy outside the star is negative and
identical to that of the black hole in the Boulware vacuum state.
The discontinuity in the vacuum energy is an effect of the star
having a very abrupt change in the mass density when crossing
the surface of the star.  We presume that if the star makes a
continuous transition from the constant density core to the
vacuum exterior then the vacuum energy would smoothly pass 
through zero.
\begin{figure}[hp] 
\begin{center}
\leavevmode\epsfxsize=0.75\textwidth\epsffile{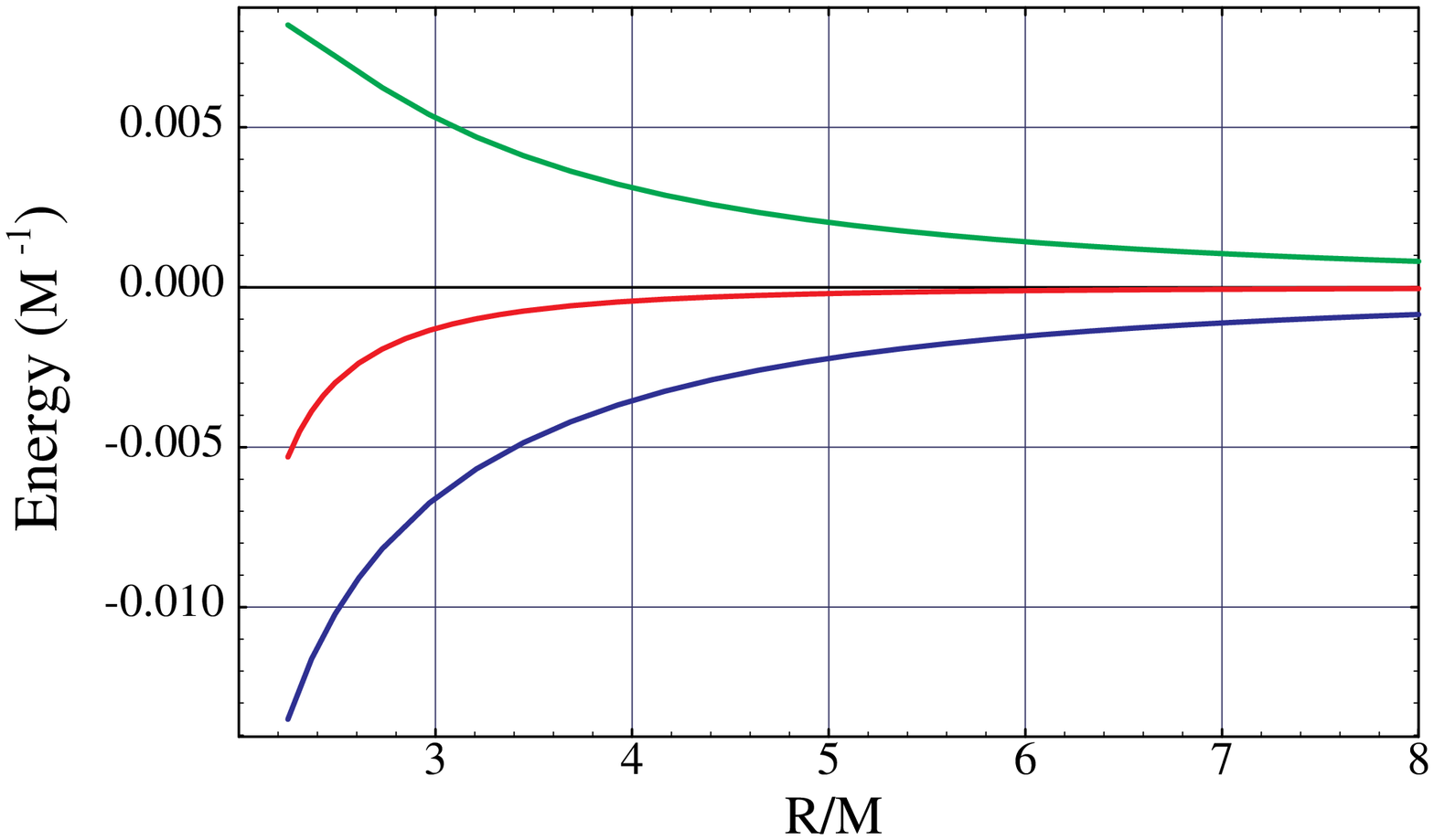}
\end{center}
\caption[Total vacuum energy for constant density stars]
{The total integrated vacuum energy density induced
inside and outside  a star as a function of the radius.  The top line is
the contribution from the positive energy in the interior of the star.
The lower line is the contribution from the exterior Boulware vacuum
energy.  The middle line is the total over all space.   The energy
is in units of inverse mass. }
\label{Fig:star_Etot}
\end{figure}
\goodbreak

The total vacuum energy of this spacetime can be found by
integrating the local energy density over all space.  The total
positive vacuum energy contained inside the star is
\begin{equation}
E_{inside} = 2 \int_0^R \rho(r)\, dr\, 
= \, {1\over 48 \pi R^2}\left( 10M-\sqrt{2MR}\,\tanh^{-1}\sqrt{2M\over R}
\right).\label{eq:E_inside}
\end{equation}
The contribution outside the star gives a total finite negative
energy 
\begin{equation}
E_{outside} = 2 \int_R^\infty \rho(r)\, dr\, 
= \,-{1\over 96 \pi }\left[{14 M\over R^2}  - {2 \over R} -{1 \over M}
\ln \left( 1 - {2 M\over R}\right) \right].\label{eq:E_outside}
\end{equation} 
The total energy due to the vacuum is the sum 
\begin{equation}
E_{Total} = E_{inside} + E_{outside}\, .
\end{equation}
This is always finite and negative, as shown in 
Figure~\ref{Fig:star_Etot} along with the individual contributions 
from the interior and exterior of the star.  
For very diffuse stars, $R \gg 2M$, the total positive interior vacuum
energy very nearly compensates for the negative energy outside the star.
In this limit the leading contributions to the total energy go as
\begin{equation}
E_{Total}\sim -{M^2\over 18 \pi R^3} - {3 M^3 \over 40 \pi R^4} - O(R^{-5})-
\cdots.
\end{equation}
In the other regime, when the star becomes very compact, the Boulware
vacuum begins to dominate and in the limit $R \rightarrow 9M/4$, the
energy density goes to its finite minimum value $E_{min} = -5.306574  
\times 10^{-3} \, M^{-1}$.  This residual energy can be considered
a mass renormalization of the star due to the vacuum polarization, 
\begin{equation}
M = M_{bare} + \delta M = M_{bare} - { 5.306574\times 10^{-3} \over M}
M_{Planck}^2.
\end{equation}
Unless the mass of the star is very close to the Planck mass, $M_{Planck}$,
this correction to the effective mass of the star is extremely small,
accounting for at most half a percent of the star's mass.  


\chapter{Derivation of the Quantum Inequalities}\label{chapt:QI_proof}

\section{General Theory for Quantized Scalar Fields}

We begin by considering the semiclassical theory of gravity, where the
classical Einstein tensor on the left-hand side of Eq.~(\ref{eq:Einstein})
is equal to the expectation value of the stress-energy tensor 
(stress-tensor) of a quantized field on the right-hand side,
\begin{equation}
G_{\mu\nu} = 8 \pi \;\langle T_{\mu\nu} \rangle \; .
\label{eq:Einstein}
\end{equation}
Here we are using Planck units, in which $\hbar = c = G = 1$.
We will take the stress-tensor to be that of the massive, minimally
coupled scalar field, $\phi(x)$, given by 
\begin{equation}
T_{\mu\nu} = \phi_{;\mu} \phi_{;\nu} - {1\over 2} 
g_{\mu\nu}
 \phi^{;\rho} \phi_{;\rho} + {1\over 2} m^2 g_{\mu\nu}\phi^2\; ,
\label{eq:Stress_tensor}
\end{equation}
where $\phi_{;\alpha}$ denotes the covariant derivative of $\phi$ on the
classical background metric and $m$ is the mass of the field.  We shall
develop quantum inequalities in globally static spacetimes, those in which
$\partial_t$ is a timelike Killing vector. Such a metric can be written
in the form
\begin{equation}
{ds}^2 = -|g_{tt}({\bf x})|{dt}^2 + g_{ij}({\bf x})dx^i dx^j \;,
\label{eq:metric}
\end{equation}
where the function $g_{tt}({\bf x})$ is related to the red or blue shift
dependent only on the observer's position in space,
and $g_{ij}({\bf x})$ is the metric of the spacelike hypersurfaces that are
orthogonal to the Killing vector in the time direction. With this metric, 
the wave equation
\begin{equation}
\Box\phi - m^2 \phi  = 
{1\over \sqrt{|g|}}\left(\partial_\alpha \sqrt{|g|} g^{\alpha\beta}
\partial_\beta \phi\right)- m^2\phi = 0
\end{equation}
becomes
\begin{equation}
 -{1\over |g_{tt}|}\partial_t^2\phi +
\nabla^i \nabla_i \phi - m^2\phi = 0,
\end{equation}
where $g={\rm det}(g_{\mu\nu})$. The timelike Killing vector allows us
to use a separation of variables to find solutions of the wave equation.
The positive frequency mode function solutions can be written as
\begin{equation}
f_\lambda({\bf x},t) = \,U_\lambda ({\bf x}) \, e^{-i\omega t}.
\end{equation}
The label $\lambda$ represents the set of quantum numbers necessary to
specify the mode. Additionally, the mode functions should have unit 
Klein-Gordon norm
\begin{equation}
\left(f_\lambda, f_{\lambda'}\right) = \delta_{\lambda\lambda'}.
\end{equation} 

A general solution of the scalar field $\phi$ can then be 
expanded in terms of creation and annihilation operators as
\begin{equation}
\phi = \sum_{\lambda} {\bigl(a_{\lambda}f_{\lambda} + a^\dagger_{\lambda}
f^\ast_{\lambda}\bigr)},
\end{equation}
when quantization is carried out over a finite box or universe.  If 
the spacetime is itself infinite, then we replace the summation by
an integral over all of the possible modes.  The creation and annihilation
operators satisfy the usual commutation relations \cite{Brl&Dv}.

In principle, quantum inequalities can be found for any geodesic 
observer \cite{F&Ro95}. In many curved spacetimes, static observers
often view the universe as having certain symmetries. By making use
of these symmetries, we can often simplify calculations or equations.
To observers moving along timelike geodesics the symmetries of the
rest observers may not be ``observable.''  In the most general sense,
the mode functions of the wave equation in a moving frame may become
quite complicated. Thus, we will concern ourselves only with static
observers, whose four-velocity, $u^\mu = ( {1\over \sqrt{|g_{tt}|}} ,
{\bf 0 })$, is in the direction of the timelike Killing vector. 
The energy density that such an observer measures is given by
\begin{equation}
\rho = T_{\mu\nu} u^\mu u^\nu = {1\over |g_{tt}|}T_{tt} = 
{1\over 2}\left[ {1\over |g_{tt}|}(\partial_t\phi)^2 +\nabla^j\phi \nabla_j\phi
+ m^2\phi^2\right]\,. \label{eq:rho}
\end{equation}
Upon substitution of the mode function expansion into  
Eq.~(\ref{eq:rho}), we find
\begin{eqnarray}
\rho & = & {\rm Re}\sum_{\lambda\lambda'} \left\{{{\omega\omega'}\over 
|g_{tt}|}
\left[ U_\lambda^* U_{\lambda'} e^{+i(\omega-\omega')t}a_{\lambda}^\dagger
a_{\lambda'}-  U_\lambda U_{\lambda'} e^{-i(\omega+\omega')t}a_{\lambda} 
a_{\lambda'}\right]\right. \nonumber\\
&&\qquad+\left[ \nabla^j U_\lambda^* \nabla_j 
U_{\lambda'} e^{+i(\omega-\omega')t}a_{\lambda}^\dagger a_{\lambda'}+
\nabla^j U_\lambda \nabla_j U_{\lambda'} e^{-i(\omega+\omega')t}
a_{\lambda} a_{\lambda'}\right]\nonumber\\
&&\qquad +m^2 \left.\left[ U_\lambda^* U_{\lambda'} e^{+i(\omega-\omega')t}
a_\lambda^\dagger a_{\lambda'}+  U_\lambda U_{\lambda'} 
e^{-i(\omega+\omega')t}a_{\lambda} 
a_{\lambda'}\right] \right\} \nonumber\\
&& +{1\over 2} \sum_\lambda \left({\omega^2\over |g_{tt}|} U_\lambda^* 
U_\lambda + \nabla^j U_\lambda^* \nabla_j U_\lambda +m^2  U_\lambda^* 
U_\lambda\right)\, .
\end{eqnarray}

The last term is the expectation value in the vacuum state, defined by
$a_\lambda |0\rangle = 0$ for all $\lambda$, and is formally divergent.
The vacuum energy density may be defined by a suitable regularization and 
renormalization procedure, discussed in more detail later in 
Sections~\ref{subsec:RW_massless} and \ref{subsec:RW_Einstein}. In
general, however, it is not uniquely defined.
This ambiguity may be side-stepped by concentrating attention upon
the difference between the energy density in an arbitrary state and that
in the vacuum state, as was done by Ford and Roman \cite{F&Ro95}. 
We will therefore concern ourselves primarily with the difference defined
by
\begin{equation}
:\rho:\; = \rho - \langle 0 |\rho | 0\rangle,
\end{equation}
where $| 0\rangle$ represents the Fock vacuum state defined by
the global timelike Killing vector. 

The renormalized energy density as defined above is valid along the
entire worldline of the the observer.  However, let us suppose that the
energy density is only sampled along some finite interval of the
geodesic.  This may be accomplished by means of a weighting function
which has a characteristic time $t_0$. The Lorentzian function,
\begin{equation}
f(t) = {t_0 \over\pi}\,{1\over {t^2 +t_0^2}}\, , 
\label{eq:Lorentzian}
\end{equation}
is a good choice.  The integral over all time of $f(t)$ yields unity
and the width of the Lorentzian is characterized by $t_0$. Using such
a weighting function, we find that the averaged energy difference is
given by
\begin{eqnarray} 
\Delta\hat \rho &\equiv& {t_0\over\pi}\int_{-\infty}^\infty
{{\langle :T_{tt}/|g_{tt}|:\rangle dt} \over{t^2 + t_0^2}}\nonumber\\
& = & {\rm Re}\sum_{\lambda\lambda'}\left\{{{\omega\omega'}\over |g_{tt}|}
\left[ U_\lambda^* U_{\lambda'} e^{-|\omega-\omega'|t_0}\langle 
a_{\lambda}^\dagger a_{\lambda'}\rangle - U_\lambda U_{\lambda'} 
e^{-(\omega+\omega')t_0}\langle a_{\lambda} a_{\lambda'}\rangle\right]
\right.\nonumber\\
&& \qquad + \left[ \nabla^j U_\lambda^* \nabla_j 
U_{\lambda'} e^{-|\omega-\omega'|t_0}\langle a_{\lambda}^\dagger
a_{\lambda'}\rangle+\nabla^j U_\lambda \nabla_j U_{\lambda'}
e^{-(\omega+\omega')t_0}\langle a_{\lambda} a_{\lambda'}\rangle\right]
\nonumber\\
& &\qquad + \, m^2 \left.\left[ U_\lambda^* U_{\lambda'} 
e^{-|\omega-\omega'|t_0}\langle a_{\lambda}^\dagger a_{\lambda'}\rangle 
+  U_\lambda U_{\lambda'} e^{-(\omega+\omega')t_0}\langle a_{\lambda} 
a_{\lambda'}\rangle\right] \right\}\, . \label{eq:DeltaRho}
\end{eqnarray}

We are seeking a lower bound on this quantity. It has been shown
\cite{Ford91,F&Ro97} that
\begin{equation}
{\rm Re}\sum_{\lambda\lambda'} f(\lambda)^* f(\lambda')
e^{-|\omega-\omega'|t_0} \langle a_{\lambda}^\dagger a_{\lambda'}\rangle
\geq {\rm Re}\sum_{\lambda\lambda'} f(\lambda)^* f(\lambda')
e^{-(\omega+\omega')t_0}\langle a_{\lambda}^\dagger a_{\lambda'}\rangle.
\label{eq:first_inequality}\end{equation}
Upon substitution of this into Eq.~(\ref{eq:DeltaRho}) we have
\begin{eqnarray} 
\Delta\hat \rho & \geq & {\rm Re}\sum_{\lambda\lambda'}\left\{{{\omega\omega'}
\over |g_{tt}|}\left[ U_\lambda^* U_{\lambda'} \langle a_\lambda^\dagger
a_{\lambda'}\rangle - U_\lambda U_{\lambda'}\langle a_\lambda 
a_{\lambda'}\rangle\right]\right.\nonumber\\
& &\qquad +  \left[ \nabla^j  U_\lambda^* \nabla_j U_{\lambda'}
\langle a_\lambda^\dagger a_{\lambda'}\rangle+
\nabla^j U_\lambda \nabla_j U_{\lambda'}\langle a_\lambda 
a_{\lambda'}\rangle\right]\nonumber\\
& &\qquad + m^2\left. 
\left[ U_\lambda^* U_{\lambda'}\langle a_\lambda^\dagger
a_{\lambda'}\rangle +  U_\lambda U_{\lambda'}  \langle a_{\lambda} 
a_{\lambda'}\rangle\right]\right\} e^{-(\omega+\omega')t_0}\,.
\label{eq:ineq1}
\end{eqnarray}  
We may now apply the inequalities proven in Appendix~\ref{sec:appendix}.
For the first and
third term of Eq.~(\ref{eq:ineq1}), apply Eq.~(\ref{eq:lowbound3}) with
$h_\lambda = \omega/\sqrt{|g_{tt}|}\, U_\lambda {\rm e}^{-\omega t_0}$ and
$h_\lambda = m\, U_\lambda {\rm e}^{-\omega t_0}$, respectively. For the 
second term of Eq.~(\ref{eq:ineq1}), apply Eq.~(\ref{eq:lowbound}) with
$A_{ij} = g_{ij}$ and 
$h_\lambda^i = \nabla^i U_\lambda {\rm e}^{-\omega t_0}$. The result is
\begin{equation}
\Delta\hat \rho \geq -{1\over 2} \sum_\lambda \left({\omega_\lambda^2\over
|g_{tt}|} U_\lambda^* U_{\lambda} + \nabla^j U_\lambda^* \, \nabla_j U_\lambda
+ m^2 U_\lambda^* U_{\lambda}\right)e^{-2\omega_\lambda t_0}.
\end{equation}
This inequality may be rewritten using the Helmholtz equation satisfied
by the spatial mode functions:
\begin{equation}
\nabla^j\nabla_j U_\lambda +(\omega^2/|g_{tt}| - m^2)U_\lambda =0\,,
\end{equation}
to obtain
\begin{equation}
\Delta\hat \rho \geq - \sum_\lambda \left({\omega_\lambda^2\over |g_{tt}|}
 + {1\over 4}\nabla^j\nabla_j\right) |U_\lambda|^2 {\rm e}^{-2
\omega_\lambda t_0}\, .   \label{eq:qi2}
\end{equation}
This can be rewritten as
\begin{equation}
\Delta\hat \rho \geq - {1\over 4}
\left({\partial_{t_0}^2\over |g_{tt}|} + \nabla^j\nabla_j\right)
\sum_\lambda |U_\lambda({\bf x})|^2 {\rm e}^{-2\omega_\lambda t_0}\, .
\label{eq:general_qi}
\end{equation}
There is a more compact notation in which Eq.~(\ref{eq:general_qi})
may be expressed.  If we take the original metric, Eq.~(\ref{eq:metric}),
and Euclideanize the time by allowing $t \rightarrow it_0$, then the
Euclidean box operator is defined by
\begin{equation}
\Box_E \equiv {\partial_{t_0}^2\over |g_{tt}|} + \nabla^j\nabla_j\, .
\end{equation}
The analytic continuation of the Feynman Green's Function to imaginary
time yields the Euclidean two-point function.  The two are related by
\begin{equation}
G_E({\bf x},t;{\bf x}',t') = i G_F({\bf x},-it;{\bf x}',-it').
\label{eq:Euclideanize}
\end{equation}
In terms of the mode function expansion, the Euclidean Green's function
is given by
\begin{equation}
G_E({\bf x},-t_0;{\bf x},+t_0) = \sum_\lambda |U_\lambda({\bf x})|^2 
{\rm e}^{-2\omega_\lambda t_0} ,
\end{equation}
where the spatial separation is allowed to go to zero but the time
separation is $2t_0$.

This allows us to write the quantum inequality for a static observer
in any static curved spacetime as
\begin{equation}
\Delta\hat \rho \geq - {1\over 4}\Box_E \, G_E({\bf x},-t_0;{\bf x},+t_0).
\label{eq:QI}
\end{equation}
Given a metric which admits a global timelike Killing vector, we 
can immediately calculate the limitations on
the negative energy densities by either of the two methods.  If
we know the solutions to the wave equation, then we may construct
the inequality from the summation of the mode functions. More
elegantly, if the Feynman two-point function is known in the
spacetime, we may immediately calculate the inequality by
first Euclideanizing and then taking the appropriate derivatives. 

It is important to note that while the local energy density may be
more negative in a given quantum state than in the vacuum, the total
energy difference integrated over all space is always non-negative.
This follows because the normal-ordered Hamiltonian,
\begin{equation}
:H: \; = \int :T_{tt}:\, \sqrt{-g}\, d^nx = \sum_\lambda \omega_\lambda \,
a_\lambda^\dagger a_\lambda \, ,
\end{equation}
is a positive-definite operator; so $\langle :H: \rangle \geq 0$.

  
\section{Quantum Averaged Weak Energy Condition}\label{sec:QAWEC}
Let us return to the form of the quantum
inequality given by Eq.~(\ref{eq:qi2}),
\begin{equation}
\Delta\hat \rho \geq - \sum_\lambda \left({\omega_\lambda^2\over |g_{tt}|}
 + {1\over 4}\nabla^j\nabla_j\right) |U_\lambda({\bf x})|^2 {\rm e}^{-2
\omega_\lambda t_0}\, .                           
\end{equation}
Since we are working in static spacetimes, the vacuum energy 
does not evolve with time, so we can rewrite this equation simply
by adding the renormalized vacuum energy density $\rho_{vacuum}$
to both sides. This is the vacuum in which the mode functions are
defined to have positive frequency with respect to the timelike 
Killing vector.  We then have
\begin{equation}
\hat\rho_{Ren.} \geq - \sum_\lambda \left({\omega_\lambda^2\over |g_{tt}|}
 + {1\over 4}\nabla^j\nabla_j\right) |U_\lambda({\bf x})|^2 {\rm e}^{-2
\omega_\lambda t_0} + \rho_{vacuum}({\bf x})\, ,                           
\end{equation}
where $\hat\rho_{Ren.}$ is the sampled, renormalized energy density
in any quantum state.  Taking the limit of the sampling time
$t_0 \rightarrow \infty$, we find (under the assumption that there exist
no modes which have $\omega_\lambda = 0$) that
\begin{equation}
\lim_{t_0\rightarrow\infty} {t_0\over\pi} \int_{-\infty}^\infty 
{\langle T_{tt}/|g_{tt}| \rangle_{Ren.} \over  t^2 + t_0^2} dt \geq
 \rho_{vacuum}({\bf x}).
\end{equation}
This leads directly to the {\it Quantum Averaged Weak Energy Condition}
for static observers \cite{F&Ro95},
\begin{equation}
\int_{-\infty}^{+\infty} \left( \langle \psi | T_{tt}/|g_{tt}| 
|\psi\rangle_{Ren.} - \rho_{vacuum} \right) dt \geq 0 .
\end{equation} 
This is a departure from the classical averaged weak energy condition,
\begin{equation}
\int_{-\infty}^{+\infty} \langle \psi | T_{tt}/|g_{tt}|\,
 |\psi\rangle_{Ren.} dt \geq 0 .
\end{equation}
We see that the derivation of the QAWEC leads to the measured energy
density along the observer's worldline being bounded below by the vacuum
energy.  

Recently, there has been considerable discussion about the vacuum
energy and to what extent it violates the classical energy conditions.
For example, Visser looked at the specific case of the violation
of classical energy conditions for the Boulware, Hartle-Hawking, and
Unruh vacuum states \cite{Viss96a,Viss96b,Viss96c,Viss97a} around a 
black hole.  However the vacuum energy is not a classical phenomenon,
so we should not expect it to obey the classical energy constraints.
From the QAWEC we see that the sampled energy density is bounded below
by the vacuum energy in the long sampling time limit.  

\section{Expansion of the QI for Short Sampling Times}\label{sec:expans}

We now consider the expansion of the two-point function
for small times.  We assume that the two-point function has the
Hadamard form 
\begin{equation}
G(x, x') = {i\over 8\pi^2} \left[ {\Delta^{1/2} \over \sigma + i\epsilon}
+ V \ln(\sigma + i\epsilon) +W\right]\, ,
\end{equation}
where $2\sigma(x, x')$ is the square of the geodesic distance between
the spacetime points $x$ and $x'$,
\begin{equation}
\Delta\equiv - g^{-1/2}(x)\, {\rm det}(\sigma_{;ab'})\: g^{-1/2}(x'),
\end{equation}
is the Van Vleck-Morette determinant, and $V(x, x')$ and $W(x, x')$ 
are regular biscalar functions. In general, these functions can be
expanded in a Taylor series in powers of $\sigma$ \cite{B&OT86},
\begin{equation}
V(x,x') = \sum_{n=0}^\infty V_n(x,x') \sigma^n \, ,
\end{equation}
where $V_n$, and similarly $W_n$, are also regular biscalar functions with
\begin{eqnarray}
V_0 &=& v_0 - {1\over 2} v_{0;a}\,\sigma^a + {1\over 2}v_{0ab}\,\sigma^a 
\sigma^b + {1\over 6} (-{3\over 2}v_{0ab;c}+ {1\over 4}v_{0;(abc)})\sigma^a 
\sigma^b \sigma^c +\cdots ,\\[8pt]
V_1 &=& v_1 - {1\over 2} v_{1;a}\,\sigma^a + \cdots,
\end{eqnarray}
and $\sigma^\nu = \sigma^{;\nu}$.  The coefficients, $v_0$, $v_{0ab}$,
\ldots are strictly geometrical objects given by
\begin{eqnarray}
v_0 &=& {1\over 2} \left[ (\xi - {1\over 6}) R + m^2 \right] ,\\[8pt]
v_{0ab} &=& -{1\over 180} R_{pqra} {R^{pqr}}_{b}  -{1\over 180} R_{apbq}R^{pq}
+{1\over 90} R_{ap}{R_b}^p - {1\over 120}\Box R_{ab} \nonumber\\
&&+ ({1\over 6}\xi -
{1\over 40}) R_{;ab} + {1\over 12}(\xi - {1\over 6}) R R_{ab} +
{1\over 12} m^2 R_{ab}\, ,\\[8pt]
v_1 &=& {1\over 720} R_{pqrs}R^{pqrs} - {1\over 720} R_{pq}R^{pq}
-{1\over 24}(\xi - {1\over 5})\Box R + {1\over 8}(\xi -{1\over 6})^2 R^2
\nonumber\\ &&+ {1\over 4}m^2 (\xi-{1\over 6})R + {1\over 8} m^4.
\end{eqnarray}
We can then express the Green's function as \cite{B&OT86}
\begin{eqnarray}
G(x,x') &=& {i\over 8\pi^2} \left\{ {1+{1\over 12}R_{ab}\sigma^a \sigma^b\
- \cdots \over \sigma + i\epsilon} + \left[ (v_0 - {1\over 2}v_{0;a}
\sigma^a + {1\over 2}v_{0ab} \sigma^a \sigma^b\ + \cdots ) \right. \right. 
\nonumber\\ &&\qquad \left. \left. +( v_1 -
{1\over 2} v_{1;a}\sigma^a + \cdots)\sigma + \cdots \right] \ln 
( \sigma + i\epsilon ) + W \right\},
\end{eqnarray}
where we have also used the Taylor series expansion of the 
Van Vleck-Morette determinant \cite{B&OT86},
\begin{equation}
\Delta^{1/2} = 1 + {1\over 12}R_{ab}\sigma^a \sigma^b
- {1\over 24}R_{ab;c}\sigma^a \sigma^b \sigma^c + \cdots \, .
\end{equation}
The state-dependent part of the Green's function, $W$, is neglected 
because it is regular as $\sigma \rightarrow 0$.  The dominant
contributions to the quantum inequality come from the divergent
portions of the Green's function in the $\sigma \rightarrow 0$
limit.

\begin{figure}
\begin{center}
\leavevmode\epsffile{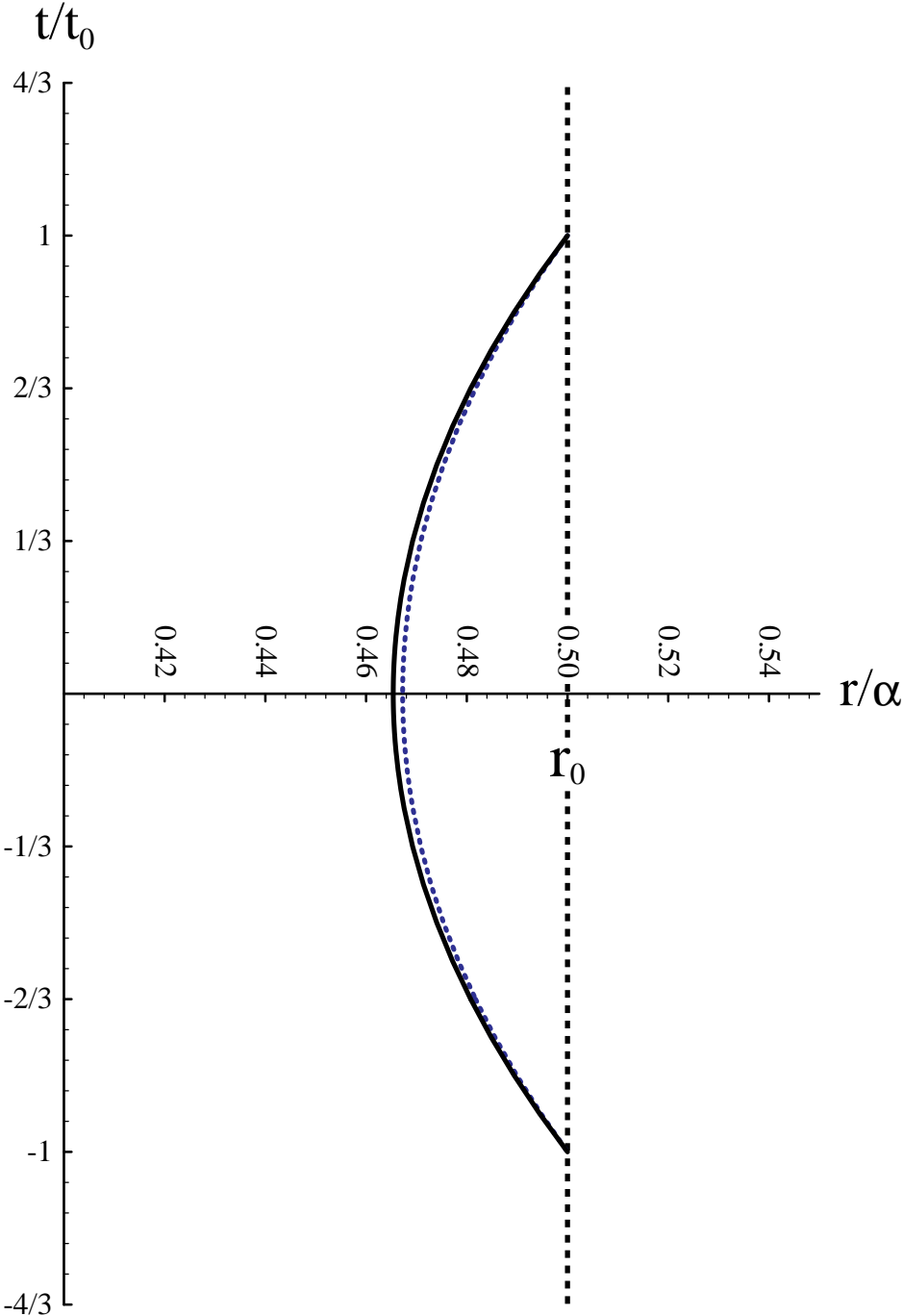}
\end{center}
\caption[Exact and approximate geodesics in de~Sitter spacetime]
{The exact geodesic path (dotted line) and
the parabolic approximation (solid line) in de~Sitter spacetime.
The coordinate distance from $r=0$ to the horizon is $\alpha$ in
the static coordinates. }
\label{fig:geodesicpath}
\end{figure}

We must now determine the geodesic distance between two spacetime
points, along a curve starting at $({\bf x}_0,-t_0)$ and ending at
$({\bf x}_0,+t_0)$.  For spacetimes in which $|g_{tt}| = 1$,
the geodesic path between them is a straight line.
Therefore, the geodesic distance is simply $2t_0$.  However, in a more
generic static spacetime where $g_{tt}({\bf x})$ is not constant,
the geodesic path between the  points is a curve,
with the observer's spatial position changing throughout time.  Thus,
we must now solve the equations of motion to find the geodesic distance
between the spacetime points.  In terms
of an affine parameter $\lambda$, the geodesic equations are found to be
\begin{equation}
{dt\over d\lambda} - {a_t \over |g_{tt}({\bf x}(\lambda) )|}  =  0
\end{equation}
and
\begin{equation}
{d^2x^i \over d\lambda^2} + {\Gamma^i}_{\mu\nu} {dx^\mu \over d\lambda}
{dx^\nu \over d\lambda}  =  0 \, ,
\end{equation}
where $a_t$ is an unspecified constant of integration.  The Christoffel
coefficients are
\begin{eqnarray}
{\Gamma^i}_{tt} & = & {1\over 2} g^{ij}\, |g_{tt}|_{,j} \, ,\cr
{\Gamma^i}_{tj} & = & 0 \, ,\cr
{\Gamma^i}_{jk} & = & {1\over 2} g^{im}\, \left( g_{mj,k} + g_{mk,j} -
g_{jk,m} \right)\, .
\end{eqnarray}
It is possible to eliminate $\lambda$ from the position equations,
and write
\begin{equation}
{d^2 x^i \over dt^2} + {1\over 2}  |{g_{tt}}|^{,i} + 
{\Gamma^i}_{jk} {dx^j \over dt} {dx^k \over dt} +  {|g_{tt}|_{,k} \over
|g_{tt}| }{dx^i \over dt} {dx^k \over dt} = 0.
\end{equation}
Now if we make the assumption that the velocity of the observer moving
along this geodesic is small, then to lowest order the second term can
be considered nearly constant, and all the velocity-dependent terms are
neglected.  It is then possible to integrate the equation exactly,
subject to the above endpoint conditions, to find 
\begin{equation}
x^i(t) \approx  -{1\over 4} {|g_{tt}|}^{,i}_{{\bf x} = {\bf x_0}} 
(t^2 -t_0^2) + x_0^i \, .
\label{eq:geodesicpath}
\end{equation}
We see that the geodesics are approximated by parabol\ae, as would be 
expected in the Newtonian limit.  A comparison of the exact solution to
the geodesic equations and the approximation is shown in
Figure~\ref{fig:geodesicpath} for the specific case of de~Sitter spacetime.
We see that the approximate path very nearly fits the exact path in the
range $-t_0$ to $+t_0$.

The geodesic distance between two spacetime points,
where the starting and ending spatial positions are the same,
is given by
\begin{equation}
\Delta s = \int_{-t_0}^{+t_0} \sqrt{ -|g_{tt}\left({\bf x}(t)\right)| +
 g_{ij}\left({\bf x}(t)\right) {dx^i\over dt}
{dx^j\over dt}} dt\, .
\end{equation}
In order to carry out the integration, let us define 
\begin{equation}
f(t) \equiv \sqrt{ -|g_{tt}\left({\bf x}(t)\right)| + 
g_{ij}\left({\bf x}(t)\right) {dx^i\over dt}{dx^j\over dt}}\, .
\end{equation}
We can expand $f(t)$ in powers of $t$ centered around $t = 0$, 
and then carry out the integration to find the geodesic distance.
The parameter $\sigma$ can now be written as
\begin{eqnarray}
\sigma({\bf x}_0,t_0) & = &
\sigma({\bf x}_0,-t_0;{\bf x}_0,+t_0) = {1\over 2}\, {\Delta s}^2\nonumber\\
& = & 2 f^2(0)\, t_0^2 + {2\over 3}
f(0)f^{''}(0) \, t_0^4 + {1\over 6}\left[ {1\over 5} f(0) f^{(IV)}(0) + 
{1\over 3} f^{''}(0)^2\right] \, t_0^6 + \cdots.\nonumber\\
\end{eqnarray}
However, we do not know the values of the metric at the 
time $t = 0$, but we do at the initial or final positions.
Therefore,  we expand the functions $f(0)$ around the time $-t_0$.
Upon using Eq.~(\ref{eq:geodesicpath}), we find
\begin{equation}
\sigma({\bf x_0},t_0) \approx -2 |g_{tt}({\bf x_0})|\, t_0^2 - {1\over 6}\,
{g_{tt}}^{,i}({\bf x_0}) \, g_{tt,i}({\bf x_0})\, t_0^4+\cdots,
\end{equation}
and
\begin{equation}
\sigma^t({\bf x_0},t_0) \approx 2 t_0 + {1\over 3}
{{g_{tt}}^{,i}({\bf x_0}) \, g_{tt,i}({\bf x_0}) \over
|g_{tt}({\bf x_0})|} \, t_0^3 + \cdots.
\end{equation}
In any further calculations, we will drop the ${\bf x_0}$ notation, with
the understanding that all of the further metric elements are evaluated at
the starting point of the geodesic.  
Using Eq.~(\ref{eq:Euclideanize}) we can then write the Euclidean
Green's function needed to derive the quantum inequality in increasing
powers of $t_0$ as
\begin{eqnarray}
G_E(x,t_0) &=& {1\over 8 \pi^2} \left[ { {1 - O(t_0^2) + \cdots}\over 
{2 |g_{tt}|\, t_0^2 - {1\over 6} {g_{tt}}^{,i} g_{tt,i}\, t_0^4}}
+ v_0 \ln ( 2 |g_{tt}|\, t_0^2 - {1\over 6} {g_{tt}}^{,i} g_{tt,i}\,
t_0^4)\right.\nonumber\\ 
&&+ \left.\left( v_{0,k}\, |{g_{tt}}|^{,k} +  2 v_1 |g_{tt}|
- 2 v_{000} \right) t_0^2 \ln ( 2 |g_{tt}|\, t_0^2 - {1\over 6}
{g_{tt}}^{,i} g_{tt,i}\, t_0^4)+\cdots \right].\nonumber\\
\end{eqnarray} 
Notice that none of the geometric terms, such as
$v_0$, changes during  Euclideanization because they are time independent.
The quantum inequality, (\ref{eq:QI}), can be written as
\begin{equation}
\Delta\hat \rho \geq - {1\over 4} \left( {1\over |g_{tt}|} 
\partial_{t_0}^2 + \nabla^i \nabla_i \right) \,
 G_E({\bf x},-t_0;{\bf x},+t_0). 
\end{equation}
If we insert the Taylor series expansion for the Euclidean Green's
function into the above expression and collect terms in powers of
the proper sampling time $\tau_0$, related to $t_0$
by $\tau_0 = |g_{tt}|^{1/2} t_0$, we can write the above expression as
\begin{eqnarray}
\Delta\hat\rho &\geq& -{3\over 32 \pi^2 \tau_0^4} \left[ \rule{0in}{.25in} 
1+{1\over 3}\left( {1\over 2}g_{tt}\, \nabla^j \nabla_j\; g_{tt}^{-1}
+{1\over 6} R - m^2 \right)\,\tau_0^2 \right. \nonumber \\
&&\qquad +{1\over 3}\left.\rule{0in}{.25in} \left( {1\over 6} R_{,k} 
{{g_{tt}}^{,k}
\over g_{tt}} -{1\over 12} \nabla^j \nabla_j \; R + 4 v_1 - 4 {v_{000}
\over |g_{tt}|} \right) \tau_0^4 \ln (2\tau_0^2) +O(\tau_0^4) + \cdots\right].
\nonumber\\
\label{eq:QI_expansion}
\end{eqnarray}
In the limit of $\tau_0 \rightarrow 0$, the dominant term of the above
expression reduces to 
\begin{equation}
\Delta\hat\rho \geq - {3\over 32 \pi^2 \tau_0^4} \, ,\label{eq:asympt_flat}
\end{equation}
which is the quantum inequality in four-dimensional Minkowski space
\cite{F&Ro95,F&Ro97}.  Thus, the term in the square brackets in 
Eq.~(\ref{eq:QI_expansion}) is the short sampling time expansion of
the ``scale'' function \cite{Pfen97a}, and does indeed reduce to one in
the limit of the sampling time tending to zero.  The range of sampling
times for which a curved spacetime can be considered ``roughly'' flat
is determined by the first non-zero correction term in 
Eq.~(\ref{eq:QI_expansion}). If this happens to be the $\tau_0^2$
term, then for the first correction to be small compared to one
implies
\begin{equation}
\tau_0 \ll \left| {1\over 2}g_{tt}\, \nabla^i \nabla_i\; g_{tt}^{-1}
+{1\over 6} R - m^2 \right|^{-1/2}\, .\label{eq:little_t}
\end{equation}
Each of the three terms on the right-hand side of this relation
has a different significance.
The $m^2$ term simply reflects that for a massive scalar field,
Eq.~(\ref{eq:QI_expansion}) is valid only when the sampling time is small
compared to the Compton time.  If we are interested in the massless scalar
field, this term is absent.  The scalar curvature term, if it is dominant,
indicates that the flat space inequality is valid only on scales which are
small compared to the local radius of curvature.
This was argued on the basis of the equivalence principle by various
authors \cite{F&Ro96,Pfen97b,E&Ro97}, but is now given a more rigorous
demonstration.  The most mysterious term in Eq.~(\ref{eq:little_t})
is that involving $g_{tt}$.  Typically, this term dominates when 
the observer sits at rest near a spacetime horizon. In this case,
the horizon acts as a boundary, so Eq.~(\ref{eq:little_t}) requires
$\tau_0$ to be small compared to the proper distance to the boundary.

In the particular case of $|g_{tt}| = 1$, we have $\tau_0 = t_0$ and 
Eq.~(\ref{eq:QI_expansion}) reduces to
\begin{equation}
\Delta\hat\rho \geq -{3\over 32 \pi^2 t_0^4} \left[ 1 
+{1\over 3}\left({1\over 6} R - m^2 \right)\,t_0^2 +{1\over 3}
\left( -{1\over 12} \nabla^i \nabla_i \; R + 4  v_1 - 4 v_{000}
\right) t_0^4 \ln (2 t_0^2) + \cdots\right].
\label{eq:gtt_equals_1}
\end{equation}
This result has also been obtained by Song \cite{Song97}, who uses
a heat kernel expansion of the Green's function to develop a short
sampling time expansion.  We can now apply Eq.~(\ref{eq:gtt_equals_1})
to a massless scalar field in the four-dimensional static Einstein
universe. The metric is given by
\begin{equation}
ds^2 = - dt^2 + a^2 \left[ d\chi^2 + \sin^2 \chi ( d\theta^2 +
\sin^2 \theta d\varphi^2 ) \right],
\end{equation}
and the scalar curvature $R = 6/a^2$ is a constant. It can be shown
that $v_1 - v_{000} = 1/8a^4$.  This leads to a quantum inequality
in Einstein's universe of the form
\begin{equation}
\Delta\hat\rho \geq -{3\over 32 \pi^2 t_0^4} \left[
1+ {1\over 3} \left({t_0\over a}\right)^2 + {1\over 3}
\left({t_0\over a}\right)^4 \ln ( t_0 / a) + O({t_0^4 \over a^4})
+ \cdots\right].
\label{eq:QI_Einstein_short}
\end{equation}
In Section~\ref{subsec:RW_Einstein}, an exact quantum inequality
valid for all $t_0/a$ will be derived.  In the limit $t_0 \ll a$, this
inequality agrees with Eq.~(\ref{eq:QI_Einstein_short}).  Similarly,
the exact inequality for the static, open Robertson-Walker universe
will be obtained in Section~\ref{subsec:RW_flat}, and in the limit 
$t_0 \ll a$ agrees with Eq.~(\ref{eq:gtt_equals_1}).


\section{Electromagnetic Field Quantum Inequality}\label{sec:EM_QI}
In this section, we derive a quantum inequality for the quantized
electromagnetic field in a static spacetime.  It has been shown
\cite{Volk71, Mash73} that the covariant Maxwell's equations for the
electromagnetic field in a curved spacetime, 
\begin{equation}
{F^{\mu\nu}}_{;\nu} = 0
\end{equation}
and
\begin{equation}
F_{\mu\nu ;\sigma} + F_{\nu\sigma ;\mu}+ F_{\sigma\mu ;\nu} = 0\, ,
\end{equation}
can be recast into the form of Maxwell's equations inside an
anisotropic material medium
in Cartesian coordinates, by using the constitutive relations
\begin{equation}
D_i = \epsilon_{ik}\,E_k - ({\bf G} \times {\bf H})_i
\end{equation}
and
\begin{equation}
B_i = \mu_{ik}H_k + ({\bf G} \times {\bf E})_i\, ,
\end{equation}
where
\begin{equation}
\epsilon_{ik} =\mu_{ik} = - (-g)^{1/2} {g^{ik}\over g_{00}}
\qquad
{\rm and}
\qquad
G_i = - {g_{0i}\over g_{00}}\, .
\end{equation}
Here the effects of the gravitational field are described by an
anisotropic dielectric and permeable medium. However, when we 
consider the metric
\begin{equation}
{ds}^2 = -|g_{tt}({\bf x})|{dt}^2 + g_{ij}({\bf x})dx^i dx^j \:,
\end{equation}
there is considerable simplification.  Because $g_{0i} = 0$ for all
$i$, the vector $G_i$ is always zero.  The constitutive relations
are then simply given by
\begin{equation}
{\bf D} = \hat\epsilon\, {\bf E}
\qquad
{\rm and}
\qquad
{\bf B} = \hat\epsilon\, {\bf H},
\label{eq:constit}
\end{equation}
where
\begin{equation}
\hat\epsilon = \hat\epsilon({\bf x}) = {\sqrt{-g}\over |g_{tt}|}\left( 
  \begin{array}{lll}
         g^{11} & g^{12} & g^{13} \\
         g^{21} & g^{22} & g^{23} \\
         g^{31} & g^{32} & g^{33}
  \end{array}\right)\, .
\end{equation}
The source-free Maxwell equations, in terms of $({\bf E},{\bf H})$, are
given by
\begin{eqnarray}
\nabla\times{\bf E} = - {\partial\, {\bf B}\over\partial t}\, ,
\quad&\quad&\quad\nabla\cdot{\bf D}=0\, ,\\[8pt]
\nabla\times{\bf H} =  {\partial\, {\bf D}\over\partial t}\, ,
\quad&\quad&\quad\nabla\cdot{\bf B}=0\, . 
\end{eqnarray}
We may also define the source-free vector potential $A_\mu ({\bf x},t)
= (0,{\bf A})$ with the relations to the electric and magnetic fields
given by  
\begin{equation}
{\bf E} = - {\partial\, {\bf A}\over\partial t}\qquad
\mbox{ and }\qquad
{\bf B} =  \nabla\times{\bf A}.
\end{equation}
It is straightforward to show that the vector potential satisfies
the wave equation
\begin{equation}
\hat\epsilon^{-1} \nabla \times \hat\epsilon^{-1} (\nabla \times
{\bf A} ) =   -{\partial^2\, {\bf A}\over\partial t^2}\, .
\label{eq:em_wave}
\end{equation}
The left-hand side of Eq.~(\ref{eq:em_wave}) involves only derivatives
with respect to the position coordinates, and the right-hand side,
temporal derivatives.  This is a clear separation of variables,
allowing us to write the positive frequency  solutions as
\begin{equation}
{\bf f}_{\bf k}^\lambda ({\bf x},t) = {\bf U}_{\bf k}^\lambda ({\bf x})
\; e^{-i\omega t},
\end{equation}
where ${\bf k}$ is the mode label for the propagation vector and $\lambda$
is the polarization state.  The vector functions, ${\bf U}_{\bf k}
^\lambda ({\bf x})$, are the solutions of
\begin{equation}
\hat\epsilon^{-1} \nabla \times \hat\epsilon^{-1} (\nabla \times
{\bf U}_{\bf k}^\lambda ) - \omega^2 {\bf U}_{\bf k}^\lambda = 0,
\end{equation}
and carry all the information about the curvature of the spacetime.
The mode functions for the vector potential are normalized such that
\begin{equation}
\left( {\bf f}_{\bf k}^\lambda, {\bf f}_{\bf k'}^{\lambda'} \right)
= -i \int d^3x \left[ {\bf f}_{\bf k}^\lambda \cdot \partial_t
{{\bf f}_{\bf k'}^{\lambda'}}^* - (\partial_t {\bf f}_{\bf k}^\lambda)
\cdot {{\bf f}_{\bf k'}^{\lambda'}}^* \right]
= \delta_{\bf k k'} \delta_{\lambda \lambda'}\, .
\end{equation}
The general solution to the vector potential can then be expanded as
\begin{equation}
{\bf A}({\bf x},t) = \sum_{{\bf k},\lambda} \left(  a_{{\bf k}\lambda}
{\bf U}_{\bf k}^\lambda ({\bf x}) e^{-i\omega t} + a_{{\bf k}\lambda}^\dagger
{{\bf U}_{\bf k}^\lambda}^* ({\bf x}) e^{+i\omega t} \right)\,.
\end{equation}
When we go to second quantization, the coefficients, 
$a_{{\bf k}\lambda}^\dagger$ and $a_{{\bf k}\lambda}$ become the 
creation and annihilation operators for the photon.  We will again
develop the quantum inequality for the electromagnetic field for
a static observer, as was done above for the scalar field.  The 
observed energy density, $\rho$, is given by \cite{Mash73} 
\begin{equation}
\rho = {1\over 2} (-g)^{-1/2}\left({\bf E \cdot D} + {\bf B \cdot H}\right).
\end{equation}
Upon substitution of the mode function expansion, and making use of
constitutive relations (\ref{eq:constit}) we find
\begin{eqnarray}
\rho & = & {1\over |g_{tt}|}\, {\rm Re}\!\!\sum_{{\bf k\, k'},\lambda\lambda'}
\omega\omega' \left[ a_{{\bf k}\lambda}^\dagger a_{{\bf k}'\lambda'}\:
({{\bf U}_{\bf k}^\lambda}_i^* \; g^{ij} \; {{\bf U}_{{\bf k}'}^{\lambda'}}_j)\,
e^{i(\omega-\omega')t} - a_{{\bf k}\lambda} a_{{\bf k}'\lambda'}\:
({{\bf U}_{\bf k}^\lambda}_i \; g^{ij} \;{{\bf U}_{{\bf k}'}^{\lambda'}}_j)\,
e^{-i(\omega+\omega')t} \right] \nonumber\\
&& + |g_{tt}|\, {\rm Re}\!\!\sum_{{\bf k\, k'},\lambda\lambda'}
\left[ a_{{\bf k}\lambda}^\dagger a_{{\bf k}'\lambda'}\:
(\nabla \times {{\bf U}_{\bf k}^\lambda}^* )_i \; g^{ij} \;
(\nabla \times {\bf U}_{{\bf k}'}^{\lambda'} )_j \, e^{i(\omega-\omega')t}
\right.\nonumber\\ 
&&\left.\qquad\qquad+ a_{{\bf k}\lambda} a_{{\bf k}'\lambda'}\:
(\nabla \times {\bf U}_{\bf k}^\lambda)_i \; g^{ij} \;
(\nabla \times {\bf U}_{{\bf k}'}^{\lambda'} )_j \, e^{-i(\omega+\omega')t}
\right] \nonumber\\
&& + {1\over 2} \sum_{{\bf k},\lambda} \left[ {\omega^2\over |g_{tt}|}  
({{\bf U}_{\bf k}^\lambda}_i \; g^{ij} \;{{\bf U}_{\bf k}^\lambda}_j^*) + 
|g_{tt}|\:(\nabla \times {\bf U}_{\bf k}^\lambda )_i \; g^{ij} \; 
(\nabla \times {{\bf U}_{\bf k}^\lambda}^* )_j \right]. 
\end{eqnarray}
The last term of the above expression is the vacuum self-energy
of the photons.  As was the case for the scalar field, we will
look at the difference between the energy in an arbitrary
state and the vacuum energy, {\it i.e.},
\begin{equation}
:\rho: = \rho - \langle 0 | \rho | 0 \rangle.
\end{equation}
Again we integrate the energy density along the worldline of the
observer, weighted by the Lor\-entz\-ian sampling function.  We may then
apply the first inequality, Eq.~(\ref{eq:first_inequality}) proven
in \cite{F&Ro97}.  To the result
we apply the inequality proven in Appendix~\ref{sec:appendix}.
We then find that the difference inequality on the energy density for
a quantized electromagnetic field is given by
\begin{equation}
\Delta\hat\rho \geq -{1\over 2} \sum_{{\bf k},\lambda} 
\left[ {\omega^2\over |g_{tt}|}  
({{\bf U}_{\bf k}^\lambda}_i \; g^{ij} \;{{\bf U}_{\bf k}^\lambda}_j^*) + 
|g^{tt}|\:(\nabla \times {\bf U}_{\bf k}^\lambda )_i \; g^{ij} \; 
(\nabla \times {{\bf U}_{\bf k}^\lambda}^* )_j \right] e^{-2\omega t_0}.
\label{eq:em_QI}
\end{equation}
This expression is similar in form to the mode function expansion
of the scalar field quantum inequality, and also reduces to an
averaged weak energy type integral in the infinite sampling time
limit.  As was the case for the scalar field, the electromagnetic
field quantum inequality (\ref{eq:em_QI}) tells us how much negative
energy an observer may measure with respect to the vacuum energy of
the electromagnetic field.  In order to find the absolute lower
bound on the negative energy density we would have to add the
renormalized vacuum energy into the above expression.

This quantum inequality
can be  easily evaluated in Minkowski spacetime.  In a box
of volume $L^3$ with periodic boundary conditions, the mode
functions are given by
\begin{equation}
{{\bf U}_{\bf k}^\lambda} = \left( 2\omega L^3 \right) ^{1/2}\,
e^{i {\bf k\cdot x}}\, \hat\epsilon_{\bf k}^\lambda,
\end{equation}
where $\hat{\epsilon}_{\bf k}^\lambda$ is a unit polarization vector
and $\omega = \sqrt{\bf k \cdot k}$.
Inserting the mode functions into Eq.~(\ref{eq:em_QI}), and using
the fact that ${\bf k} \cdot \hat{\epsilon}_{\bf k}^\lambda = 0$, we
find
\begin{equation}
\Delta\hat{\rho} \geq - {1\over 2 L^3} \sum_{{\bf k},\lambda} \omega \,
e^{-2\omega t_0}.
\end{equation}
The summation over the spin degrees of freedom yields
\begin{equation}
\Delta\hat\rho \geq - {1\over  L^3} \sum_{\bf k} \omega \,
e^{-2\omega t_0}.
\end{equation}
In the continuum limit, $L \rightarrow \infty$, the vacuum energy
density vanishes, and the renormalized quantum inequality is found
to be
\begin{eqnarray}
\hat\rho &\geq& -{1\over (2 \pi)^3}\int_{-\infty}^{\infty} d^3 k \:
\omega\, e^{-2\omega t_0},\nonumber\\[8pt]
&=& -{3\over 16 \pi^2 \, t_0^4}.
\end{eqnarray}
This quantum inequality for Minkowski spacetime was originally
proven by Ford and Roman \cite{F&Ro97} using an alternative method.
Comparison  with the quantum inequality for the scalar field
in Minkowski space, Eq.~(\ref{eq:qi1}), shows that the electromagnetic
field quantum inequality differs by a factor of 2.  This is a result
of the electromagnetic field having two polarization degrees of 
freedom, unlike the scalar field which has only one.

\chapter{Scalar Field Examples}\label{Chapt:examp}
\section{Two-Dimensional Spacetimes}\label{sec:2D_conform}

There are a number of interesting results for two-dimensional
spacetimes that we will discuss. The unique conformal properties
in two dimensions allow the quantum inequality for all 
two-dimensional static spacetimes to be written in the form
\begin{equation}
\Delta\hat\rho\geq -{1\over 8\pi \tau_0^2}\, ,
\end{equation}
where $\tau_0$ is the proper time of the stationary observer.
Similarly, the quantum inequality for the flux traveling in
one direction is
\begin{equation}
\hat f \equiv {t_0\over \pi}\int_{-\infty}^\infty {\langle T^{xt}
\rangle \over t^2+t_0^2} \,dt \,\geq -{1\over 16\pi t_0^2}\, .
\end{equation}
We emphasize that these results are only for static two-dimensional
spacetimes. The original derivation of the quantum inequality of
the preceding chapter relies on the metric having a natural choice 
of a timelike Killing vector field.  This allows us to clearly 
define the notion of positive frequency for the mode functions.
The conformal property of two-dimensional spacetimes is more
general than the quantum inequality, in that even spacetimes which
are time evolving will be conformal to some part of two-dimensional
Minkowski spacetime.

Recently, Flanagan \cite{Flan97} has demonstrated another interesting
aspect of the quantum inequalities in two-dimensional spacetimes.
He has shown that a general
form of the quantum inequality can be developed in two-dimensional
Minkowski spacetime which is independent of the choice of the sampling
function.  We will summarize his results below, but first we will
look at the conformal triviality of all static two-dimensional spacetimes.

\subsection{Conformal Properties}\label{subsec:2D_conform}

It is a unique property of two-dimensional spacetimes that any metric
can be cast into the form
\begin{equation}
g_{\mu\nu}(x)\, \rightarrow \, g_{\mu\nu}(y) = \Omega^2(y) \eta_{\mu\nu}
\label{eq:2dMetric}
\end{equation}
by choosing an appropriate coordinate transform, $x \rightarrow y = y(x)$.
For the massless scalar field, the wave equation
\begin{equation}
\Box \phi(x) = 0
\end{equation}
then becomes
\begin{equation}
\Omega^{-2}(y) \eta^{\mu\nu} \partial_\mu  \partial_\nu \bar\phi(y) =0.
\end{equation}
The mode functions in the two different coordinates are related by
\begin{equation}
\phi(x) = \bar\phi ( y(x) ).
\end{equation} 
However, $\bar\phi(y)$ is the scalar field solution to the wave equation
in Minkowski spacetime, composed from the standard mode functions
\begin{equation}
\bar f_k(y) = (4\pi\omega)^{-1/2} e^{i k\cdot y}\, ,
\end{equation}
where $k_0 = \omega = |k_1|$.  In two dimensions, the Green's functions
are found to transform as
\begin{equation}
D(x,x')  = \bar D(y(x),y(x')) \, ,
\end{equation}
where $\bar D(y,y')$ is the standard Minkowski Green's functions.
For example, the Wightman Green's function in two-dimensional Minkowski
spacetime is
\begin{equation}
D^+(y,y') = \langle 0 | \bar\phi(y) \bar\phi(y') | 0\rangle = {1\over 4\pi}
\int_{-\infty}^\infty {dk\over \omega} e^{ik\cdot(y-y')}\, .
\end{equation}
This would also be the Wightman Green's function in an arbitrary
curved two-dimensional spacetime expressed in the coordinates chosen
from the metric (\ref{eq:2dMetric}).  It is evident from the expression
above that the massless Green's functions have an infrared divergence.
It is possible to carry out the integration of the above Green's 
function, and renormalize away the infrared divergence.  However,
for our purposes, the integral representation is sufficient.

We have already shown for static spacetimes that the QI is given by
\begin{equation}
\Delta\hat\rho \geq - {1\over 4} \Box_E D_E(x,-it_0; x, +it_0).
\end{equation}
If we now make use of the conformal relations for the box operator
and the Green's function, we find
\begin{equation}
\Delta\hat\rho \geq - {1\over 4} \Omega^{-2}(y) \bar{\Box}_E
\bar D_E(y,-it_0; y, +it_0)
=  - {1\over 16 \pi} \Omega^{-2}(y) \partial^2_{t_0} 
\int_{-\infty}^\infty {dk\over \omega} e^{-2\omega t_0}.
\end{equation}
We see that while the Green's function has an infrared divergence,
the quantum inequality does not.  We can interchange the order of the
integration and differentiation with the resulting integral being
well defined for all values of $k$,  and find 
\begin{equation}
\Delta\hat\rho\geq -{1\over 8 \pi \left[ \Omega(y) t_0 \right]^2}\, .
\end{equation}
However, in static spacetimes the coordinate time and the proper time
of an observer are related by $\tau_0 = \Omega(y) t_0$. We can then write
the quantum inequality in a more covariant form as
\begin{equation}
\Delta\hat\rho \geq -{1\over 8\pi \tau_0^2}\, .
\end{equation}
This is exactly the form of the quantum inequality found by Ford and
Roman in Minkowski spacetime \cite{F&Ro95}.  The difference inequality
above is applicable to all static two-dimensional spacetimes.  However
the renormalized quantum inequalities in any two-dimensional spacetimes
will be different because the vacuum energies in these spacetimes are
not identical.  We can write the renormalized two-dimensional quantum
inequality as
\begin{equation}
\hat\rho_{Ren.} \geq -{1\over 8\pi \tau_0^2} + \rho_{vacuum},
\end{equation}
where $\rho_{vacuum}$ is the vacuum energy that is defined with respect
to the timelike Killing vector field.

\subsection{Sampling Functions in Two-Dimensional Spacetimes}

In the previous chapter, we have proven quantum inequalities
using the Lorentzian sampling function, Eq.~(\ref{eq:Lorentzian}).
The Lorentzian was chosen to facilitate the proof of the inequality.
Recently, Flanagan \cite{Flan97} has shown that quantum inequalities
for the massless scalar field in two-dimensional Minkowski spacetime
can be developed for arbitrary sampling functions.  (Note: Flanagan
uses the term ``smearing functions.'') We summarize his results below.

We begin with a smooth, nonnegative function $f(v)$ with unit
normalization,
\begin{equation}
\int_{-\infty}^\infty \, f(v) \, dv = 1.
\end{equation}
It is not necessary that the sampling function have compact support
so long as the function approaches zero sufficiently quickly to 
ensure convergence of the above integral. 
Using the stress-tensor in the standard $(x,t)$ coordinates,
we define three smeared integral quantities:  the time-smeared
energy density
\begin{equation}
\Delta\hat\rho = {\cal E}_T \equiv \int_{-\infty}^\infty dt 
\, f(t)\, \langle : T_{tt}(0,t) :\rangle,
\label{eq:Flanagan_1}\end{equation}
the time-smeared flux
\begin{equation}
\Delta\hat F = {\cal E}_F \equiv \int_{-\infty}^\infty dt 
\, f(t) \, \langle : T^{xt}(0,t): \rangle,
\end{equation}
and the spatially-smeared energy density
\begin{equation}
{\cal E}_S \equiv \int_{-\infty}^\infty dx \, f(x)\,  
\langle : T_{tt}(x,0) :\rangle.
\end{equation}
Flanagan then shows that all three of the above integral
quantities are bounded below by
\begin{equation}
{\cal E}_{T,min} = {\cal E}_{S,min} = 2 {\cal E}_{F,min} =
- {1\over 24 \pi} \int_{-\infty}^\infty dv \, {f'(v)^2\over f(v)}.
\end{equation}
This is achieved by use of a Bogolubov transformation which 
converts the quadratic form of the integral in Eq.~(\ref{eq:Flanagan_1})
into a simple form.  (Details of the proof can be found in \cite{Flan97}.)
Flanagan also shows that this is an optimum lower bound for a given
sampling function. This lower bound is achieved by the vacuum state
associated with the new coordinates used in the Bogolubov transform.
This new vacuum state is a generalized multi-mode squeezed state of
the original vacuum defined in the $(x,t)$ coordinates.

For arbitrary sampling functions, it is now straightforward to 
find the optimum quantum inequality in two-dimensional Minkowski
spacetime.  For example, if we use the Lorentzian sampling function, 
Eq.~(\ref{eq:Lorentzian}), in the time-smeared energy density, we find
\begin{equation}
\Delta\hat\rho \geq -{1\over 48 \pi t_0^2}\, .
\label{eq:Flan_QI}
\end{equation}
This is six times more restrictive than the quantum inequality
derived in the preceding section.  We knew that the quantum
inequality developed using the Green's function was a lower
bound, but it was not necessarily known if it was the optimum
lower bound. We now see that Eq.~(\ref{eq:Flan_QI}) represents
the optimum quantum inequality for the Lorentzian sampling
function.   

\section{Three-Dimensional Spacetimes}\label{sec:3D_spaces}
We now discuss examples of three-dimensional spacetimes in which it
is straightforward to find the quantum inequalities.  First we will
develop the quantum inequality in Minkowski spacetime for a massive
scalar field. The addition of the mass makes the quantum inequality
more restrictive.  In order to generate negative energy densities,
we must overcome both the momentum of the particle and its rest mass.
We will then look at the three-dimensional equivalent of the static
Einstein universe. Unlike the two-dimensional examples of the preceding
section, we will see that in three dimensions the form of the quantum
inequality is modified from its flat space form because of the spacetime
curvature. In addition, we will be able to renormalize the quantum
inequality in the three-dimensional closed universe.  Here there exists
a non-zero vacuum energy that must be taken into consideration.
In this case the quantum inequality tells us how much negative energy
we can measure relative to the background vacuum energy.
  
\subsection{Minkowski Spacetime}
\begin{figure} 
\begin{center}
\leavevmode\epsfxsize=0.8\textwidth\epsffile{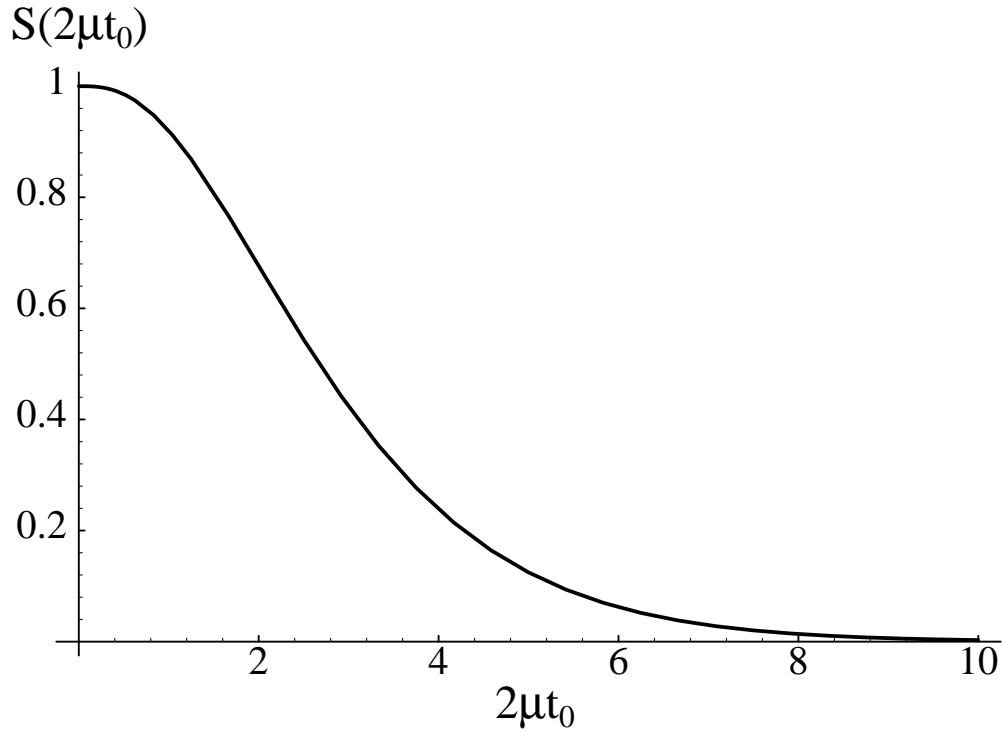}
\end{center}
\caption[Three-dimensional Minkowski scale function]
{The scale function $S(2\mu t_0)$ for three-dimensional Minkowski 
spacetime.}
\label{fig:Massive3D}
\end{figure}

\index{Minkowski spacetime}
We begin with the easiest example, three-dimensional flat spacetime,
where we know that the positive frequency massive mode functions
are given by
\begin{equation}
f_{\bf k}(x,y,t) = [2\omega (2\pi)^2]^{-1/2} 
e^{i{\bf k} \cdot {\bf x}-i\omega t}\, ,
\end{equation}
with a frequency $\omega = \sqrt{|\bf k|^2 + \mu^2}$ and $-\infty < k_i
< \infty$. The required two-point function is then defined by
\begin{equation}
G_E(2t_0) = \int_{-\infty}^\infty dk_x\,dx_y\; \left|
{1\over\sqrt{2\omega(2\pi)^2}} e^{i{\bf k} \cdot {\bf x}}\right|^2
e^{-2\omega t_0}\, .
\end{equation}
It is rather straightforward to carry out the required integrations to
find
\begin{equation}
G_E(2t_0) = {1\over 8\pi t_0} \, e^{-2mt_0}.
\end{equation}
Now the difference inequality is given by Eq.~(\ref{eq:QI}), where the 
Euclidean box operator is 
\begin{equation}
\Box_E = \partial_{t_0}^2 + \partial_x^2 + \partial_y^2 \, .
\end{equation}
One thing that should be pointed out before proceeding is that
the vacuum energy in three-dimensional Minkowski spacetime is defined to
be zero.  Under such circumstances, we are placing a lower
bound on the energy density of the state itself, and not just on the
difference of the energy density between the state and the vacuum energy.
The quantum inequality becomes
\begin{equation}
\hat\rho \geq - {1\over 32\pi} \partial_{t_0}^2 \left[
{1\over t_0} \, e^{-2\mu t_0}\right] \, ,
\end{equation}
which upon taking the time derivatives can be written as
\begin{equation}
\hat\rho \geq - {1\over16\pi t_0^3} S(2\mu t_0)\, .
\end{equation}
The ``scale function,'' $S(x)$, is defined as
\begin{equation}
S(x) = \left( 1 + x +{1\over2}x^2 \right)\,e^{-x}\, .
\end{equation}
The scale function represents modifications to the quantum inequality
due to the mass of the the particle.
It is plotted in Figure~\ref{fig:Massive3D}. When $\mu = 0$, the 
scale function becomes one and the quantum inequality reduces to the
massless form
\begin{equation}
\hat\rho \geq -{1\over 16\pi t_0^3}\, .\label{eq:3D_QI_Mink}
\end{equation}
This is what would have been found if we had chosen the massless mode
functions at the beginning of this derivation.  It is interesting to note
that in the limit of the mass of the scalar particle becoming very large,
the scale function will tend to zero.  Thus, if the particle
has a mass, it becomes more difficult to observe negative energy 
densities.

Since the mode functions are invariant under Lorentz transformations,
\index{Lorentz transform}
we can always choose to develop the inequality in the observer's rest
frame and then transform to any other frame. The quantum inequality can 
therefore be written in a more covariant form in flat spacetime as
\begin{equation}
\hat\rho \,\equiv\,{\tau_0\over\pi}{\int_{-\infty}^\infty}
{\langle :T_{\alpha\beta} u^\alpha u^\beta :\rangle \over 
{\tau^2+\tau_0^2}}d\tau \,\geq\, -{1\over 16\pi \tau_0^3}\, S(2\mu\tau_0).
\end{equation}
Here $\tau$ is the observer's proper time, $\tau_0$ is the sampling time,
and $u^\alpha = d\,x^\alpha(\tau)/d\tau$ is the observer's four-velocity.

\subsection{The Closed Universe}\index{3-D Closed Universe}
Let us consider the three-dimensional spacetime with a length element given
by 
\begin{equation}
ds^2 = - dt^2 + a^2 \left(d\theta^2 + \sin^2\theta d\varphi^2\right).
\label{eq:Sphere}
\end{equation}
Here constant time-slices of this universe are two spheres of radius
$a$. The wave equation on this background with a coupling 
of strength $\xi$ to the Ricci scalar $R$ is
\begin{equation}
\Box\phi -(\mu^2+\xi R)\phi = 0.
\end{equation}
For the metric~(\ref{eq:Sphere}), the wave equation becomes
\begin{equation}
-\partial_t^2\phi + {1\over{a^2\sin\theta}}\partial_\theta\left(\sin\theta
\partial_\theta\phi\right) + {1\over{a^2\sin^2\theta}}\partial_\varphi^2\phi
-[\mu^2 + \xi\left(2\over a^2\right)]\phi = 0,
\end{equation}
which has the solutions
\begin{equation}
f_{lm}(t,\theta,\varphi) = {1\over{\sqrt{2a^2\omega}}} {\rm Y}_{lm}(\theta
,\varphi) e^{-i\omega_l t}.
\end{equation}
The ${\rm Y}_{lm}$ are the usual spherical harmonics\index{spherical harmonics}
with unit normalization.  The mode labels take the values, $l = 0,1,2,
\cdots$ and $-l\leq m\leq l$.  For each state of the primary quantum number
$l$, there is a degeneracy of $2l+1$ associated with the $m$, each
having the same eigenfrequency
\begin{equation}
 \omega_l = a^{-1} \sqrt{ l(l+1) + 2\xi + \mu^2 a^2} \, .
\end{equation}
The coupling parameter $\xi$ here can be seen to contribute
to the wave functions as a term of the same form as the mass.
However, we will look only at minimal coupling ($\xi = 0$).
\index{minimal coupling}
The lower bound on the energy density is given by  
\begin{equation}
\Delta\hat \rho \geq - {1\over {2a^2} }
\sum_{l= 0}^\infty \sum_{m=-l}^{+l} \omega_l |{\rm Y}_{lm}(\theta,\varphi)|^2 
e^{-2\omega_l t_0} - {1\over 8 a^2}\nabla^i \nabla_i
\sum_{l= 0}^\infty \sum_{m=-l}^{+l} {1\over \omega_l}|{\rm Y}_{lm}
(\theta,\varphi)|^2 e^{-2\omega_l t_0}.
\label{eq:spherebound}
\end{equation}
However the spherical harmonics obey a sum rule \cite{Jackson}
\begin{equation}
\sum_{m=-l}^{+l} |{\rm Y}_{lm}(\theta,\varphi)|^2  = {{2l+1}\over{4\pi}}.
\label{eq:sum_rule}
\end{equation}
We immediately see that the difference inequality is independent of
position, as expected from the spatial homogeneity, and that the second
term of Eq.~(\ref{eq:spherebound}) does not contribute.  We have
\begin{equation}
\Delta\hat \rho \geq - {1\over {8\pi a^2} }
\sum_{l= 0}^\infty  (2l+1)\, \omega_l  
e^{-2\omega_l t_0}. 
\end{equation}
This summation is finite due to the exponentially decaying term. We are
now left with evaluating the sum for a particular set
of values for the mass $\mu$, the radius $a$, and the sampling time $t_0$.  
Let $\eta = {t_0 / a}$.  Then
\begin{equation} 
\Delta\hat \rho \geq  -{1\over{16\pi t_0^3}}\left[2\eta^3\sum_{l=1}^\infty 
(2l+1)\, \tilde\omega_l \, 
    {\rm e}^{-2\eta\tilde\omega_l}\right]= -{1\over{16\pi t_0^3}}
F(\eta,\mu),                   \label{eq:qi3d}
\end{equation}
where
\begin{equation}\tilde\omega_l = \sqrt{l(l+1)+a^2 \mu^2}\, .
\end{equation}
The coefficient $-{1/(16\pi t_0^3)}$ is the right-hand side of the inequality
for the case of a massless field in an infinite three-dimensional Minkowski
space. The ``scale function,'' $F(\eta,\mu)$, represents how the mass and the 
curvature of the closed spacetime affects the difference inequality.    

\subsection{Massless Case}\label{subsec:RW_massless}
\begin{figure}
\begin{center}
\leavevmode\epsfxsize=0.8\textwidth\epsfbox{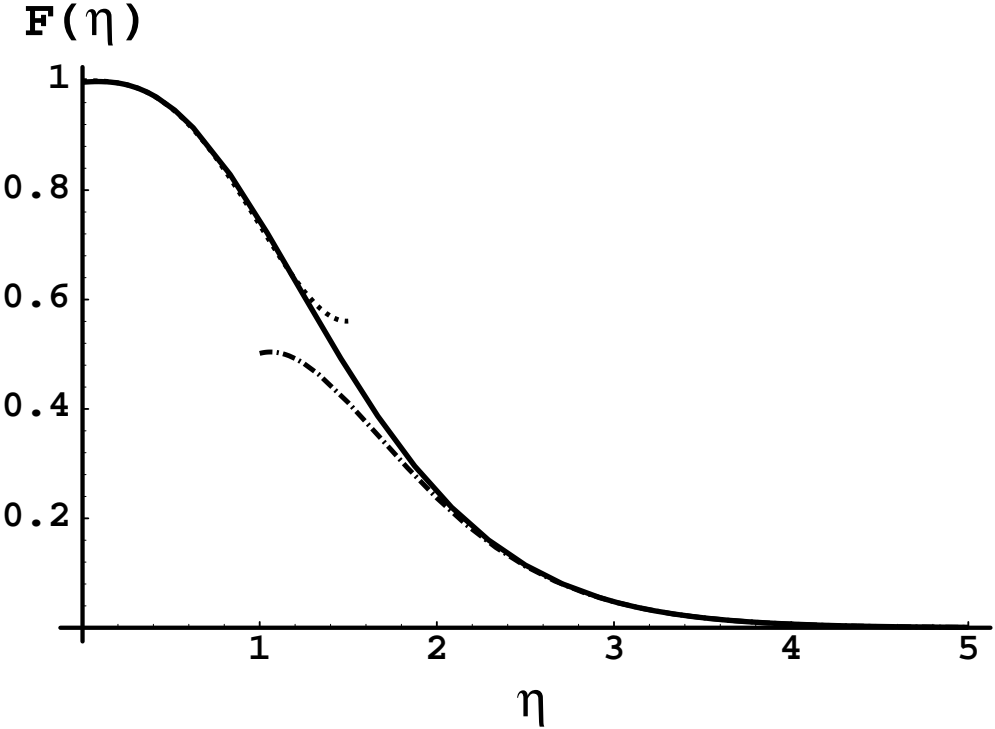}
\end{center}
\caption[Three-dimensional closed universe scale function]
{The scale function $F(\eta)$ and its asymptotic forms
for the three-dimensional closed universe.
The solid line is the exact form of the function.  The dotted
line is the small $\eta$ approximation, while the dot-dash 
line is the large $\eta$ approximation. }
\label{fig:3D_Closed}
\end{figure}

In terms of the variable $\eta = {t_0 / a}$, which is the ratio of
the sampling time to the radius of the universe, we can
write the above expression for $F(\eta,0)$ as
\begin{equation} 
F(\eta) = F(\eta,0) = 2\eta^3\sum_{l=1}^\infty 
(2l+1)\sqrt{l(l+1)}\, {\rm e}^{-2\eta\sqrt{l(l+1)}}.
\end{equation}
A plot of $F(\eta)$ is shown in Figure~\ref{fig:3D_Closed}.  In the limit of 
$\eta\rightarrow 0$, when the sampling time is very small or 
the radius of the universe has become so large that it approximates 
flat space, the function $F(\eta)$ approaches one, yielding
the flat space inequality.

We can look at the inequality in the two asymptotic regimes of $\eta$.
In the large $\eta$ regime each term of greater $l$ in the exponent
will decay faster than the previous term.  The $l = 1$ term yields a
good approximation.  In the other regime, when the sampling time is
small compared to the radius, we can use the Plana summation formula
to calculate the summation explicitly:
\begin{equation}
\sum_{n=1}^\infty f(n) + {1\over 2} f(0) = \int_0^\infty f(x)dx + i
\int_0^\infty {f(ix)-f(-ix) \over {{\rm e}^{2\pi x} -1}} dx \,,
                                              \label{eq:Plana}
\end{equation}
where
\begin{equation}
f(x) = (2x+1)\sqrt{x(x+1)}\, {\rm e}^{-2\eta\sqrt{x(x+1)}}.
\end{equation}
Immediately we see that for our summation $f(0) = 0$. The first integral
can be done with relative ease yielding
\begin{equation}
\int_0^\infty f(x) dx =
\int_0^\infty (2x+1)\sqrt{x(x+1)}\, 
             {\rm e}^{-2\eta\sqrt{x(x+1)}}dx = \frac{1}{2\eta^3} \,.
\end{equation}
This term reproduces the flat space inequality. The second integral 
in Eq.~(\ref{eq:Plana}) therefore contains all the corrections due to 
non-zero curvature  of the spacetime.  Since $\eta$ is small,
we can expand the exponent in a Taylor series around $\eta = 0$. 
Keeping the lowest order terms, we find
\begin{equation}
i\int_0^\infty {f(ix)-f(-ix) \over {e^{2\pi x} -1}} dx \sim
- I_0 -\eta I_1 + \cdots \,,
                                     \label{eq:expans}
\end{equation}
where
\begin{equation}
I_0 = \int_0^\infty \sqrt{2x}\; {\left( 2x\sqrt{\sqrt{x^2+1}-x} +
\sqrt{\sqrt{x^2+1}+x}\right) \over {{\rm e}^{2\pi x} -1}}\; dx 
   \approx 0.265096,
\end{equation}
and
\begin{equation}
I_1 = \int_0^\infty {4x(2x^2-1) \over {{\rm e}^{2\pi x} -1}} dx 
= -\frac{2}{15} \,.
\end{equation}

  Therefore the function $F(\eta)$ in the small $\eta$ limit is given by
\begin{equation}
F(\eta) \simeq 1 - 0.530192\, \eta^3 + {4 \over 15} \eta^4 + O(\eta^5)
         + \cdots\, ,
\end{equation}
and in the large $\eta$ limit by
\begin{equation}
F(\eta) \simeq 6\sqrt{2}\, \eta^3\, {\rm e}^{-2\sqrt{2}\eta} + \cdots \,.
                              \label{eq:large_eta}
\end{equation}
Both of these asymptotic forms are plotted along with the exact form of
$F(\eta)$ in Figure~\ref{fig:3D_Closed}. The graph shows that the
two asymptotic limits follow the exact graph very precisely
except in the interval $ 1 < \eta <2$.  
These results for the function $F(\eta)$, combined with 
Eq.~(\ref{eq:qi3d}), yield the difference inequality for the massless
scalar field.

The difference inequality does not require knowledge of
the actual value of the renormalized vacuum energy, $\rho_{vacuum}$. However, 
if we wish to obtain a bound on the energy density itself, we must combine 
the difference inequality and $\rho_{vacuum}$. One procedure for computing 
$\rho_{vacuum}$
is analogous to that used to find the Casimir energy\index{Casimir vacuum}
in flat spacetime
with boundaries: One defines a regularized energy density, subtracts the 
corresponding flat space energy density, and then takes the limit in 
which the regulator is removed. A possible choice of regulator is to insert
a cutoff function, $g(\omega)$, in the mode sum and define the regularized
energy density as
\begin{equation}
\rho_{reg} =  \frac{1}{8 \pi a^2} \sum_\lambda \omega_\lambda \,
               g(\omega_\lambda) =
  \frac{1}{8 \pi a^2}  \sum_{l= 0}^\infty  (2l+1)\, \omega_l\, g(\omega_l)
        \,.
\end{equation}
In the limit that $a \rightarrow \infty$, we may replace the sum by an
integral and obtain the regularized flat space energy density:
\begin{equation}
\rho_{FS reg} =  \frac{1}{4 \pi}  \int_0^\infty {\rm d} \omega \, \omega\,
               g(\omega)        \,.
\end{equation}
The renormalized vacuum energy density may then be defined as
\begin{equation}
\rho_{vacuum} = \lim_{g \rightarrow 1} (\rho_{reg} - \rho_{FS reg}) \,.
\end{equation}
The vacuum energy density so obtained will be denoted by $\rho_{Casimir}$.
An analogous procedure was used in \cite{Ford75b} to obtain the vacuum
energy density for the conformal scalar field in the four-dimensional
Einstein universe.
An explicit calculation for the present case, again using the Plana
summation formula, yields
\begin{equation}
\rho_{Casimir} =  - {I_0 \over{8\pi a^3}} 
               \approx - { 0.265096 \over{8\pi a^3}} .
\end{equation}
This agrees with the result obtained by Elizalde \cite{Elizalde} using the
zeta function technique. In this case $\rho_{Casimir} < 0$,
whereas the analogous calculation for the conformal scalar field in this
three-dimensional spacetime yields a positive vacuum energy density,
$\rho_{Casimir} = 1/{96\pi a^3}$. It should be noted that this procedure
works for the closed universe because the divergent part of $\rho_{reg}$
is independent of $a$. More generally, there may be curvature-dependent
divergences which must also be removed. 

Let us now return to the explicit forms of the difference inequality. In
the limit that $t_0 \gg a$, Eqs.~(\ref{eq:qi3d}) and (\ref{eq:large_eta})
yield
\begin{equation}
\Delta\hat \rho \geq 
-{{3\sqrt{2}}\over{8\pi a^3}}\, {\rm e}^{-2\sqrt{2}\, t_0/a}
    \,.                               
\end{equation}
Similarly, in the limit that $t_0 \ll a$, we find
\begin{equation}
\Delta\hat \rho \geq  -{1\over{16\pi t_0^3}} - \rho_{Casimir} -
{{t_0}\over{60\pi a^4}} + \cdots  \, . \label{eq:qi3}
\end{equation}
Thus the bound on the renormalized energy density in an arbitrary
quantum state in the latter limit becomes
\begin{equation}
\hat \rho_{Ren.} \geq  -{1\over{16\pi t_0^3}} - 
{{t_0}\over{60\pi a^4}} + \cdots  \, .  \label{eq:qi4}
\end{equation}

In the $a\rightarrow\infty$ limit, both Eqs.~(\ref{eq:qi3}) and 
(\ref{eq:qi4}) reduce to the quantum inequality for three-dimensional
Minkowski space, Eq.~(\ref{eq:3D_QI_Mink}).  In this limit $\rho_{Casimir}
= 0$ so $\Delta\hat\rho = \hat\rho$.  
In the other limit when the sampling time becomes long, we find
that $\Delta\hat \rho$ decays exponentially as a function of the sampling 
time. This simply reflects the fact that the difference in energy density
between an arbitrary state and the vacuum state satisfies the  quantum 
averaged weak energy condition, Eq.~(\ref{eq:QAWEC}).  

\section{Four-Dimensional Robertson-Walker Universe}

Now we will apply the same method to the case of the three homogeneous
and isotropic universes given by the four-dimensional static 
Robertson-Walker metrics.  Here we have\footnote{We are using $\epsilon$
as a free parameter which takes the value $\epsilon = -1$ in the open
universe, $\epsilon = 0$ in the flat universe, and $\epsilon = +1$ in
the closed universes.} 
\begin{equation}
[\epsilon = 0]: \qquad g_{ij}dx^i dx^j = dx^2 + dy^2 + dz^2,
\end{equation}
for the flat universe with no curvature (Minkowski spacetime).  
For the closed universe with constant radius $a$, {\it i.e.},
the universe of constant positive curvature, the spatial length
element is given by
\begin{equation}
[\epsilon = 1]: \qquad  g_{ij}dx^i dx^j = a^2 \left[ d\chi^2 + \sin^2\chi
(d\theta^2+\sin^2\theta d\varphi^2)\right] \, ,  \label{eq:openmetric}
\end{equation}
where $0 \leq \chi \leq \pi$, $0 \leq \theta \leq \pi$, and 
$ 0 \leq \varphi < 2\pi$. 
The open universe $[\epsilon = -1]$ is given by making the replacement 
$\sin\chi \rightarrow \sinh\chi$ in Eq.~(\ref{eq:openmetric}) and now allowing
$\chi$ to take on the values $0 \leq \chi < \infty$.  To find the lower
bound of the quantum inequalities above we must solve for the eigenfunctions
of the covariant Helmholtz equation
\begin{equation}
\nabla_i \nabla^i U_\lambda({\bf x}) + (\omega^2_\lambda - \mu^2)
U_\lambda({\bf x}) = 0,
\end{equation}
where $\mu$ is the mass of the scalar field and $\omega_\lambda$ is the
energy. A useful form of the solutions for this case is given by Parker
and Fulling \cite{Parker}. 

\subsection{Flat and Open Universes in Four Dimensions}\label{subsec:RW_flat}

In the notation developed in Chapter~\ref{chapt:QI_proof}, the spatial
portion of the wave functions for flat (Euclidean) space is given by 
(continuum normalization)
\begin{eqnarray}
[\epsilon = 0]:\qquad\quad U_{\bf k}({\bf x}) &=& [2\omega (2\pi)^3]^{-1/2} 
e^{i{\bf k} \cdot {\bf x}}, \\
\omega &=& \sqrt{|\bf k|^2 + \mu^2}\, ,\\
 {\bf k} &=& (k_1,k_2,k_3)\qquad (-\infty < k_j < \infty).\nonumber
\end{eqnarray}
It is evident that $|U_{\bf k}|^2$ will be independent
of position. This immediately removes the second term of the 
inequality in Eq.~(\ref{eq:qi2}).

In the open universe, the spatial functions are given by
\begin{eqnarray}
[\epsilon = -1]:\qquad\quad U_{\lambda}({\bf x}) &=& (2a^3\omega_q)^{-1/2} 
 \Pi^{(-)}_{ql}(\chi) {\rm Y}_{lm}(\theta,\varphi)\,, \\
\omega_q &=& \sqrt{{(q^2+1)\over a^2} + \mu^2},\\
 \lambda &=& (q,l,m).\nonumber
\end{eqnarray}
Here $0 <q<\infty;\quad  l = 0, 1, \cdots;$ and $m = -l, -l+1, \cdots, +l$.
The sum over all modes involves an integral over the radial momentum $q$. 
The functions $\Pi^{(-)}_{ql}(\chi)$ are given in Eq.~(5.23) of 
Birrell and Davies \cite{Brl&Dv}. Apart from the normalization factor,
they are 
\begin{equation}
\Pi^{(-)}_{ql}(\chi) \propto \sinh^l\chi
\left( d\over d\cosh\chi\right)^{l+1} \cos q\chi.
\end{equation} 
As with the mode functions of the three-dimensional closed spacetime above,
the mode functions of the open four-dimensional universe satisfy an addition
theorem \cite{Parker,Lif&Khal,Bunch77}
\begin{equation}
\sum_{lm} |\Pi^{(-)}_{ql}(\chi) {\rm Y}_{lm}(\theta,\varphi)|^2 =
{q^2\over 2\pi^2}\,.
\end{equation}
Since the addition theorem removes any spatial dependence, we again get no
contribution from the Laplacian term of the quantum inequality,
Eq.~(\ref{eq:qi2}). Upon
substitution of the mode functions for both the flat and open universes
into the quantum inequality and using the addition theorem in the open
spacetime case we have
\begin{equation}
[\epsilon = 0]:\qquad\Delta\hat \rho \geq -{1\over 16 \pi^3}
\int_{-\infty}^\infty d^3k\, \omega_{\bf k}\, 
                                       {\rm e}^{-2\omega_{\bf k}t_0} ,
\end{equation}
and 
\begin{equation}
[\epsilon = -1]:\qquad\Delta\hat \rho \geq - {1\over 4\pi^2 a^3}
\int_0^\infty dq\, q^2\, \omega_q\, {\rm e}^{-2\omega_q t_0} , 
\end{equation}
respectively.  The three-dimensional integral in momentum space can be carried
out by making a change to spherical momentum coordinates. The two cases can
be written compactly as
\begin{equation}
\Delta\hat\rho \geq -{1\over 4 \pi^2} \int_0^\infty dk\, k^2\, 
\tilde\omega\, {\rm e}^{-2\tilde\omega t_0},
\end{equation}
where 
\begin{equation}
\tilde\omega = \tilde\omega(k,a,\mu) = \left(k^2 - {\epsilon/ a^2} 
+\mu^2 \right)^{1/2} .
\end{equation}
Note that $\epsilon = 0$ for flat space and $k = {q/ a}$
for the open universe.  This integral can be carried out explicitly in
terms of modified Bessel functions $K_n(z)$.  
The result is
\begin{equation}
\Delta\hat\rho \geq -{3\over 32 \pi^2 t_0^4} \left[ 
{1\over6}\left( z^3 K_3(z) - z^2 K_2(z)
\right) \right]  = -{3\over 32 \pi^2 t_0^4} G(z),  
\end{equation}\label{eq:flatmassiveQI}
where
\begin{equation}
z = 2 t_0 \sqrt{  \mu^2 - {\epsilon\over a^2} }\, .
\end{equation}

\begin{figure}
\begin{center}
\leavevmode\epsfxsize=0.8\textwidth\epsffile{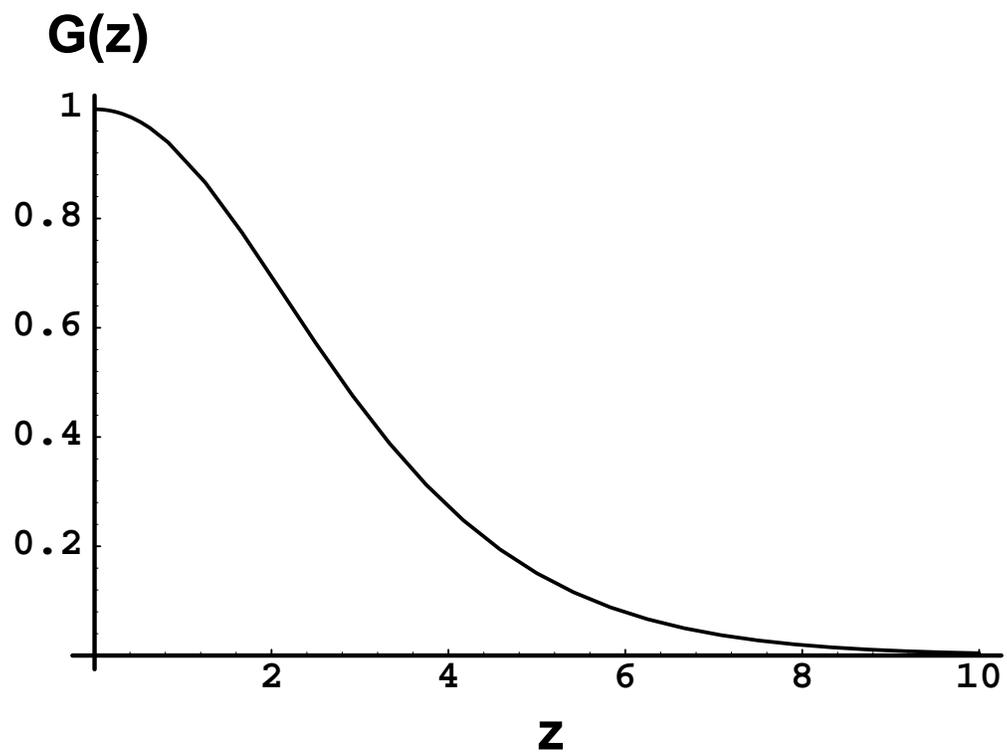}
\end{center}
\caption[Four-dimensional open and flat universe scale function]
{The scale function $G(z)$ for the open and flat Universes.}
\label{fig:open&flat}
\end{figure}

The coefficient $ -3/ (32\pi^2 t_0^4)$ is the lower bound on $\hat \rho$
found in \cite{F&Ro95,F&Ro97} for a massless scalar field in 
Minkowski spacetime. The function $G(z)$ is the ``scale function,''
similar to that found in the case of the three-dimensional closed universe.
It is the same function  found in \cite{F&Ro97} for Minkowski spacetime
($\epsilon=0$), and is plotted in Figure~\ref{fig:open&flat}.  Again
we see that in the limit of $z \rightarrow 0$ the scale function
approaches unity, returning 
the flat space massless inequality in four dimensions, Eq.~(\ref{eq:qi1}).  

\subsection{The Einstein Universe}\label{subsec:RW_Einstein}

In the case of the closed universe, the spatial mode functions are the 
four-dimensional spherical harmonics,\index{spherical harmonics}
which have the form
\cite{Parker,Ford76} 
\begin{eqnarray}
[\epsilon = 1]:\qquad\quad U_{\lambda}({\bf x}) &=& (2\omega_n a^3)^{-1/2}\, 
 \Pi^{(+)}_{nl}(\chi)\, {\rm Y}_{lm}(\theta,\varphi)\, , \\
\omega_n &=& \sqrt{{n(n+2)\over a^2} + \mu^2},\\
\lambda &=& (n,l,m).\nonumber
\end{eqnarray}
Here, $n = 0, 1, 2, \cdots;\quad  l = 0, 1, \cdots,n;$ and 
$m = -l, -l+1, \cdots, +l$.
The function  $\Pi^{(+)}$ is found from $\Pi^{(-)}$ by replacing
$\chi$ by $-i\chi$ and $q$ by $-i(n+1)$ \cite{Parker}.  Alternatively,
they can be written in terms of Gegenbauer \index{Gegenbauer polynomials} 
polynomials \cite{Ford76,Erdelyi} as
\begin{equation}
\Pi^{(+)}_{nl}(\chi) \propto \sin^l\chi \;{\rm C}^{l+1}_{n-l}(\cos \chi).
\end{equation}
In either case, the addition theorem\index{addition theorem} is found
from Eq.~11.4(3) of \cite{Erdelyi} (with $p=2$ and $\xi = \eta$).
This reduces to 
\begin{equation}
\sum_{lm} |\Pi^{(+)}_{nl}(\chi) {\rm Y}_{lm}(\theta,\varphi)|^2 
= {(n+1)^2\over 2\pi^2} \,,
\end{equation}
from which it is easy to show that the energy density inequality,
Eq.~(\ref{eq:qi2}), becomes
\begin{equation}
\Delta\hat \rho \geq - {1\over 4\pi^2 a^3}
\sum_{n=0}^\infty (n+1)^2\, \omega_n\, {\rm e}^{-2\omega_n t_0}. 
\end{equation}

If we use the variable $\eta = {t_0 / a}$ in the above
equation, we can simplify it to
\begin{equation}
\Delta\hat \rho \geq - {3\over 32\pi^2 t_0^4} H(\eta,\mu).
\end{equation}
Again we find the flat space solution in four dimensions, multiplied by
the scale function $H(\eta,\mu)$, which is defined as
\begin{equation}   
H(\eta,\mu) \equiv {8\over 3} \eta^4 \sum_{n=1}^\infty (n+1)^2 
\sqrt{n(n+2)+a^2\mu^2}\; {\rm e}^{-2\eta \sqrt{n(n+2)+a^2\mu^2}},
\label{Eq:closed_Universe_scale_function}
\end{equation}
and is plotted in Figure~\ref{fig:Einstein} for $\mu = 0$.  The scale
function here has a small bump occurring at roughly $\eta \approx
0.5$ with a height of 1.03245.  This may permit the magnitude of the
negative energy to be slightly greater for a massless scalar field in
the Einstein universe than is allowed in a flat universe for comparable
sampling times. A similar result was shown to exist for massive fields
in two-dimensional Minkowski spacetime \cite{F&Ro97}. 

\subsection{Massless Asymptotic Limits in the Einstein Universe}
\label{sec:massless}
As with the three-dimensional closed universe, we can find the 
asymptotic limits of this function. We again follow the method
of the previous section, assuming the scalar field is
massless, and making use of the Plana summation formula to
find
\begin{equation}
H(\eta,0) = {8\over 3}\eta^4\, \bigl( I_2 + I_3 \bigr) \, , 
\end{equation}
where
\begin{equation}
I_2 = \int_0^\infty (x+1)^2\sqrt{x(x+2)}
\, {\rm e}^{-2\eta\sqrt{x(x+2)}}\, dx\, ,
\end{equation}
and
\begin{eqnarray}
I_3 &=& 2 {\rm Re} \left[ i\int_0^\infty {(ix+1)^2\sqrt{ix(ix+2)}
{\rm e}^{-2\eta\sqrt{ix(ix+2)}}\over {\rm e}^{2\pi x} -1} \, dx \right], 
                                                        \nonumber \\
&=& \int_0^\infty \sqrt{2x}\: \frac{ (x^2-1)\sqrt{\sqrt{x^2+4}+x}
    -2x \sqrt{\sqrt{x^2+4}-x} }{{\rm e}^{2\pi x} -1} \:{\rm d}x \, ,\nonumber \\
 &\approx& -0.356109  \,.
\end{eqnarray}
The first integral can be done in terms 
of Struve ${\bf H}_n(z)$ and Neumann $N_n(z)$ functions, with the result
\begin{equation}
I_2 = {\pi\over 16} {d^2\over d\eta^2}\left[{1\over\eta}
\left( {\bf H}_1(2\eta) - N_1(2\eta)\right)\right] + \frac{4}{15} \eta \,.
                                   \label{eq:I2}
\end{equation}

If we follow the same procedure as in the previous section for defining
the renormalized vacuum energy density for the minimally coupled scalar
field in the Einstein universe, then we obtain
\begin{equation}
\rho_{Casimir} = {1\over{4\pi^2 a^4}}\, I_3 
       \approx - { 0.356109 \over{4\pi^2 a^4}} \,. \label{eq:4dCasimir}
\end{equation}
The same method yields $\rho_{Casimir} = 1/{480\pi^2 a^4}$ for the massless
conformal scalar field \cite{Ford75b}. Here our result for the minimal
field, Eq.~(\ref{eq:4dCasimir}), differs from that obtained by Elizalde
\cite{Elizalde} using the zeta function method, $\rho'_{Casimir} = 
- 0.411502/{4\pi^2 a^4}$. This discrepancy probably reflects  that
the renormalized vacuum energy density is not uniquely defined. The 
renormalized stress-tensor in a curved spacetime is only defined up to
additional finite renormalizations of the form of those required to
remove the infinities. In general this includes the geometrical tensors
$^{(1)}H_{\mu\nu}$ and $^{(2)}H_{\mu\nu}$. (See any of the references in
\cite{Brl&Dv, Fulling, Ford94} 
for the definitions of these tensors and a discussion of
their role in renormalization.) In the Einstein universe, both of
these tensors are nonzero and are proportional to $1/a^4$. Thus the
addition of these tensors to $\langle T_{\mu\nu} \rangle$ will change
the numerical coefficient in $\rho_{Casimir}$. The logarithmically
divergent parts of $\langle T_{\mu\nu} \rangle$ which are proportional
to $^{(1)}H_{\mu\nu}$ and $^{(2)}H_{\mu\nu}$ happen to vanish in the
Einstein universe, but not in a more general spacetime. In principle,
we should imagine that the renormalization procedure is performed in
an arbitrary spacetime, and only later do we specialize to a
specific metric. Unfortunately, it is computationally impossible to do
this explicitly. Thus, the fact that a particular divergent term
happens to vanish in a particular spacetime does not preclude the 
presence of finite terms of the same form.

\begin{figure}[hbtp]
\begin{center}
\leavevmode\epsfxsize=0.7\textwidth\epsffile{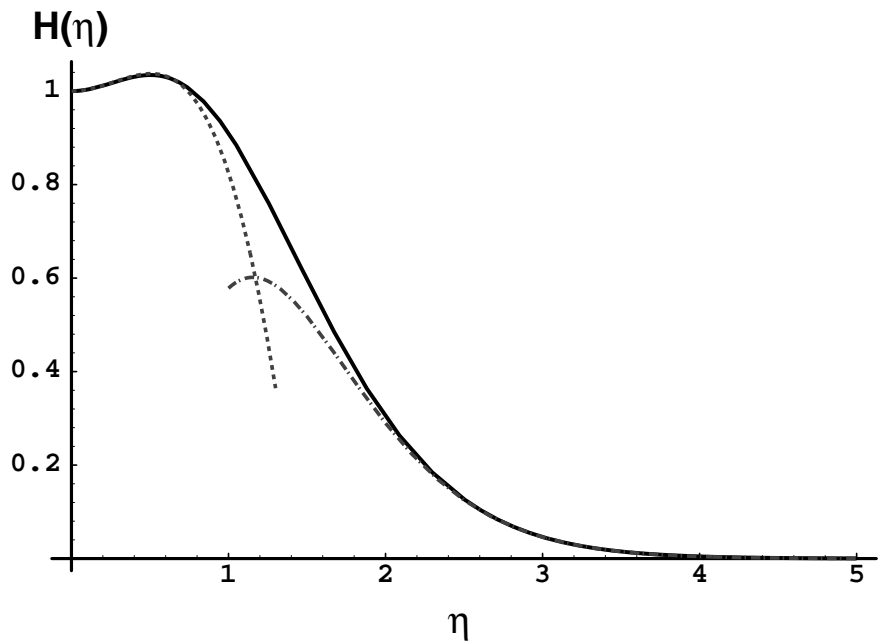}
\end{center}
\caption[Four-dimensional Einstein universe scale function ]  
{The scale function $H(\eta)$ for the Einstein universe.  The
solid line is the exact result.  The dotted line is the 
asymptotic expansion for small $\eta$,  while the dot-dash
curve is the large $\eta$ approximation.  The maximum occurs
at $\eta\approx 0.5$.}
\label{fig:Einstein}
\end{figure}

In the small $\eta$ limit, we can expand the Struve and the Neumann
functions in Eq.~(\ref{eq:I2}) in a Taylor series to obtain
\begin{equation}   
H(\eta,0) = 1 + {1\over 3}\eta^2 + {(3+4\gamma)\over 12} \eta^4
+{8\over 3}(-0.356109)\eta^4 + {1\over 3}\eta^4 \ln \eta 
   + O(\eta^6) + \cdots  ,
\end{equation}
where $\gamma$ is Euler's constant, which arises in the Taylor series
expansion of the Neumann function.  This is similar to that of the
three-dimensional universe and again contains a term of the form of
the Casimir energy.  In the large $\eta$ limit we again keep just the
first term of the series~(\ref{Eq:closed_Universe_scale_function}).
Both asymptotic forms are plotted with the exact solution in
Figure~\ref{fig:Einstein}.  We see that the asymptotic form is
again a very good approximation except in the interval $1 < \eta
<2$, as was the case for the three-dimensional closed universe.

The difference inequalities for massless fields are then given by
\begin{equation}
\Delta\hat \rho \geq  -{3\over{32\pi^2 t_0^4}} \left[ 1 
 + {1\over 3}\eta^2 + {(3+4\gamma)\over 12} \eta^4
 + {1\over 3}\eta^4 \ln \eta + \cdots \right] - \rho_{Casimir}\, ,
\end{equation}
for $t_0 \ll a$ and
\begin{equation}
\Delta\hat \rho \geq -{\sqrt{3}\over{\pi^2 a^4}}\, 
{\rm e}^{-2\sqrt{3}\, t_0/a}, \qquad {\rm for}\quad t_0 \gg a\,.
\end{equation}
Using $\Delta\hat\rho =\hat\rho_{Ren.} - \hat\rho_{Casimir}$,
where $\hat\rho_{Casimir}$ is the expectation value of $\rho_{Ren.}$
in the vacuum state, we can calculate the renormalized energy density
that would be constrained by the quantum inequalities, subject to
renormalization ambiguities. For example, in the four-dimensional
Einstein universe the renormalized energy density inequality is 
\begin{equation}
\hat \rho_{Ren.} \geq  -{3\over{32\pi^2 t_0^4}} \left[ 1 
 + {1\over 3}\eta^2 + {(3+4\gamma)\over 12} \eta^4
 + {1\over 3}\eta^4 \ln \eta + O(\eta^6) + \cdots \right]\, ,
 \label{Eq:einstein_ren}
\end{equation}
for $t_0 \ll a$.  Here the coefficient of the $\eta^4$ term could
be altered by a finite renormalization, but the rest of the expression 
is unambiguous. This result is identical to the short sampling time
expansion of the quantum inequality in the Einstein universe, 
Eq.~(\ref{eq:QI_Einstein_short}), developed in Section~\ref{sec:expans}.
Expressions similar to Eq.~(\ref{Eq:einstein_ren}) could be found for 
any of the other cases above.
In the case of the flat Robertson-Walker universe, there is no Casimir
vacuum energy.  Under such circumstances the difference inequality and
the renormalized energy density inequality are the same, and are free
of ambiguities.

\section{Quantum Inequalities Near Planar Mirrors}\label{sec:planar}

\subsection{Single Mirror}
Consider four-dimensional Minkowski spacetime with a perfectly
reflecting boundary at $z=0$, located in the $x-y$ plane, at which
we require the scalar field to vanish.  The two-point function can be
found by using the standard Feynman Green's function in Minkowski space, 
\begin{equation}
G^{(0)}_F(x,x') ={-i\over 4\pi^2 [ (x-x')^2 + (y-y')^2 +(z-z')^2 -(t-t')^2]},
\end{equation}
and applying the method of images to find the required Green's function
when the boundary is present. For a single conducting plate we have
\begin{eqnarray}
G_F(x,x') &=& {-i\over 4\pi^2}\left[ {1\over{ (x-x')^2 + (y-y')^2 +(z-z')^2
-(t-t')^2}} \right.\nonumber\\
&& \qquad - \left. {1\over{ (x-x')^2 + (y-y')^2 +(z+z')^2 -(t-t')^2}}
\right].
\end{eqnarray}
If we Euclideanize by allowing $t\rightarrow -it_0$, $t'\rightarrow it_0$
and then take $x' \rightarrow x$, we find
\begin{equation}
G_E(2t_0) = {1\over 16\pi^2}\left( {1\over t_0^2} -
{1\over {t_0^2 + z^2}} \right) .
\end{equation}
In addition, the Euclidean box operator is given by
\begin{equation}
\Box_E = \partial^2_{t_0} +\partial^2_x+\partial^2_y+\partial^2_z.
\end{equation}
It is easily shown that the quantum inequality is given by
\begin{equation}
\Delta\hat\rho \geq -{1\over 4}\, \Box_E \, G_E(2t_0)
= -{3\over 32\pi^2\, t_0^4} + {1\over 16\pi^2 (t_0^2 + z^2)^2}\, .
\end{equation}

The first term of this inequality is identical to that for Minkowski
space.  The second term represents the effect of the mirror on the
quantum inequality. For the minimally coupled scalar field 
we know from Eq.~(\ref{eq:exterior_ST}) that there is
a non-zero, negative vacuum energy density which diverges
as the mirror is approached.  Adding this vacuum term to both the left-
and right-hand sides of the above expression allows us to find the
renormalized quantum inequality for this spacetime,
\begin{equation}
\hat\rho_{Ren.} \geq -{3\over 32\pi^2\, t_0^4} +
{1\over 16\pi^2 (t_0^2 + z^2)^2} - {1\over 16\pi^2  z^4}\, .
\end{equation}

There are two limits in which the behavior of the renormalized
quantum inequality can be studied.  First consider $z \gg t_0$.
In this limit, the correction term due to the mirror, and the
vacuum energy very nearly cancel and the quantum inequality
reduces to
\begin{equation}
\hat\rho_{Ren.} \geq -{3\over 32\pi^2\, t_0^4}\,.
\end{equation}
This is exactly the expression for the quantum inequality in Minkowski
spacetime.  Thus, if an observer samples the energy density on time
scales which are small compared to the light travel time to the boundary,
then the Minkowski space quantum inequality is a good approximation.

The other important limit is when $z \ll t_0$.  This is the case
for observations made very close to the mirror, but for very long
times.  The quantum inequality then reduces to
\begin{equation}
\hat\rho_{Ren.} \geq -{1\over 16\pi^2\, z^4}\,.
\end{equation}
Here, we see that the quantum field is satisfying the quantum
averaged weak energy condition.  Recall that throughout the present
paper, we are concerned with observers at rest with respect to the
plate.  If the observer is moving and passes through the plate,
then it is necessary to reformulate the quantum inequalities in
terms of sampling functions with compact support \cite{F&P&R97}.
It should be noted that the divergence of the vacuum energy on the
plate is due to the unphysical nature of perfectly reflecting
boundary conditions.  If the mirror becomes transparent at high
frequencies, the divergence is removed.  Even if the mirror is 
perfectly reflecting, but has a nonzero position uncertainty, the
divergence is also removed \cite{F&SV97b}.

\subsection{Two Parallel Plates}

Now let us consider the case of two parallel plates, one located in
the $z = 0$ plane and another located in the $z=L$ plane.  We are 
interested in finding the quantum inequality in the region between
the two plates, namely $0\leq z \leq L$.  We can again use the method
of images to find the Green's function.  In this case, not only do we
have to consider the reflection of the source in each mirror, but 
we must also take into account the reflection of one image in the
other mirror, and then the reflection of the reflections. This
leads to an infinite number of terms that must be summed to find the
exact form of the Green's function.  If we place a source at
$(t', x', y', z')$, then there is an image of the source  at
$(t', x', y', -z')$ from the mirror at $z=0$ and a second image
at $(t', x', y', 2L-z')$ from the mirror at $z=L$.  Then, we must
add the images of these images to the Green's function, continuing
{\it ad infinitum} for every pair of resulting images.  If we use
the notation
\begin{equation}
G_F(z, a \pm z') \equiv G^{(0)}_F(t, x , y, z; t', x', y', a \pm z'),
\end{equation}
where $a$ is a constant, then we can write the Green's functions
between the plates as
\begin{eqnarray}
G(x,x') &=& G_F(z, z') - G_F(z, -z') + \sum_{n=1}^\infty\left[
G_F(z, 2nL+z') - G_F(z, -2nL-z')\right. \nonumber\\
&&\left. + G_F(z, -2nL+z') - G_F(z, 2nL-z') \right].
\end{eqnarray}
We Euclideanize as above, and let the spatial separation
between the source and observer points go to zero, to find
\begin{eqnarray}
G_E(2t_0) &=& {1\over 16\pi^2}\left( {1\over t_0^2} -{1\over {t_0^2 + z^2}}
\right)\nonumber\\ 
&&\qquad + {1\over 16\pi^2} \sum_{n=1}^\infty\left[{2\over {t_0^2 + (nL)^2}}
-{1\over {t_0^2 + (nL+z)^2}}-{1\over {t_0^2 +(nl-z)^2}}
\right].
\end{eqnarray}
It is now straightforward to find the quantum inequality,
\goodbreak
\begin{eqnarray}
\Delta\hat\rho &\geq& -{3\over 32\pi^2\, t_0^4}\, + 
{1\over 16\pi^2 (t_0^2 + z^2)^2}\nonumber\\ 
&&\qquad + {1\over 16\pi^2 }\sum_{n=1}^\infty\left\{
{(nL)^2 - 3t_0^2 \over [t_0^2 + (nL)^2]^3} + {1\over [t_0^2 + (nL+z)^2]^2}
+ {1\over [t_0^2 + (nL-z)^2]^2} \right\}.
\end{eqnarray}
The first
term in the expression above is identical to that found for Minkowski
space.  The second term is the modification of the quantum inequality
due to the mirror at $ z = 0$.  The modification due to the presence
of the second mirror is contained in the summation, as well as all
of the multiple reflection contributions. When the Casimir vacuum
energy, given by \cite{Fulling}
\begin{equation}
\rho_{vacuum} = -{\pi^2\over 48 L^4}\,{3-2\sin^2(\pi z/L) \over 
\sin^4(\pi z/L)}\, - \, {\pi^2 \over 1440 L^4}\, ,
\end{equation}
is added back into this equation for renormalization, we
find, as with a single mirror, that close to  the
mirror surfaces the vacuum energy comes to dominate and the quantum
inequality becomes extremely weak. 

\section{Spacetimes with Horizons}\label{sec:horizons}

We will now change from flat spacetimes with boundaries
to spacetimes in which there exist horizons.  We
will begin with the two-dimensional Rindler spacetime
to develop the quantum inequality for uniformly accelerating
observers.  For these observers, there
exists a particle horizon along the null rays $x = \pm t$ 
(see Figure~\ref{fig:rindler}).

We will then look at the static coordinatization of 
de~Sitter spacetime.  Again there exists a particle horizon 
in this spacetime, somewhat similar to that of the Rindler
spacetime. The two problems differ because Rindler spacetime
is flat while the de~Sitter spacetime has constant, positive
spacetime curvature.
   
\subsection{Two-Dimensional Rindler Spacetime}
\begin{figure}
\begin{center}
\leavevmode\epsffile{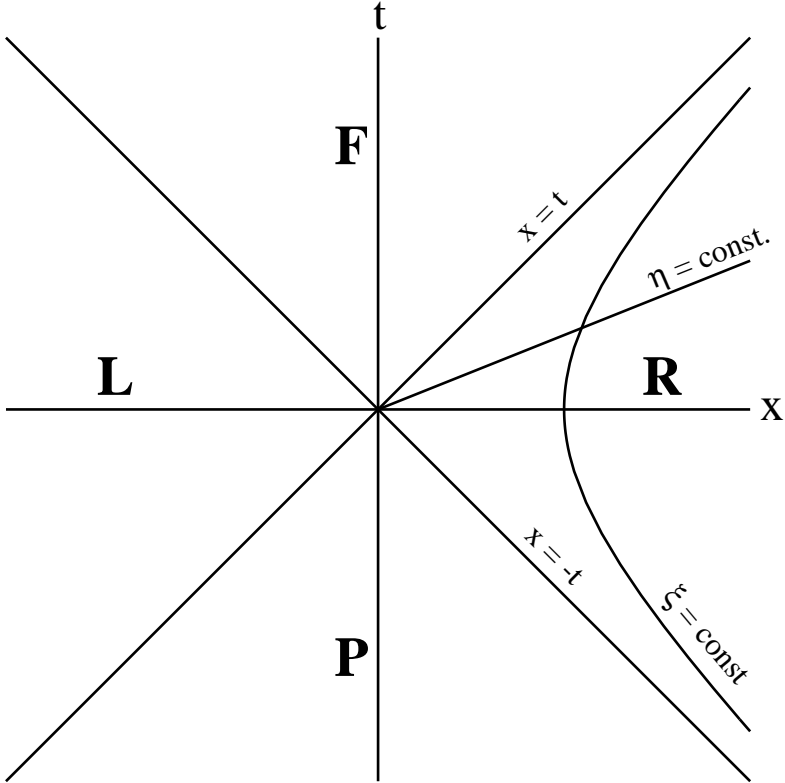}
\end{center}
\caption[Rindler coordinatization of two-dimensional Minkowski spacetime]
{The Rindler coordinatization of two-dimensional Minkowski
spacetime.  The time coordinate $\eta =$ constant are straight lines
passing through the origin, while the space coordinate $\xi =$ constant
are hyperbol\ae.  The Minkowski spacetime is covered by four separate
coordinate patches, labeled by {\bf L}, {\bf R}, {\bf F}, and {\bf P}.
The two null rays ($x=t$ and $x=-t$) act as horizons. }
\label{fig:rindler}
\end{figure}

We begin with the usual two-dimensional Minkowski metric
\begin{equation}
ds^2 =-dt^2+dx^2 .
\end{equation}
Now let us consider an observer who is moving with constant acceleration.
We can transform to the observer's rest frame (Section~4.5 of \cite{Brl&Dv})
by 
\begin{eqnarray}
t &=& a^{-1} e^{a\xi} \sinh a\eta\,,\\
x &=& a^{-1} e^{a\xi} \cosh a\eta\,, 
\end{eqnarray}
where $a$ is a constant related to the acceleration by
\begin{equation}
a\;e^{-a\xi} = {\rm proper\ acceleration}.
\end{equation}
The metric in the rest frame of the observer is then given by
  \begin{equation}
ds^2 = e^{2a\xi}( -d\eta^2 + d\xi^2 ).\label{eq:rindler_metric}
\end{equation}
The accelerating observer's coordinates $(\eta,\xi)$ only cover
one quadrant of Minkowski spacetime, where $x>|t|$.  This is
shown in Figure~\ref{fig:rindler}.  Four different coordinate
patches are required to cover all of Minkowski spacetime in the
regions labeled {\bf L}, {\bf R}, {\bf F}, and {\bf P}.  For the
remainder of the paper we will be working specifically in the
left and right regions, labeled {\bf L} and {\bf R} respectively.
In these two regions, uniformly
accelerating observers in Minkowski spacetime can be represented
by observers at rest at constant $\xi$ in Rindler coordinates, as
shown by the hyperbola in Figure~\ref{fig:rindler}.

The massless scalar wave equation in Rindler spacetime is given by
\begin{equation}
e^{-2a\xi} \left(-{d^2\over d\eta^2} + {d^2\over d\xi^2}\right)
\phi(\eta ,\xi) = 0,
\end{equation}
which has the positive frequency mode function solutions
\begin{equation}
f_k(\eta ,\xi)= (4\pi\omega)^{-1/2}\;e^{ik\xi \pm i\omega\eta}.
\end{equation}
Here $-\infty<k<\infty$ and $\omega = |k|$. The plus and minus
signs correspond to the left and right Rindler wedges, respectively.
Using the above mode functions, we can expand the general solution
as
\begin{equation}
\phi(\eta ,\xi) = \int_{-\infty}^\infty dk \left[ b_k^L f_k(\eta ,\xi) +
{b_k^L}^\dagger f_k^*(\eta ,\xi) + b_k^R f_k (\eta ,\xi) + 
{b_k^R}^\dagger f_k^*(\eta ,\xi) \right],
\end{equation}
where ${b_k^L}^\dagger$ and ${b_k^L}$ are the creation and annihilation
operators in the left Rindler wedge and similarly for ${b_k^R}^\dagger$
and ${b_k^R}$ in the right Rindler wedge.  We also need to define two
vacua, $|0_L\rangle$ and $|0_R\rangle$, with the properties
\begin{equation}
{b_k^L}^\dagger|0_R\rangle ={b_k^R}^\dagger|0_L\rangle =
b_k^L |0_L\rangle = b_k^L |0_R\rangle =b_k^R |0_L\rangle 
=b_k^L |0_R\rangle  = 0.
\end{equation}
The Rindler particle states are then excitations above the vacuum given by
\begin{eqnarray}
|\{1_k\}_L\rangle &=& {b_k^L}^\dagger|0_L\rangle,\\
|\{1_k\}_R\rangle &=& {b_k^R}^\dagger|0_R\rangle.
\end{eqnarray} 
With this in hand, we can find the two-point function in either the
left or right regions. Let us consider the right Rindler wedge,
where
\begin{eqnarray}
G^+(x,x') &=& \langle 0_R| \phi(x) \phi(x') |0_R\rangle,\nonumber\\
&=& \int_{-\infty}^\infty dk \, f_k(x) f_k^*(x')\, ,\nonumber\\
&=& {1\over 4\pi} \int_{-\infty}^\infty {dk\over\omega}
e^{ik (\xi-\xi') - i\omega (\eta-\eta')}.
\end{eqnarray}
To find the Euclidean two-point function required for the
quantum inequality, we first allow the spatial separation
to go to zero and then take $(\eta - \eta') \rightarrow -2i\eta_0$,
yielding
\begin{equation}
G_E(2\eta_0) = {1\over 2\pi}\int_0^\infty {d\omega\over\omega}
e^{-2\omega\eta_0}.
\end{equation}
In two dimensions, the Euclidean Green's function for the massless
scalar field has an infrared divergence as can be seen from the form
above, in which the integral is not well defined in the limit of
$\omega\rightarrow 0$.  However, in the process of finding the 
quantum inequality we act on the Green's function with the Euclidean
box operator. If we first take the derivatives of the Green's function,
and then carry out the integration, the result is well defined for all
values of $\omega$. In Rindler space, the Euclidean box operator is 
given by
\begin{equation}
\Box_E = e^{-2a\xi} \left({d^2\over d\eta^2} + {d^2\over d\xi^2}\right).
\end{equation}
It is now easy to solve for the quantum inequality 
\begin{equation}
\Delta\hat\rho \geq -{1\over4} \Box_E G_E(2\eta_0)
= -{1\over 2\pi} e^{-2a\xi} \int_o^\infty d\omega\;\omega
e^{-2\omega\eta_0}
= -{1\over 8\pi \left(e^{a\xi}\,\eta_0\right)^2}\, .
\end{equation}
However, the coordinate time $\eta_0$ is related to the observer's proper
time by
\begin{equation}
\tau_0 = e^{a\xi}\eta_0\, ,
\end{equation}
allowing us to rewrite the quantum inequality in a more covariant
form,
\begin{equation}
\Delta\hat\rho\geq -{1\over 8\pi\tau_0^2}\,.\label{eq:2D_QI}
\end{equation}
This is exactly the same form of the quantum inequality as found
in two-dimensional Minkowski spacetime \cite{F&Ro95,F&Ro97}.
As we have seen in Section~\ref{subsec:2D_conform}, this
is a typical property of static two-dimensional spacetimes 
which arises because all two-dimensional static spacetimes are
conformal to one another. However, we reiterate that the
renormalized quantum inequalities are not identical in
different spacetimes because of differences in the vacuum
energies.

\subsection{de~Sitter Spacetime}
Let us now consider four-dimensional de~Sitter spacetime.  The scalar
field quantum inequality, Eq.~(\ref{eq:QI}), assumes a timelike Killing
vector, so it will be convenient to use the static parametrization
of de~Sitter space,
\begin{equation}
ds^2 = -(1-{r^2\over \alpha^2})\,dt^2 + (1-{r^2\over \alpha^2})^{-1}
dr^2 + r^2(d\theta^2 + \sin^2\theta\;d\varphi^2)\; .
\end{equation}
There is a particle horizon at $r = \alpha$ for an observer sitting
at rest at $ r = 0$.  The coordinates take the values, $0 \leq r < 
\alpha$, $0\leq\theta \leq \pi$, and $0\leq\varphi < 2\pi$.  It should
be noted that this choice of metric covers one quarter of de~Sitter 
spacetime.     

The scalar wave equation is
\begin{equation}
\left(1-{r^2\over \alpha^2}\right)^{-1} \partial_t^2 \phi - {1\over r^2}
\partial_r\left[ r^2 \left(1-{r^2\over \alpha^2}\right) \partial_r 
\right]\phi -{{\bf L}^2\over r^2}\phi + \mu^2\phi  = 0\,,
\end{equation}
where ${\bf L}^2$ is the square of the standard angular momentum
operator and is defined by
\begin{equation}
{\bf L}^2 ={1\over \sin\theta}\partial_\theta \left(
\sin\theta\,\partial_\theta \right) + {1\over \sin^2\theta}
\partial_\varphi^2\, . 
\end{equation}
The unit norm positive frequency mode functions are found 
\cite{Lohiya,Lapedes,Higuchi,Sato94,Kaiser} to be of the form
\begin{equation}
\hat\phi_{\omega,l,m}(t,r,\theta,\varphi) = {1\over\sqrt{4\pi\alpha^3
\omega}}\;f_\omega^l(z) \;{\rm Y}_{lm}(\theta,\varphi)
\;e^{-i\omega t},
\end{equation}
where $z = r/\alpha$ is a dimensionless length,  ${\rm Y}_{lm}
(\theta,\varphi)$ are the standard spherical
harmonics, and the mode labels $l$ and $m$ take the values  $l = 0, 1,
2, \cdots$ and $-l\leq m\leq l$.  The radial portion of the the
solution is given by   
\begin{equation}
f_\omega^l(z) = {\Gamma(b^+_l)\Gamma(b^-_l) \over \Gamma(l+{3\over 2})
\Gamma(i\alpha\omega)} \; z^l \; (1-{z^2})^{i\alpha\omega/ 2}\; 
{\rm F}(\,b^-_l, \,b^+_l;\, l+{3\over 2} ;\, z^2)\, ,
\end{equation}
where ${\rm F}(\alpha,\beta ;\gamma ; z)$ is the hypergeometric 
function \cite{Gradshteyn} and
\begin{equation}
b^\pm_l = {1\over 2}\left(l + {3\over 2} + i\alpha\omega \pm\sqrt{{9\over 4}
-\alpha^2\mu^2}\right).\label{eq:b_pm}
\end{equation}
We can then express the two-point function as
\begin{equation}
G(x,x') = \sum_{lm}\int_0^\infty\,dk {1\over 4\pi\alpha^2 k}
{f_k^l}^*(z) \; f_k^l(z') \;{\rm Y}_{lm}^*(\theta,\varphi)\; 
{\rm Y}_{lm}(\theta',\varphi') \; e^{ik(t-t')/a},
\end{equation} 
where $k \equiv \alpha\omega$.  Now if we Euclideanize according to
Eq.~(\ref{eq:Euclideanize}) and set the spatial separation of the
points to zero, we may use the addition theorem for the
spherical harmonics, Eq.~(\ref{eq:sum_rule}), to find the Euclidean
Green's function
\begin{equation}
G_E = {1\over 16\pi^2\alpha^2}
\sum_l \int_0^\infty dk {(2l+1)\over k} \left|{\Gamma(b^+_l)\Gamma(b^-_l)
\over \Gamma(l+{3\over 2}) \Gamma(ik)} \right|^2
 z^{2l} \, \left|{\rm F}(b^-_l, b^+_l ; l+{3\over 2} ;\, z^2)
\right|^2  e^{-2kt_0 / \alpha}\, . \label{eq:deSitterGreens}
\end{equation}
This is independent of the angular coordinates, as expected,
because de~Sitter spacetime is isotropic. We now need the Euclidean box
operator.  Because of the angular independence of the Green's function,
it is only necessary to know the temporal and radial portions of the
box operator.  The energy density inequality,
Eq.~(\ref{eq:QI}), then becomes
\begin{equation}
\Delta\hat \rho \geq  -{1\over 4}\left\{ {1\over(1-z^2)}
\partial_{t_0}^2 + {1\over\alpha^2 z^2} \partial_z
\left[ z^2 \,(1-z^2) \partial_z \right] \right\}\;
G_E({\bf x},-t_0;{\bf x},+t_0).
\label{eq:open_form}
\end{equation}

The temporal derivative term in Eq.~(\ref{eq:open_form}) will simply
bring down two powers of $k/\alpha$. Using the properties of the
 hypergeometric function, it can be shown that
\begin{equation}
\left|{\rm F}(b^-_l, b^+_l; l+{3\over 2} ;\, z^2) \right|^2 =
(1-z^2)^{ik} \;{\rm F}^2(b^-_l, b^+_l ; l+{3\over 2} ;\, z^2),
\end{equation}
from which we can take the appropriate spatial derivatives. 
If we allow $z \rightarrow 0$, then we have $F \rightarrow 1$
and only the $l=0$ terms will contribute in the time derivative part of
Eq.~(\ref{eq:open_form}).  For the radial derivative, it is possible
to show that  
\begin{equation}
\lim_{z\rightarrow 0}{1\over z^2} \partial_z \left\{ z^2 \,(1-z^2) \partial_z 
\left[ z^{2l}(1-z^2)^{ik} {\rm F}^2( z^2)
\right]\right\}
=
\left\{ \matrix{  
         2(\alpha^2\mu^2-k^2)&{\rm for}\; l=0,\cr
         6&{\rm for}\;l=1,\cr
         0&{\rm otherwise.}} \right.
\end{equation}
Using this result, we find for the observer at $r = 0$ that
\begin{equation}
\Delta\hat\rho \geq -{1\over 8\pi^4 \alpha^4 } \int_0^\infty dk 
\sinh(\pi k) \left[
(k^2+\alpha^2\mu^2)\left|\Gamma(b^-_0)\Gamma(b^+_0)\right|^2+
4\left|\Gamma(b^-_1)\Gamma(b^+_1)\right|^2\right] e^{-2 t_0 k / \alpha}.
\end{equation}
There are two cases for which the right-hand side can be evaluated
analytically, $\mu = 0$ and $\mu = \sqrt{2}/\alpha$. For $\mu = 0$,
we have
\begin{eqnarray}
\Delta\hat\rho &\geq& -{1\over 8\pi^4 \alpha^4 } \int_0^\infty dk 
\sinh(\pi k) \left[
k^2\left|\Gamma(i{k\over 2})\,\Gamma({3\over 2}+i{k\over 2})\right|^2
+4\left|\Gamma({1\over 2}+i{k\over 2})\,\Gamma(2+i{k\over 2})
\right|^2\right] e^{-2 t_0 k / \alpha},\nonumber\\
&=&-{1\over 8\pi^2 \alpha^4 }\int_0^\infty dk (2k^3+5k)
e^{-2 t_0 k / \alpha}
=-{3\over 32\pi^2 t_0^4 }\left[ 1 + {5\over3}
\left({t_0\over\alpha}\right)^2 \right]\, ,
\label{eq:minimalDeSitter}\end{eqnarray}
where we have used the identities for gamma functions proven
in Appendix~\ref{chpt:gammas}. Similarly for $\mu = \sqrt{2}/\alpha$,
we find
\begin{equation}
\Delta\hat\rho \geq -{3\over 32\pi^2 t_0^4 }\left[ 1 + 
\left({t_0\over\alpha}\right)^2 \right].
\label{eq:ConformalDeSitter}\end{equation}
We can compare these results with the short sampling time approximation
from Section~\ref{sec:expans}.  Solving for the necessary geometrical
coefficients, we find
\begin{eqnarray}
v_{000} &=& \left( {29\over 60}{1\over\alpha^4} - {1\over 4}{\mu^2\over
\alpha^2}\right) |g_{tt}|\, ,\\[8pt]
v_1 & = & {29\over 60}{1\over\alpha^4} - {1\over 2}{\mu^2\over
\alpha^2} + {1\over 8} \mu^4 \, , \\[8pt]
{1\over 2}g_{tt}\nabla^j\nabla_j\, g_{tt}^{-1} &=& {1\over\alpha^2} 
{\left(3-r^2/\alpha^2\right) \over \left(1-r^2/\alpha^2\right)}
\, . \end{eqnarray}
The general short sampling time expansion, Eq.~(\ref{eq:QI_expansion}),
now becomes
\goodbreak
\begin{eqnarray}
\Delta\hat\rho &\geq&  - {3\over 32 \pi^2 \tau_0^4} \left\{ 1+ {1\over 3}
\left[ {2\over\alpha^2} - \mu^2 + {1\over\alpha^2}{(3-r^2/\alpha^2) 
\over (1-r^2/\alpha^2)} \right] \, \tau_0^2 \right.\nonumber\\
&&\qquad\qquad \left.+ {\mu^2\over 6}\left(
\mu^2 - {2\over\alpha^2}\right)\, \tau_0^4 \, \ln (2 \tau_0^2 / 
\alpha^2) + O(\tau_0^4) + \cdots\right\}, 
\end{eqnarray}
where $\tau_0 = (1 -r^2/\alpha^2 )^{1/2} t_0$.
If $r = 0 $ and $\mu$ takes the values $0$ or $\sqrt{2}/\alpha$, this
agrees with Eqs.~(\ref{eq:minimalDeSitter}) or (\ref{eq:ConformalDeSitter}),
respectively.  Note that this small $t_0$ expansion is valid for all radii, 
$0 \leq r < \alpha$.  We can also find the proper sampling time from
Eq.~(\ref{eq:little_t}) for which this expansion is valid,
\begin{equation} 
\tau_0 \ll \tau_m \equiv \alpha \sqrt{{1-r^2/\alpha^2} \over
{5 - 3r^2/\alpha^2}}.
\end{equation}    
For the observer sitting at the origin of the coordinate system,
$\tau_0 \ll \alpha/ \sqrt{5}$.  This is the scale on which
the spacetime can be considered ``locally flat.'' For observers at
$r > 0$, who do not move on geodesics, $\tau_m$ decreases and
approaches zero as $r \rightarrow \alpha$:  
\begin{equation}
\tau_m \sim \sqrt{\alpha (\alpha-r)} ,\qquad r\rightarrow\alpha.
\end{equation}
Note that the proper distance to the horizon from radius $r$ is
\begin{equation}
\ell = \int_r^\alpha {dr' \over \sqrt{1 -r'^2 /\alpha^2 }} =
\alpha\left[ \pi/2 - \arcsin (r/\alpha)\right]
\sim\sqrt{2\alpha (\alpha-r)} ,\qquad \mbox{ as } r\rightarrow\alpha.
\end{equation}
Thus, for observers close to the horizon, if the sampling time is
small compared to this distance to the horizon, {\it i.e.}, if
$\tau_0 \ll \ell$, then $\tau_0 \ll \tau_m$ and the short sampling
time expansion is valid. 

We can also obtain a renormalized quantum inequality
for the energy density at the origin for the case  $\mu = \sqrt{2}
\alpha$.  By the addition of the vacuum energy to both  sides of
Eq.~(\ref{eq:ConformalDeSitter}) we find for $r = 0$ 
\begin{equation}
\hat\rho_{Ren.} \geq -{3\over 32\pi^2 t_0^4 }\left[ 1 + 
\left({t_0\over\alpha}\right)^2 \right] - {1\over 960\pi^2\alpha^4}\, .
\end{equation}

We can now predict what will happen in the infinite
sampling time limit of the renormalized quantum inequality for any
observer's position. We know from Eqs.~(\ref{eq:deSitterGreens})
and (\ref{eq:open_form}) that the difference inequality will always
go to zero, yielding a QAWEC in static de~Sitter spacetime of
\begin{equation}
\lim_{t_0\rightarrow\infty} {t_0\over\pi} \int_{-\infty}^\infty 
{\langle T_{tt}u^0 u^0 \rangle_{Ren.} \over  t^2 + t_0^2} dt \geq
{1\over 480\pi^2\alpha^4}\left[ -{\alpha^2\over (\alpha^2-r^2)}+
{1\over2}(1-{r^2\over\alpha^2}) \right].
\label{eq:QAWECdeSitter}
\end{equation}
We immediately see that for a static observer who is arbitrarily close
to the horizon in de~Sitter spacetime, the right hand side of
Eq.~(\ref{eq:QAWECdeSitter}) becomes extremely negative, and diverges
on the horizon itself.  This is similar to the behavior found
for static observers located near the perfectly reflecting mirror
discussed earlier and for Rindler space.


\section{Black Holes}\label{sec:black_holes}
We now turn our attention to an especially
interesting spacetime in which quantum inequalities can be
developed, the exterior region of a black hole in two
and four dimensions.     
 
\subsection{Two-Dimensional Black Holes}\label{sec:2D_Black_holes}
Let us consider the metric
\begin{equation}
ds^2 = - C(r)\,dt^2 + C(r)^{-1}\, dr^2\, ,
\end{equation}
where $C(r)$ is a function chosen such that $C\rightarrow1$ and 
$\partial C / \partial r \rightarrow 0$ as $r\rightarrow \infty$.
Additionally, there is an event horizon at some value $r_0$
where $C(r_0) = 0$.  For example, in the Schwarzschild
spacetime, $C(r) = 1 -2Mr^{-1}$, there is a horizon at $r = 2M$.
Another choice for $C$ is that of the Reissner-Nordstr\"om black
hole, where $C(r) = 1 -2Mr^{-1} + Q^2 r^{-2}$.  In general,
we will leave the function $C$ unspecified for the remainder
of the derivation.  The above metric leads to the massless,
minimally coupled scalar wave equation
\begin{equation}
-{1\over C(r)}\partial_t^2\phi(r,t) + \partial_r\left[C(r)
\partial_r \phi(r,t) \right] = 0\, .
\end{equation}
Unlike in four dimensions, the two-dimensional wave equation can be
analytically solved everywhere.  If we use the standard definition
of the $r^*$ coordinate,
\begin{equation}
r^* \equiv \int {dr\over C(r)}\, ,
\end{equation}
then it is convenient for us to take as the definition
of the positive frequency mode functions
\begin{equation}
f_k(r,t) = i\left( 4\pi\omega \right)^{-1/2} \;
e^{ikr^*-i\omega t},\qquad \omega = |k|,
\end{equation}
where $-\infty < k < \infty$.  

The problem of finding the quantum inequality simply reduces to using the
mode functions to find the Euclidean Green's function.  We have 
\begin{equation}
G_E(2t_0) = \int_{-\infty}^\infty dk\;\left|{i\over \sqrt{4\pi\omega}}
e^{ikr^*}\right|^2 \, e^{-2\omega t_0}
= {1\over 2\pi} \int_0^\infty d\omega\;\omega^{-1}
e^{-2\omega t_0}. 
\end{equation}
As in the case of two-dimensional Rindler space, the Euclidean
Green's function has an infrared divergence.  We can again apply
the Euclidean box operator first and then do the integration to
obtain the quantum inequality, 
\begin{equation}
\Delta\hat\rho \geq -{1\over 2\pi C(r)}\int_0^\infty d\omega\;
\omega\,e^{-2\omega t_0}  =  -{1\over 8\pi C(r) t_0^2}\, . 
\end{equation}
However, the observer's proper time is related to the coordinate time
by $\tau = C(r)^{1/2} t$, such that we can write the difference
inequality as
\begin{equation}
\Delta\hat\rho \geq  -{1\over 8\pi \tau_0^2}.
\end{equation}
This is the same form as found for two-dimensional Minkowski and
Rindler spacetimes. This is the expected result because of the conformal
equivalence of all two-dimensional spacetimes.

This now brings us to the matter of renormalization.  There exist
three candidates for the vacuum state of a black hole: the Boulware
vacuum, the Hartle-Hawking vacuum, and the Unruh vacuum.  However
the derivation of the difference inequality relies on the mode
functions being defined to have positive frequency with respect to
the timelike Killing vector $\partial_t$, and the vacuum state
being destroyed by the annihilation operator, {\it i.e.},
\begin{equation}
a_k \; |0_k\rangle = 0, \qquad\mbox{for all } k\, .
\end{equation}
In Schwarzschild spacetime, this defines the Boulware vacuum.
Thus, we can solve for the renormalized quantum inequality,
\begin{equation}
\hat\rho_{Ren.}\equiv{t_0\over\pi} \int_{-\infty}^\infty 
{\langle{T_{tt}/ | g_{tt} |} \rangle_{Ren.}\over{t^2+t_0^2}}dt
\geq  -{1\over 8\pi \tau_0^2} + \rho_{B}(r)\; .
\end{equation}
The Boulware vacuum energy density in two dimensions
for the Reissner-Nordstr\"om black hole is given explicitly by 
(see Section~8.2 of \cite{Brl&Dv})
\begin{equation}
\rho_{B}(r) = {1\over 24\pi}\left(1 -{2M\over r} + {Q^2\over r^2}
\right)^{-1}\left[ -{4M\over r^3}+
{7 M^2\over r^4} + {6 Q^2\over r^4} - {14 MQ^2\over r^5} +
{5 Q^4\over r^6}\right].
\end{equation}
In the limit $\tau_0 \rightarrow \infty$, we recover a QAWEC
condition on the energy density
\begin{equation}
\lim_{t_0\rightarrow\infty} {t_0\over \pi} \int_{-\infty}^\infty 
{\langle{T_{tt}/g_{tt}}\rangle_{Ren.}\over t^2+t_0^2} dt
\geq \rho_{B}(r).\label{eq:AWEC_boulware}
\end{equation}
This has the interpretation that the integrated energy density in
an arbitrary  particle state can never be more negative than that
of the Boulware vacuum state. In particular, this will be true for
the Hartle-Hawking and Unruh vacuum states.  

\subsection{Four-Dimensional Schwarzschild Spacetime}

Now let us turn to the four-dimensional Schwarzschild  spacetime with
the metric
\begin{equation}
ds^2 = -\left(1-{2M\over r}\right) dt^2 + \left(1-{2M\over r}\right)^{-1} dr^2
+ r^2\,(d\theta^2 + \sin^2\theta\, d\varphi^2)\, .
\end{equation}
The normalized mode functions for a massless scalar field in the
exterior region ($r>2M$) of Schwarzschild spacetime can be written as 
\cite{DeWitt}
\begin{eqnarray}
\stackrel{\rightarrow}{f}_{\omega l m}(x) &=& (4\pi\omega)^{1/2} 
e^{-i\omega t} \stackrel{\rightarrow}{R}_l(\omega | r)
{\rm Y}_{lm}(\theta,\varphi),\nonumber\\
\stackrel{\leftarrow}{f}_{\omega l m}(x) &=& (4\pi\omega)^{1/2}
e^{-i\omega t} \stackrel{\leftarrow}{R}_l(\omega | r)
{\rm Y}_{lm}(\theta,\varphi),
\end{eqnarray}
where $\stackrel{\rightarrow}{R}_l(\omega | r)$ and 
$\stackrel{\leftarrow}{R}_l(\omega | r)$ are the
outgoing and ingoing solutions to the radial portion
of the wave equation, respectively.  Although they cannot be written
down analytically, their asymptotic forms are
\begin{equation}
\stackrel{\rightarrow}{R}_l(\omega | r) \sim \left\{ 
\begin{array}{ll}
         r^{-1} e^{i\omega r^*} + \stackrel{\rightarrow}{\rm A}_l
         (\omega) r^{-1} e^{-i\omega r^*}\, ,& \qquad r\rightarrow 2M,\\
        {\rm B}_l(\omega) r^{-1} e^{i\omega r^*}\, ,
         & \qquad r\rightarrow \infty, 
\end{array}\right.
\end{equation}
for the outgoing modes and
\begin{equation}
\stackrel{\leftarrow}{R}_l(\omega | r) \sim \left\{
\begin{array}{ll}
        {\rm B}_l(\omega) r^{-1} e^{-i\omega r^*}\, ,
        & \qquad r\rightarrow 2M,\\
        r^{-1} e^{-i\omega r^*} + \stackrel{\leftarrow}{\rm A}_l(\omega)
        r^{-1} e^{i\omega r^*}\, ,& \qquad r\rightarrow \infty, 
\end{array}\right.
\end{equation}
for the ingoing modes.  The normalization factors ${\rm B}_l(\omega)$, 
$\stackrel{\rightarrow}{\rm A}_l(\omega)$, and $\stackrel{\leftarrow}{\rm A}_l
(\omega)$ are the transmission and reflection coefficients
for the scalar field with an angular momentum-dependent potential
barrier.

Now let us consider the two-point function in the Boulware vacuum.  It is
given by 
\begin{equation}
G_B(x,x') =   \sum_{lm} \int_0^\infty {d\omega \over 4\pi\omega}
e^{-i\omega(t-t')}\,{\rm Y}_{lm}(\theta,\varphi) {\rm Y}_{lm}^*
(\theta',\varphi')\,
\left[\stackrel{\rightarrow}{R}_l(\omega | r)
\stackrel{\rightarrow}{R}_l^*(\omega | r') +\stackrel{\leftarrow}{R}
_l(\omega | r) \stackrel{\leftarrow}{R}_l^*(\omega | r')\right]\,.
\end{equation}
We are interested in the two-point function when the spatial separation
goes to zero, {\it i.e.}, letting $r'\rightarrow r$, $\theta' \rightarrow \theta$,
and $\varphi' \rightarrow \varphi$.  We can again make use of the 
addition theorem for the spherical harmonics, Eq.~(\ref{eq:sum_rule}).
Let us also Euclideanize, by taking $(t-t')\rightarrow
-2it_0$.  The Euclidean two-point function then reduces to
\begin{equation}
G_{BE}(2t_0) = {1\over 16\pi^2} \sum_l \int_0^\infty {d\omega\over\omega}\,
e^{-2t_0}\,(2l+1)\, \left[|\stackrel{\rightarrow}{R}_l(\omega | r)|^2
+|\stackrel{\leftarrow}{R}_l(\omega | r)|^2\right]\,.
\end{equation}
In the two asymptotic regimes, close to the event
horizon of the black hole ($r\rightarrow 2M$), or far from the
black hole ($r\rightarrow\infty$), the
radial portion of the wave equation also satisfies a sum rule.
It was found by Candelas \cite{Candelas} that
\begin{equation}
\sum_{l=0}^\infty (2l+1) |\stackrel{\rightarrow}{R}_l(\omega | r)|^2
\sim \left\{
\begin{array}{ll}
             4\omega^2(1- 2M/ r)^{-1},& \qquad 
             r\rightarrow 2M, \\
             r^{-2}\sum_{l=0}^\infty (2l+1) |{\rm B}_l(\omega)|^2  ,
             & \qquad r\rightarrow \infty, 
\end{array}\right.
\end{equation}
and
\begin{equation}
\sum_{l=0}^\infty (2l+1) |\stackrel{\leftarrow}{R}_l(\omega | r)|^2
\sim \left\{ 
\begin{array}{ll} 
             (2M)^{-2}\sum_{l=0}^\infty (2l+1) |{\rm B}_l(\omega)|^2,
             & \qquad  r\rightarrow 2M, \\
             4\omega^2  ,&  \qquad r\rightarrow \infty,
\end{array}\right.
\end{equation}
with the coefficient ${\rm B}_l(\omega)$ given, in the case
$2M\omega \ll 1$, by \cite{Jens92}
\begin{equation}
{\rm B}_l(\omega)\approx {(l!)^3\over (2l+1)!\,(2l)!}\,
(-4iM\omega)^{l+1}\, . \label{eq:Jenson} \end{equation}
If we insert these relations into the Green's functions,  it is
possible to carry out the integration in $\omega$.
In the near field limit
\begin{equation}
G_{BE}(2t_0) \sim {1\over 16\pi^2} \left[ {1\over(1-2M/ r) t_0^2}
+{1\over 4M^2} \sum_{l=0} {(l!)^6\over [(2l)!]^3}\,
\left({2M\over t_0}\right)^{2l+2}\right] , \qquad r\rightarrow 2M,
\label{eg:green_near}\end{equation}
and in the far field limit,
\begin{equation}
G_{BE}(2t_0) \sim {1\over 16\pi^2} \left[ {1\over t_0^2}+ {1\over r^2}
\sum_{l=0}{(l!)^6\over [(2l)!]^3}\,\left({2M\over t_0}\right)^{2l+2}
\right] , \qquad r\rightarrow \infty. \label{eq:green_far}
\end{equation}

We immediately see that the Green's function is independent of the
angular coordinates, as expected from the spherical symmetry.
Note that the maximum value of $l$ for which the expansion in
Eqs.~(\ref{eg:green_near}) and (\ref{eq:green_far}) can be used
depends upon the order of the leading terms which have been dropped
in Eq.~(\ref{eq:Jenson}).  If this correction is $O\left( (M\omega)^{l+2}
\right)$, then only the $l=0$ terms are significant, as $B_0$ would
then contain subdominant pieces which yield a contribution to
$G_{BE}(2t_0)$ larger than the leading contribution from $B_1$.
In what follows, we will explicitly retain only the $l=0$ contribution.
In order to find the quantum inequality around a black hole we must
evaluate
\begin{equation}
\Delta\hat\rho \geq -{1\over 4}\, \Box_E \; G_E(2t_0).
\end{equation}
However, the only parts of the Euclidean box operator
that are relevant are the temporal and radial terms, {\it i.e.},
\begin{equation}
\Box_E \Rightarrow (1-{2M/r})^{-1}\partial_{t_0}^2 + 
r^{-2}\partial_r [r^2(1-{2M/r})\partial_r].
\end{equation}
Upon taking the appropriate derivatives, and using the relation of
the proper time of a stationary observer to the coordinate time,
\begin{equation}
\tau_0 = t_0\,\sqrt{1-{2M/r}}\, , 
\end{equation}
we find that the quantum inequality is given by
\begin{equation}
\Delta \hat\rho \geq -{3\over 32\pi^2 \tau_0^4 }\left\{  
{1\over 6} \left(2M\over r\right)^2 \left( \tau_0 \over r \right)^2
\left(1-{2M \over r}\right)^{-1} + 1 +\left(1-{2M \over r}\right) 
+ O\left[ \left(1-{2M \over r}\right)^2 \right] + \cdots\right\},
\label{eq:near}
\end{equation}
as $ r\rightarrow 2M$ and
\begin{eqnarray}
\Delta \hat\rho &\geq& -{3\over 32\pi^2 \tau_0^4} \left\{
1 - {2M\over r} + \left({2M\over r}\right)^2 \left[ 1 +{1\over 3}
\left({\tau_0\over r}\right)^2\right] \right.\nonumber\\
&&\qquad\qquad \left. - \left({2M\over r}\right)^3
\left[ 1 +\left({\tau_0\over r}\right)^2\right] + 
 O\left[ \left({2M\over r}\right)^4\right] +\cdots \right\},
\label{eq:far}  
\end{eqnarray}
as $r\rightarrow\infty$.
An alternative approach to finding the quantum inequality is to use
the short time expansion from Section~\ref{sec:expans}, which yields
\begin{equation}
\Delta \hat\rho \geq -{3\over 32\pi^2 \tau_0^4} - {1\over 16\pi^2\tau_0^2}
\left[{M^2\over  r^4 (1- 2M/ r)} + O(\tau_0^2) + \cdots\right].
\label{eq:shortSchwarz}
\end{equation}

Note that this short time expansion coincides with the first two terms
of the $r\rightarrow 2M$ form, Eq.~(\ref{eq:near}).  This is somewhat
unexpected, as Eq.~(\ref{eq:near}) is an expansion for small $r-2M$ with
$\tau_0$ fixed, whereas Eq.~(\ref{eq:shortSchwarz}) is an expansion for
small $\tau_0$ with $r$ fixed.

We immediately see from Eq.~(\ref{eq:far}) 
that we recover the Minkowski space quantum inequality in the $r
\rightarrow \infty$ limit.   If we consider experiments performed
on the surface of the earth, where the radius of the earth
is several orders of magnitude larger than its equivalent Schwarzschild
radius, then the flat space inequality is an exceptionally good 
approximation. From Eq.~(\ref{eq:little_t}), we can also find the
proper sampling time for which the inequality Eq.~(\ref{eq:shortSchwarz})
holds,
\begin{equation} 
\tau_0 \ll {r^2\over 2M} \sqrt{2\left(1-{2M\over r}\right)}.
\end{equation}

As was the case in two dimensions, if we allow the sampling time
to go to infinity in the exact quantum inequality, we recover the 
QAWEC, Eq.~(\ref{eq:AWEC_boulware}), for the four-dimensional black
hole.  The QAWEC says that the renormalized energy density for an
arbitrary particle state, sampled over the entirety of the rest
observer's worldline, can never be more negative than the
Boulware vacuum energy density.

\chapter{A Dynamic Spacetime: The Warp Drive}\label{Chapt:Warp}

In the preceding chapters, we looked at various static spacetimes
in which quantum inequalities could be developed. Now we will apply
the quantum inequality restrictions for a scalar field to 
Alcubierre's warp drive metric  on a  scale in which a local region
of spacetime can be considered ``flat.'' This is accomplished by
taking the sampling time of the quantum inequality to be very small.
Over short timescales, the equivalence principle tells us
that the spacetime should be nearly static and almost flat. The short
sampling time expansion of the quantum inequality developed in
Section~\ref{sec:expans} tells us that on short enough time scales
we can apply the Minkowski space quantum inequality with reasonable
reliability.  From this we are able to place limits on the parameters
of the ``warp bubble.'' We will show that the bubble wall thickness
is on the order of only a few hundred Planck lengths. Additionally, we
will show that the total integrated energy density needed to maintain
the warp metric with such thin walls is physically unattainable.

\section{Introduction}
In both the scientific community and pop culture, humans have been
fascinated with the prospect of being able to travel between the stars
within their own lifetimes. Within the framework of special relativity,
the space-going traveler may move with any velocity up to, but not
including, the speed of light.  The astronauts would experience
a time dilation which would allow them to make the round trip from earth
to any star, and back to earth, in an arbitrarily short elapsed
time.  However, upon returning to earth such
observers would find that their family and friends would have aged 
considerably.  This is the well-known twin paradox
\cite{MTW, Wheeler90, Taylor}. 

Recently, Miguel Alcubierre proposed a metric  \cite{Alcu94}, fondly
called the warp drive, in which a spaceship could travel to a 
star a distance $D$ away and return home, such that the elapsed time for
the stationary observers on earth would be less than $2D/c$, where $c$ is
the velocity of light.  What is most surprising about this spacetime
is that the proper time of the space going traveler's trip is identical
to the elapsed time on earth.  However, the spaceship never
{\it locally} travels faster than the speed of light.  In fact, the
spaceship can sit at rest with respect to the interior of the warp bubble.
The ship is carried along by the spacetime, much in the same way that
the galaxies are receding away from each other at extreme speeds due to
the expansion of the universe, while locally they are at rest.  The
warp drive utilizes this type of expansion and contraction in order
to achieve the ability to travel faster than light.

Although warp drive sounds appealing, it does have one serious drawback.
As with traversable wormholes, in order to achieve warp drive we must
employ exotic matter, that is, negative energy densities, which are a
violation of the classical energy conditions. We have seen that the 
quantum inequality restrictions do allow negative energy densities to
exist \cite{F&Ro92, F&Ro95, Ford78, Ford91, F&Ro97}. However, they
place serious limitations on their magnitude and duration. Even before
the short sampling time expansion of the quantum inequality was
developed, researchers had applied the flat space quantum inequality
to the curved spacetime geometries of wormholes \cite{F&Ro96} with 
the restriction that the negative energy be sampled on timescales
smaller than the minimum local radius of
curvature.  It was argued that over such small sampling times, the 
spacetime would be locally flat and the inequalities would be valid.
This led to the conclusion that static wormholes must either be on
the order of several Planck lengths in size, or there would be large
discrepancies in the length scales that characterize the wormhole.  

As we have seen in the preceding chapter, exact quantum inequalities 
can be developed for the static spacetimes in two, three, and four
dimensions \cite{Pfen97a}.  It was found that the quantum inequalities
take the flat space form modified by a scale function which depends
on the sampling time, the local radius of curvature, and the mass of 
the scalar field.  In the limit of the sampling time being smaller
than the local radius of curvature, the quantum inequalities reduce to
the flat space form, often accompanied by higher order corrections due
to the curvature \cite{Pfen97a, Pfen98a}. In the limit of the radius of
curvature going to infinity,  the flat space inequalities are recovered.

We would like to apply the same method to the warp drive metric, but
such an exercise would require that we know the solutions to the
Klein-Gordon equation for the mode functions of the scalar field.  Such
an approach, although exact, would be exceptionally difficult.  In this
section we will therefore apply the flat space inequality directly to the
warp drive metric
but restrict the sampling time to be small.  We will thereby be
able to show that the walls of the warp bubble must be exceedingly
thin compared to its radius.  This constrains the negative energy
to an exceedingly thin band surrounding the spaceship. Similar results
have been demonstrated for wormholes \cite{F&Ro96}, where the negative
energy is concentrated in a thin band around the throat.  Recently, 
it has been shown for the Krasnikov metric \cite{Kras95}, which also
allows superluminal travel, that the required negative energy is also
constrained to a very thin wall \cite{E&Ro97}. We will then 
calculate the total negative energy that would be required to generate
a macroscopic sized bubble capable of transporting humans.  As we will
see, such a bubble would require physically unattainable energies.
 
\section{Warp Drive Basics}
Let us discuss some of the basic principles of the warp drive spacetime.
We begin with a flat (Minkowski) spacetime and then consider a small
spherical region, which we will call the bubble, inside this spacetime.
On the forward edge of the bubble, we cause spacetime to contract, and
on the trailing edge is an equal spacetime expansion.  The region inside
the bubble, which can be flat, is therefore transported forward with
respect to distant objects.  Objects at rest inside the bubble are
transported forward  with the bubble, even though they have no (or
nominal) local velocity.  Such a spacetime is described by the 
Alcubierre warp drive metric
\begin{equation}
ds^2 = -dt^2 + [dx - v_s(t)f(r_s(t))dt]^2 +dy^2+dz^2,
\label{eq:warp_metric}
\end{equation}
where $x_s(t)$ is the trajectory of the center of the bubble and
$v_x(t)= dx_s(t) / {dt}$ is the bubble's velocity. The variable 
$r_s(t)$ measures the distance outward from the center of the bubble
given by
\begin{equation}
r_s(t) = \sqrt{(x-x_s(t))^2 +y^2+z^2}.
\end{equation}
The shape function of the bubble is given by $f(r_s)$, which
Alcubierre originally chose to be
\begin{equation}
f(r_s) = {{\tanh[\sigma(r_s -R)] - \tanh[\sigma(r_s +R)]}\over 
2 \tanh[\sigma\; R]}. \label{eq:alcub}
\end{equation}
The variable $R$ is the radius of the warp bubble, and $\sigma$ is a 
free parameter which can be used to describe the thickness of the
bubble walls. In the large $\sigma$ limit, the function $f(r_s)$ quickly
approaches that of a top hat function, where $f(r_s) = 1$ for $r_s 
\leq R$ and zero everywhere else.  It is not necessary to choose
a particular form of $f(r_s)$. Any function will suffice so long as it
has the value of approximately 1 inside some region  $r_s<R$ and
goes to zero rapidly outside the bubble, such that as $r_s \rightarrow 
\infty$ we recover Minkowski space. In order to make later calculations
easier, we will use the piecewise continuous function
\begin{equation}
f_{p.c.}(r_s) = \left\{\begin{array}{ll}
     1 & r_s < R-{\Delta\over2},\\
     -{1\over\Delta}(r_s -R-{\Delta\over2})\qquad & R-{\Delta\over2}<r_s<
      R+{\Delta\over2},\\
      0&r_s > R+{\Delta\over2},
\end{array} \right.
\label{eq:f_approx}
\end{equation}
where $R$ is the radius of the bubble.  The variable $\Delta$ is the
bubble wall thickness.  It is chosen to relate to the parameter
$\sigma$ for the Alcubierre form of the shape function by setting the
slopes of the functions $f(r_s)$ and $f_{p.c.}(r_s)$ to be equal at
$r_s=R$. This leads to
\begin{equation}
\Delta = {\left[1+\tanh^2(\sigma R)\right]^2\over
 2\;\sigma\;\tanh(\sigma R)}\;,
\end{equation}
which in the limit of large $\sigma R$ can be approximated
by $\Delta \simeq 2/\sigma$.  

\begin{figure}
\begin{center}
\leavevmode\epsfxsize=0.6\textwidth\epsffile{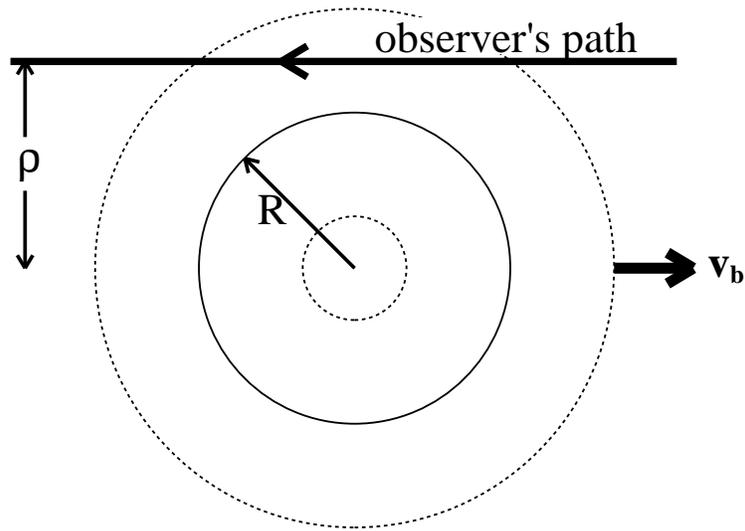}
\end{center}
\caption[Path through a warp bubble]
{The path of an observer who passes through the outer region
of the bubble, shown in the bubble's rest frame.  As viewed from the 
interior of the bubble, the observer is moving to the left.}
\label{fig:path}
\end{figure}

\begin{figure}
\begin{center}
\leavevmode\epsfxsize=0.5\textwidth\epsffile{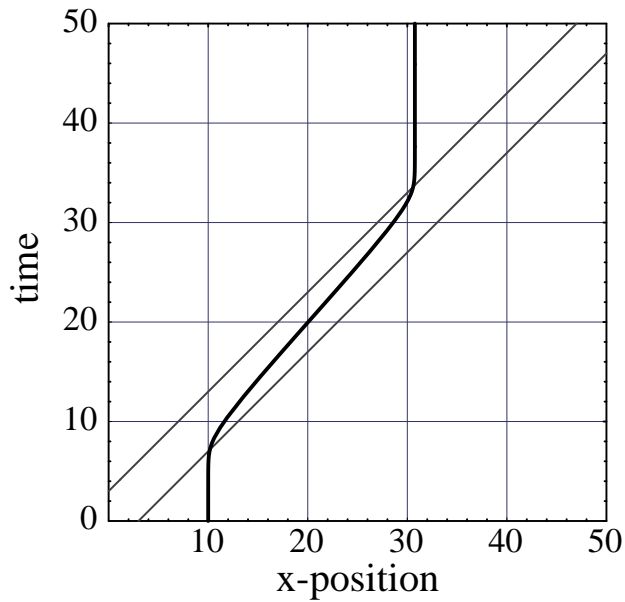}
\end{center}
\caption[Worldline of an observer passing through a warp bubble wall]
{The worldline (the dark line) of the geodesic observer
passing through the outer region of a warp bubble, plotted in the
observer's initial rest frame.  The two lighter diagonal lines are
the worldlines of the center of the bubble wall on the front and
rear edges of the bubble, respectively. The bubble has a radius
of 3, a velocity of 1, and a $\sigma$ of 1. The
plot shows an observer who begins at rest at $x=10$, $y^2+z^2 =
\rho^2 = 4$.  The shape function is of the form given by Alcubierre,
Eq.~(\ref{eq:alcub}).}
\label{fig:geodesics}
\end{figure}

We now turn our attention to the solutions of the geodesic equation.
It is straightforward to show that  
\begin{equation}
{dx^\mu \over dt }= u^\mu = \left( 1, v_s(t)f(r_s(t)),0,0 \right),\qquad
u_\mu = (-1,0,0,0),
\end{equation}
is a first integral of the geodesic equations.  Observers with this
four-velocity are called the Eulerian observers by Alcubierre.  We see
that the proper time and the coordinate time are the same for all 
observers. Also, the y and z components of the four-velocity are zero.
The bubble therefore exerts no ``force'' in the directions perpendicular
to the direction of travel. In Figure~\ref{fig:path}, we have plotted one
such trajectory for a observer that passes through the wall of a warp bubble
at a distance $\rho$ away from the center of the bubble. The x-component
of the four-velocity is dependent on the shape function, and solving this
explicitly for all cases can be rather difficult due to the time
dependence of $r_s(t)$. A spacetime plot of an observer with the
four-velocity given above is shown in Figure~\ref{fig:geodesics} 
for a bubble with constant velocity.


We see that the Eulerian observers are initially at rest.  As the
front wall of the bubble approaches, the observer begins to
accelerate in the direction of travel of the bubble, relative
to observers at large distances.  Once inside the bubble
the observer moves with a nearly constant velocity given by
\begin{equation}
\left.{dx(t)\over dt}\right|_{max.} = v_s(t_\rho) f(\rho),
\label{eq:Vmax}
\end{equation}
which will always be less than the bubble's velocity unless $\rho 
= (y^2+z^2)^{1/2} = 0$. The time $t_\rho$ is defined by $r_s(t_\rho)
= \rho$, {\it i.e.}, it is the time at which the observer reaches the bubble 
equator.  Such observers then decelerate, and are left at rest as
they pass from the rear edge of the bubble wall.  In other words no
residual momentum is imparted to these observers during the ``collision.''
However, they have been displaced forward in space along the trajectory
of the bubble.

There is also another interesting feature of these geodesics.
As already noted, the observers will move with a nearly constant
velocity through the interior of the bubble. This holds true for
any value of $\rho$.  However, the velocity is still dependent
upon the value of $\rho$, so observers at different distances 
from the center of the bubble will be moving with different
velocities relative to one another. 
If a spaceship of finite size is placed inside the bubble
with its center of mass coincident with the center
of the bubble, then the ship will experience a net ``force'' pushing it
opposite to the direction of motion of the bubble so long as $1-f(r_s)$
is nonzero at the walls of the ship. 
The ship would therefore have to use its engines to maintain its
position inside the bubble.  In addition, the ship would be subject
to internal stresses on any parts that extended sufficiently far away  
from the center of the bubble

In the above discussion we have used the Alcubierre form of the shape
function, $f(r_s)$.  If the piecewise continuous form,
Eq.~(\ref{eq:f_approx}),  is used, one
finds similar results with some modification.  Inside the bubble, where
$r_s < (R-{\Delta/2})$, every observer would move at exactly the speed
of the bubble.  So any observer who reached the bubble interior would
continue on with it forever.  This arises because everywhere
inside the bubble, spacetime is perfectly flat due to $f(r_s) = 1$.
For observers whose geodesics pass solely through the bubble walls,
so $(R-{\Delta/2}) < \rho < (R+{\Delta/2})$, the result is more or
less identical to that of the geodesics found with the Alcubierre shape
function.  This is the most interesting region because it 
contains the largest magnitude of negative energy.
 
\begin{figure}
\begin{center}
\leavevmode\epsfxsize=0.8\textwidth\epsffile{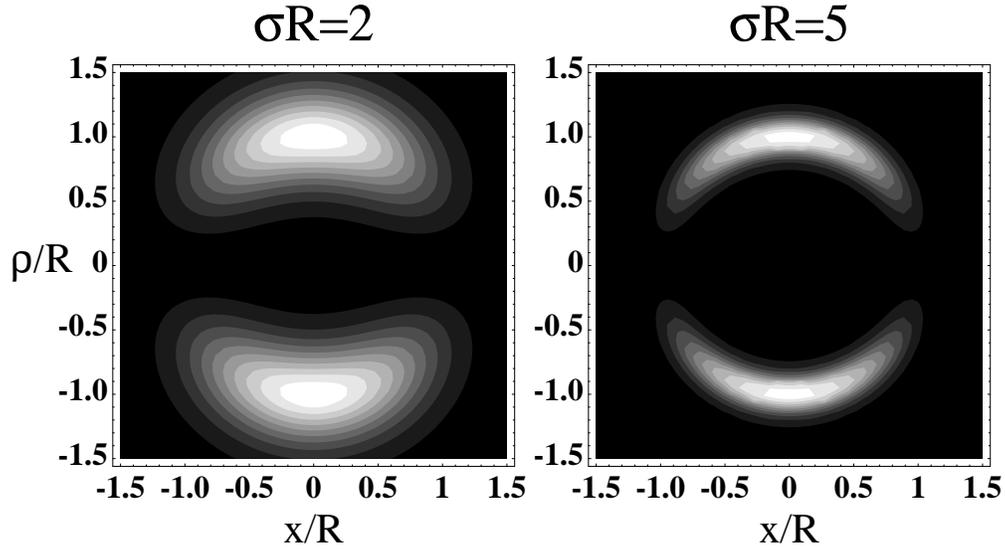}
\end{center}
\caption[Plot of the negative energy in a warp drive]
{The negative energy density is plotted for a longitudinal
cross section of the warp metric traveling at constant velocity
$v_s = 1$ to the right for the Alcubierre shape function.  Black
regions are devoid of matter, while white regions are maximal
negative energy.}
\label{fig:Energy_density}
\end{figure}

We now turn our attention to the energy density distribution of the
warp drive metric.  Using the first integral of the geodesic equations,
it is easily shown that
\begin{equation}
\langle T^{\mu\nu}u_\mu u_\nu\rangle 
 = \langle T^{tt} \rangle = {1\over 8\pi}G^{tt} = -{1\over 8\pi}
{v_s^2(t) \rho^2 \over 4 r_s^2(t) }\left(d f(r_s)\over dr_s
\right)^2,
\label{eq:EnergyDensity}
\end{equation}
where $\rho = [y^2 + z^2]^{1/2}$, is the radial distance perpendicular
to the $x$-axis defined above. We immediately see that the energy
density measured by any Eulerian observer is always negative, as 
shown in Alcubierre's original paper \cite{Alcu94}.  In 
Figure~\ref{fig:Energy_density}, we
see that the distribution of negative energy is concentrated in a
toroidal region perpendicular to the direction of travel.

In Section~\ref{sec:tot_Energy} we will integrate the energy density 
over all of space to obtain the total negative energy required to
maintain the bubble, under the restrictions of the quantum inequalities. 
As we will show, the total energy is physically unrealizable in the
most extreme sense.  
 
\section{Quantum Inequality Restrictions}
We begin with the quantum inequality (QI) for a free, massless
scalar field in four-dimensional Minkowski spacetime originally 
derived by Ford and Roman \cite{F&Ro95},  
\begin{equation}
{\tau_0\over\pi}\int_{-\infty}^\infty {{\langle T_{\mu\nu}u^\mu u^\nu
 \rangle}\over{\tau^2 + \tau_0^2}}d\tau\geq -{3\over{32\pi^2 \tau_0^4}}
 \; ,\label{eq:original}
\end{equation}
where $\tau$ is an inertial observer's proper time, and $\tau_0$ 
is an arbitrary sampling time. It has been argued by Ford and
Roman \cite{F&Ro96} that one may apply the QI to non-Minkowski
spacetimes if the sampling time is smaller than the local radius
of curvature. We know this to be true. It comes from the small sampling
time expansion of the QI developed in Section~\ref{sec:expans}.

We begin by taking the expression for the energy density (\ref
{eq:EnergyDensity}), and inserting it into the quantum inequality,
Eq.~(\ref{eq:original}), yielding
\begin{equation}
t_0 \int_{-\infty}^{+\infty} {v_s(t)^2 \over r_s^2} \left({df(r_s)\over
dr_s}\right)^2 {dt\over {t^2+t_0^2}} \leq {3\over \rho^2 t_0^4} \, .
\label{eq:explt_form}
\end{equation}
If the time scale of the sampling is sufficiently small compared
to the time scale over which the bubble's velocity is changing, then
the warp bubble's velocity can be considered roughly constant,
$v_s(t) \approx v_b$, during the sampling interval. 
We can now find the form of the geodesic at the time the sampling is
taking place. Because of the small sampling time, the
$[t^2 +t_0^2]^{-1}$ term becomes strongly peaked, causing the QI
integral to sample only a small portion of the geodesic.  We therefore
place the observer at the equator of the warp bubble at
$t = 0$. Then the geodesic is well-approximated by
\begin{equation}
x(t) \approx f(\rho) v_b t \; ,
\end{equation}
which results in
\begin{equation}
r_s(t) = \left[ (v_b t)^2 (f(\rho) - 1)^2 + \rho^2 \right]^{1/2}\;.
\end{equation}

Finally, we must specify the form of the shape function of the bubble.
We can expand any shape function as a Taylor series about the sampling
point, $r_s(t) \rightarrow \rho$. Then we take the appropriate
derivatives to obtain the term needed for the quantum inequality. 
We find
\begin{equation}
{df(r_s)\over dr_s} \approx f'(\rho) + f''(\rho)[r_s(t) - \rho]+\cdots.
\end{equation}
The leading term is the slope of the shape function at the sampling
point, which is in general roughly proportional to the  inverse of
the bubble wall thickness.  We can therefore use, with no loss of
generality, the piecewise continuous form of the shape function 
(\ref{eq:f_approx}) to obtain a good order of magnitude 
approximation for any choice of shape function.
The quantum inequality (\ref{eq:explt_form}) then becomes
\begin{equation}
t_0 \int_{-\infty}^{+\infty} {dt \over (t^2 + \beta^2) (t^2+t_0^2)} 
\leq {3 \Delta^2 \over v_b^2 \, t_0^4 \, \beta^2} ,
\end{equation}
where
\begin{equation}
\beta = {\rho \over {v_b (1 - f(\rho) ) }}\, .
\end{equation}
Formally the integral should not be taken over all time but just the
time the observer is inside the bubble walls.  However, the sampling 
function rapidly approaches zero.  Therefore contributions to the
integral from the distant past or the far future are negligible.  The
integral itself can be performed as the principal value of a contour that
is closed in the upper half of the complex plane, 
\begin{equation}
\int_{-\infty}^{+\infty} {dt \over (t^2 + \beta^2) (t^2+t_0^2)}
 = {\pi \over t_0 \;\beta\; (t_0 + \beta)},
\end{equation}
yielding an inequality of
\begin{equation}
{\pi \over 3} \leq {\Delta^2 \over {v_b^2 \; t_0^4}} 
\left[{v_b t_0 \over \rho} (1-f(\rho)) +1\right] \; .
\end{equation}

The above inequality is valid only for sampling times on which the
spacetime may be considered approximately flat.  We must therefore
find some characteristic length scale below which this occurs.  For
an observer passing through the bubble wall at a distance $\rho$ from
the center, the Riemann tensor in the static background frame can be
calculated. Then we transform the components to the observer's
frame by use of an orthonormal tetrad of unit vectors.  In this
frame, the tetrad is given by the velocity vector $u^\mu (t)$ and
three unit vectors $\hat x$, $\hat y$, and $\hat z$. The largest
component of the Riemann tensor in the orthonormal frame is given by
\begin{equation}
|R_{{\hat t}{\hat y}{\hat t}{\hat y}}| = 
{3 v_b^2\; y^2 \over 4\;\rho^2}\left[ {d f(\rho)
\over d\rho} \right]^2.
\end{equation}
This yields
\begin{equation}
r_{min} \equiv {1\over \sqrt{|R_{{\hat t}{\hat y}{\hat t}{\hat y}}|}}
\sim {2\Delta \over {\sqrt 3}\; v_b}\; ,
\end{equation}
when $y= \rho$ and the piecewise continuous form of
the shape function is used.  The sampling time must be smaller than
this length scale, so we take
\begin{equation}
t_0 = \alpha {2\Delta \over {\sqrt 3}\; v_b}, \qquad 0<\alpha\ll 1 .
\end{equation}
Here $\alpha$ is the ratio of the sampling time to the
minimal radius of curvature. If we insert this into the quantum
inequality and use
\begin{equation}
{\Delta \over \rho} \sim {v_b t_0 \over \rho} \ll 1 \; ,
\end{equation}
we may neglect the term involving $1-f(\rho)$ to find
\begin{equation}
\Delta \leq {3\over 4}\sqrt{3\over\pi}\;{v_b \over \alpha^2}\, .
\end{equation}
Now as an example, if we let $\alpha = 1/10$, then
\begin{equation}
\Delta \leq 10^2\, v_b\; L_{Planck}\, ,
\label{eq:wall_thickness}
\end{equation}
where $L_{Planck}$ is the Planck length.  Thus, unless $v_b$ is
extremely large, the wall thickness cannot be much above the Planck
scale. Typically, the walls of the warp bubble are so thin that the
shape function could be considered a ``step function'' for most purposes.

\section{Total Energy Calculation}\label{sec:tot_Energy}
We will now look at the total amount of negative energy that is
involved in the maintenance of a warp metric.  For simplicity, let us
take a bubble that moves with constant velocity such that $x_s(t)
= v_b \,t$. Because the total energy is constant, we can calculate it
at time $t=0$. We then have
\begin{equation}
r_s(t=0) = [x^2+y^2+z^2]^{1\over 2} = r.
\end{equation}
With this in mind we can write the integral of the local matter energy
density over proper volume as
\begin{equation}
E = \int dx^3 \sqrt{|g|}\;\, \langle T^{tt}\rangle = -{v_b^2 \over 32\pi}
\int {\rho^2 \over r^2} \left(d f(r)\over dr \right)^2 dx^3 \; ,
\end{equation}
where $g = {\rm Det}|g_{ij}|$ is the determinant of the spatial metric
on the constant time hypersurfaces. Portions of this integration can be
carried out by making a transformation to spherical coordinates.  By
doing so, we find
\begin{equation}
E = - {1\over 12} v_b^2 \int_0^\infty r^2
\left( d\;f(r)\over dr\right)^2 dr\,.
\end{equation}
Since we are making only order of magnitude estimates of the
total energy, we will use a piecewise continuous approximation
to the shape function given by Eq.~(\ref{eq:f_approx}). 
Taking the derivative of this shape function, we find that the
contributions to the energy come only from the bubble wall region,
and we end up evaluating
\begin{eqnarray}
E &=&- {1\over 12} v_b^2 \int_{R-{\Delta\over 2}}^{R+{\Delta\over 2}} 
r^2 \left(- 1 \over \Delta\right)^2 dr,\nonumber\\[6pt]
&=&- {1\over 12} v_b^2\left( {R^2 \over\Delta}+{\Delta\over 12}\right).
\label{eq:energy}
\end{eqnarray}
For a macroscopically useful warp drive, we want the radius of the
bubble to be at least in the range of 100 meters so that we may
fit a ship inside.  We showed in the previous section that
the wall thickness is constrained by (\ref{eq:wall_thickness}).
If we use this constraint and let the bubble radius be equal to
100 meters, then we may neglect the second term on the right-hand-side
of Eq.~(\ref{eq:energy}).  
It follows that
\begin{equation}
E \leq -6.2\times10^{70}\, v_b \; L_{Planck} \; \sim\;
-6.2\times10^{65}\, v_b \; {\rm grams}.
\end{equation}
Because a typical galaxy has a mass of approximately
\begin{equation}
M_{Milky Way} \approx 10^{12}\; M_{sun} = 2\times 10^{45} {\rm grams},
\end{equation}
the energy required for a warp bubble is on the order of 
\begin{equation}
E \leq - 3 \times 10^{20} \; M_{galaxy} \; v_b\; .
\end{equation}
This is a fantastic amount of negative energy, roughly ten orders of
magnitude greater than the total mass of the entire visible universe.

If it were possible to violate the quantum inequality restrictions and
make a bubble with a wall thickness on the order of a meter, things would
be somewhat improved.  The total energy required in the case of the
same size radius and $\Delta = 1$ meter would be on the order
of a quarter of a solar mass, which would be more practical, yet
still not attainable.

\section{Summary}
We see from Eq.~(\ref{eq:wall_thickness}), that quantum
inequality restrictions on the warp drive metric constrain the
bubble walls to be exceptionally thin.  Typically, the walls are on
the order of only hundreds or thousands of Planck lengths.
Similar constraints on the size of the negative energy region
have been found in the case of traversable wormholes \cite{F&Ro96}.

From Eq.~(\ref{eq:wall_thickness}), we may think that by making the
velocity of the bubble very large we can make the walls thicker.
However, this causes another problem.  For every order of magnitude
by which the velocity increases, the total negative energy required to
generate the warp drive metric also increases by the same magnitude.
It is evident that for macroscopically sized bubbles to be useful
for human transportation, even at subluminal speeds, the required
negative energy is physically unattainable.  

On the other hand, we may consider the opposite regime.  Warp bubbles
are still conceivable if they are very tiny, {\it i.e.}, much less than the
size of an atom.  Here the difference in length scales is not as great.
As a result, a smaller amount of negative energy is required to maintain
the warp bubble.  For example, a bubble with a radius of 
one electron Compton wavelength would require
a negative energy of order $E \sim - 400 M_{sun}$.

The above derivation assumes that we are using a quantized, massless
scalar field to generate the required negative energy.  Similar
quantum inequalities have been proven for both massive scalar fields
\cite{Pfen97a,F&Ro97} and the electromagnetic field \cite{F&Ro97}.
In the case of the massive scalar field, the quantum inequality
becomes even more restrictive, thereby requiring the bubble walls
to be even thinner.  For the quantized electromagnetic field, the
wall thickness can be made larger by a factor of $\sqrt{2}$, due
to the two spin degrees of freedom of the photon.  However this is
not much of an improvement over the scalar field case.

\appendix
\chapter{Inequalities}\label{sec:appendix}
In this appendix, we prove the following inequality: 
Let $A_{ij}$ be a real, symmetric $n \times n$ matrix with
non-negative eigenvalues. (For the purposes of this manuscript,
we may take either $n=2$, for three-dimensional spacetimes, or $n=3$, 
for four-dimensional spacetimes.)
Further let $h_\lambda^i$ be a complex $n$-vector, which is also a
function of the mode label $\lambda$. Then in an arbitrary quantum
state $|\psi \rangle$, the inequality states that
\begin{equation}
{\rm Re}\sum_{\lambda,\lambda'} A_{ij} 
\left[ h^{i\,*}_\lambda h^j_{\lambda'} 
             \langle a_\lambda^\dagger a_{\lambda'} \rangle
\pm h^i_\lambda h^j_{\lambda'} \langle a_\lambda a_{\lambda'}\rangle
\right]
\geq -{1\over 2}\sum_{\lambda} A_{ij} h^{i\,*}_\lambda h^j_\lambda \,.
\label{eq:lowbound}
\end{equation}
In order to prove this relation, we first note that 
\begin{equation}
A_{ij} = \sum_{\alpha =1}^n \kappa_\alpha V_i^{(\alpha)}\,V_j^{(\alpha)}\,,
\end{equation}
where the $V_i^{(\alpha)}$ are the eigenvectors of $A_{ij}$, and the
$\kappa_\alpha \geq 0$ are the corresponding eigenvalues. Now define the
hermitian vector operator
\begin{equation}
Q^i  =  \sum_{\lambda} \left( h^{i\,*}_\lambda a_\lambda^\dagger 
+ h^i_\lambda a_\lambda  \right) \,.
\end{equation}
Note that
\begin{equation}
\left\langle {Q^i}^\dagger A_{ij} Q^j \right\rangle =
\sum_{\alpha =1}^n \kappa_\alpha 
\left\langle Q^{i\,\dagger}  V_i^{(\alpha)}\,V_j^{(\alpha)} Q^j \right\rangle =
\sum_{\alpha =1}^n \kappa_\alpha ||V_i^{(\alpha)}\, Q^i |\psi\rangle ||^2
\geq 0 \,.
\end{equation}
Furthermore,
\begin{equation}
\left\langle Q^{i\, \dagger} A_{ij} Q^j \right\rangle =
2\,{\rm Re}\sum_{\lambda,\lambda'} A_{ij} 
\left[ h^{i\,*}_\lambda h^j_{\lambda'} \langle a_\lambda^\dagger 
a_{\lambda'}\rangle
+ h^i_\lambda h^j_{\lambda'} \langle a_\lambda a_{\lambda'}\rangle \right]
+ \sum_{\lambda} A_{ij} h^{i\, *}_\lambda h^j_\lambda \, ,
\end{equation}
from which Eq.~(\ref{eq:lowbound}) with the `$+$'-sign follows immediately.
The form of Eq.~(\ref{eq:lowbound}) with the `$-$'-sign can be obtained
by letting $h^j_\lambda \rightarrow i h^j_\lambda$.

As a special case, we may take $A_{ij} = \delta_{ij}$ and obtain
\begin{equation}
{\rm Re}\sum_{\lambda,\lambda'} 
\left[ {\bf h}_\lambda^*\cdot {\bf h}_{\lambda'} 
\langle a_\lambda^\dagger a_{\lambda'}\rangle
\pm {{\bf h}_\lambda}\cdot {\bf h}_{\lambda'}
     \langle a_\lambda a_{\lambda'}\rangle \right]
\geq -{1\over 2}\sum_{\lambda} |{\bf h}_\lambda|^2 \,.
\label{eq:lowbound2}
\end{equation}
As a further special case, we may take the vector ${\bf h}_\lambda$ to
have only one component, {\it e.g.}, ${\bf h}_\lambda = (h_\lambda,0,0)$, in
which case we obtain
\begin{equation}
{\rm Re}\sum_{\lambda,\lambda'} 
\left[ {h_\lambda}^*  h_{\lambda'} 
\langle a_\lambda^\dagger a_{\lambda'}\rangle
\pm {h_\lambda} h_{\lambda'}
     \langle a_\lambda a_{\lambda'}\rangle \right]
\geq -{1\over 2}\sum_{\lambda} |h_\lambda|^2 \,.
\label{eq:lowbound3}
\end{equation}
This last inequality was originally proven in  \cite{Ford91} for real
$h_\lambda$, and a simplified proof using the method adopted here is given
in \cite{F&Ro97}.


\chapter{Gamma Functions}\label{chpt:gammas}
Here we discuss the following set of relations for gamma
functions of complex argument,
\begin{eqnarray}
\left | \Gamma(ik/2) \right |^2 &=&{\pi\over k/2 \sinh(\pi k /2)}\, ,
\label{eq:G1} \\
\left | \Gamma(1/2 + ik/2) \right |^2 &=& {\pi\over \cosh(\pi k /2)}\, ,
\label{eq:G2}\\
\left | \Gamma(1 + ik/2) \right |^2 &=& {\pi k/2 \over \sinh(\pi k/2)}\, ,
\label{eq:G3}\\
\left | \Gamma(3/2 + ik/2) \right |^2 &=& {\pi\over 2} \left(1+k^2\right)
{\cosh(\pi k /2) \over \cosh(\pi k) +1}\, ,\label{eq:G4}\\
\left | \Gamma(2 + ik/2) \right |^2 &=& {\pi\over 4} k\,\left(4+k^2\right)
{\sinh(\pi k /2) \over \cosh(\pi k) -1}\,.\label{eq:G5}
\end{eqnarray}
In addition we will prove the last two,  Eqs.~(\ref{eq:G4})
and (\ref{eq:G5}).  

The first three gamma function relations, Eqs.~(\ref{eq:G1}-\ref{eq:G3}),
can be found in any mathematics text which includes special functions,
such as Gradshteyn and Ryzhik \cite{Gradshteyn}, Section~8.33.  Also,
we note an important property for gamma functions of complex
argument that is often not stated,
\begin{equation}
\Gamma(x+iy) \Gamma(x-iy) = \left | \Gamma(x+iy) \right |^2 ,
\qquad \mbox{for }x\mbox{ and }y\mbox{ real},
\end{equation}
namely that the complex conjugate of a gamma
function is equal to the gamma function of the complex conjugate of
the argument.

The last two gamma function relations, Eqs.~(\ref{eq:G4}) and
(\ref{eq:G5}) are found from Eq.~(8.332.4) of Gradshteyn and Ryzhik,
which we repeat here:
\begin{equation}
\Gamma(1+x+iy) \Gamma(1-x+iy) \Gamma(1+x-iy) \Gamma(1-x-iy) =
{2\pi^2 (x^2 + y^2) \over \cosh 2y\pi - \cos 2x\pi},
\end{equation}
for both $x$ and $y$ real.  Using the relation for complex
conjugates of Gamma functions we can rewrite the above expression as
\begin{equation}
\left | \Gamma(1+x+iy) \right | ^2 = {1\over \left |\Gamma(1-x+iy)
\right | ^2} \times {2\pi^2 (x^2 + y^2) \over \cosh 2y\pi - \cos 2x\pi}\,.
\label{eq:intermediate_step}
\end{equation}
In order to obtain relation~(\ref{eq:G4}), we take $x = 1/2$ and
$y = k/2$.  We then have
\begin{equation}
\left | \Gamma(3/2 + ik/2) \right |^2 =  {1\over \left |\Gamma(1/2+ik/2)
\right | ^2}\times {2\pi^2 \left[(1/2)^2 + (k/2)^2\right] \over 
1 + \cosh \pi k}\, .
\end{equation}
Inserting Eq.~(\ref{eq:G2}) into the above expression, we
arrive at relation~(\ref{eq:G4}).  Analogously, if we take
$x=1$ in Eq.~(\ref{eq:intermediate_step}) we obtain 
relation~(\ref{eq:G5}).

\bibliography{Articles,Books,Tufts}
\bibliographystyle{ieeetr}

\end{document}